\documentclass[sigconf,screen,nonacm,review=false,timestamp=false,printacmref=false]{acmart}
\AtBeginDocument{%
  \providecommand\BibTeX{{%
    \normalfont B\kern-0.5em{\scshape i\kern-0.25em b}\kern-0.8em\TeX}}}

\setcopyright{acmcopyright}
\copyrightyear{2022}
\acmYear{2022}
\acmDOI{XXXXXXX.XXXXXXX}

\acmConference[Conference acronym 'XX]{Make sure to enter the correct
  conference title from your rights confirmation emai}{June 03--05,
  2018}{Woodstock, NY}
\acmPrice{15.00}
\acmISBN{978-1-4503-XXXX-X/18/06}

\usepackage{latexsym}

\usepackage[utf8]{inputenc}
\usepackage[T1]{fontenc}

\usepackage{amsxtra}
\usepackage{thm-restate}
\usepackage{stackengine}
\usepackage{url}
\usepackage{cleveref}
\usepackage{graphicx}
\usepackage{mathtools, cuted}
\usepackage{nicefrac}       
\usepackage{microtype}      
\usepackage{enumerate,dsfont,mathrsfs}
\usepackage{comment}
\usepackage{prettyref,soul,xspace}
\usepackage{paralist}
\usepackage{caption}
\usepackage{subcaption}
\usepackage[page]{appendix}
\usepackage{booktabs}       %
\usepackage{multirow,makecell}
\usepackage{pifont}
\usepackage[frozencache,cachedir=minted_cache]{minted}
\usepackage{wrapfig}
\usepackage{dblfloatfix}
\setminted{%
    fontsize=\tiny,%
    autogobble=true,%
    breakanywhere=true,%
    breaklines=true,%
    breaksymbol=,%
    breakanywheresymbolpre=,%
    }%
\makeatletter
\newenvironment{tabminted}{%
    \let\FV@ListVSpace\relax  
    \minted
}{%
    \endminted
    \unskip   
    \aftergroup\@tabmintedend
}
\newcommand*{\tabminted@finalstrut}[1]{%
    \ifdim\prevdepth>0pt
        \ifdim\dp#1>\prevdepth
            \vskip\dimexpr(\dp#1)-\prevdepth\relax
        \fi
    \else
        \vskip\dimexpr(\dp#1)\relax
    \fi
}
\newcommand*{\@tabmintedend}{%
    \let\@finalstrut\tabminted@finalstrut
}
\makeatother

\def\textrightarrow{$\rightarrow$}
\newcommand{\reviewer}[3]{
  \expandafter\newcommand\csname #1\endcsname[1]{
    \ifthenelse{\equal{\final}{1}} {
      \textcolor{#3}{}
    } {
    \textcolor{#3}{[\textsf{\footnotesize \textbf{#2:} ##1]}}
    }
  }
}

\crefname{equation}{equation}{equations}
\Crefname{equation}{Equation}{Equations}
\crefname{theorem}{theorem}{theorems}
\Crefname{theorem}{Theorem}{Theorems}
\crefname{assumption}{assumption}{assumptions}
\Crefname{assumption}{Assumption}{Assumptions}
\crefname{lemma}{lemma}{lemmas}
\Crefname{lemma}{Lemma}{Lemmas}
\crefname{definition}{definition}{definitions}
\Crefname{definition}{Definition}{Definitions}
\crefname{corollary}{corollary}{corollaries}
\Crefname{corollary}{Corollary}{Corollaries}
\crefname{proposition}{proposition}{propositions}
\Crefname{proposition}{Proposition}{Propositions}
\crefname{claim}{claim}{claims}
\Crefname{claim}{Claims}{Claims}
\crefname{problem}{problem}{problems}
\Crefname{problem}{Problem}{Problems}
\crefname{solution}{solution}{solutions}
\Crefname{solution}{Solution}{Solutions}
\crefname{proof}{proof}{proofs}
\Crefname{proof}{Proof}{Proofs}
\crefname{proofof}{proof}{proofs}
\Crefname{proofof}{Proof}{Proofs}
\crefname{algocf}{algorithm}{algorithms}
\Crefname{algocf}{Algorithm}{Algorithms}

\newcommand{\btcb}{\begin{tcolorbox}}
\newcommand{\etcb}{\end{tcolorbox}}
\newcommand{\bbm}{\begin{bmatrix}}
\newcommand{\ebm}{\end{bmatrix}}
\newcommand{\bassume}{\begin{assumption}}
\newcommand{\eassume}{\end{assumption}}
\newcommand{\be}{\begin{equation}}
\newcommand{\ee}{\end{equation}}
\newcommand{\ben}{\begin{equation*}}
\newcommand{\een}{\end{equation*}}
\newcommand{\bea}{\begin{aligned}}
\newcommand{\eea}{\end{aligned}}
\newcommand{\ba}{\begin{equation}\begin{aligned}}
\newcommand{\ea}{\end{aligned}\end{equation}}
\newcommand{\bd}{\begin{definition}}
\newcommand{\ed}{\end{definition}}
\newcommand{\bprop}{\begin{proposition}}
\newcommand{\eprop}{\end{proposition}}
\newcommand{\bt}{\begin{theorem}}
\newcommand{\et}{\end{theorem}}
\newcommand{\bcor}{\begin{corollary}}
\newcommand{\ecor}{\end{corollary}}
\newcommand{\beg}{\begin{example}}
\newcommand{\eeg}{\end{example}}
\newcommand{\bnt}[1]{\begin{namedthm}{#1}}
\newcommand{\ent}{\end{namedthm}}
\newcommand{\blm}{\begin{lemma}}
\newcommand{\elm}{\end{lemma}}
\newcommand{\bp}{\begin{proof}}
\newcommand{\ep}{\end{proof}}
\newcommand{\bpb}{\begin{problem}}
\newcommand{\epb}{\end{problem}}
\newcommand{\benum}{\begin{enumerate}}
\newcommand{\eenum}{\end{enumerate}}
\newcommand{\bitem}{\begin{itemize}}
\newcommand{\eitem}{\end{itemize}}
\definecolor{firebrick}{RGB}{178,34,34}

\def\brst\begin{restatable}
\newcommand{\erst}{\end{restatable}}

\newcommand{\abs}[1]{\left\lvert #1 \right\rvert}

\renewcommand{\Pr}{\mathbb{P}}

\renewcommand{\phi}{\varphi}

\newcommand{\calA}{\mathcal{A}}

\newcommand{\n}{{\boldsymbol n}}

\newcommand{\A}{{\boldsymbol A}}

\reviewer{jiayao}{JZ}{NavyBlue}

\graphicspath{{figs/lowpdf/}}

\def\final{0}

\begin{document}
\title[Investigating Fairness Disparities in Peer Review]{Investigating Fairness Disparities in Peer Review:\\A Language Model Enhanced Approach}

\author{Jiayao Zhang}
\email{zjiayao@upenn.edu}
\affiliation{%
  \institution{University of Pennsylvania}
  \city{Philadelphia}
  \state{Pennsylvania}
  \country{USA}
  \postcode{19104}
}
\author{Hongming Zhang}
\email{hzhangal@cse.ust.hk}
\affiliation{%
  \institution{Tencent AI Lab}
  \city{Seattle}
  \state{Washington}
  \country{USA}
}
\author{Zhun Deng}
\email{zhundeng@g.harvard.edu}
\affiliation{%
  \institution{Columbia University}
  \city{New York}
  \state{New York}
  \country{USA}
}
\author{Dan Roth}
\email{danroth@upenn.edu}
\affiliation{%
  \institution{AWS AI Labs and University of Pennsylvania}
  \city{Philadelphia}
  \state{Pennsylvania}
  \country{USA}
}

\renewcommand{\shortauthors}{Zhang et al.}


\begin{abstract}
    Double-blind peer review mechanism has become the skeleton of academic research across multiple disciplines including computer science, yet
    several studies have questioned the quality
    of peer reviews and raised concerns on potential biases in the process. 
    In this paper, we conduct a thorough and rigorous study on fairness disparities in peer review with the help of large language models (LMs). We collect, assemble, and maintain a comprehensive relational database\footnote{See \url{https://cogcomp.github.io/iclr_database}.} for the International Conference on Learning Representations (ICLR) conference from 2017 to date
    by aggregating data from OpenReview, Google Scholar, arXiv, and CSRanking, and extracting high-level
    features using language models.
    We postulate and study fairness disparities
    on multiple protective attributes of interest, including author gender, geography, author, and institutional prestige. 
    We observe that the level of disparity differs
    and textual features are essential in reducing biases in the predictive modeling. We distill several insights
    from our analysis on study the peer review process with the help
    of large LMs.
    Our database also provides avenues for studying new 
    natural language processing (NLP) methods
    that facilitate the understanding of the peer review mechanism.
    We study a concrete example towards automatic machine review
    systems and provide baseline models for the
    review generation and scoring tasks 
    such that the database can be used as a benchmark.

\end{abstract}


%
%
%


\begin{teaserfigure}
\begin{subfigure}[t]{0.5\textwidth}
    	    \centering
    	    \includegraphics[width=0.9\textwidth]{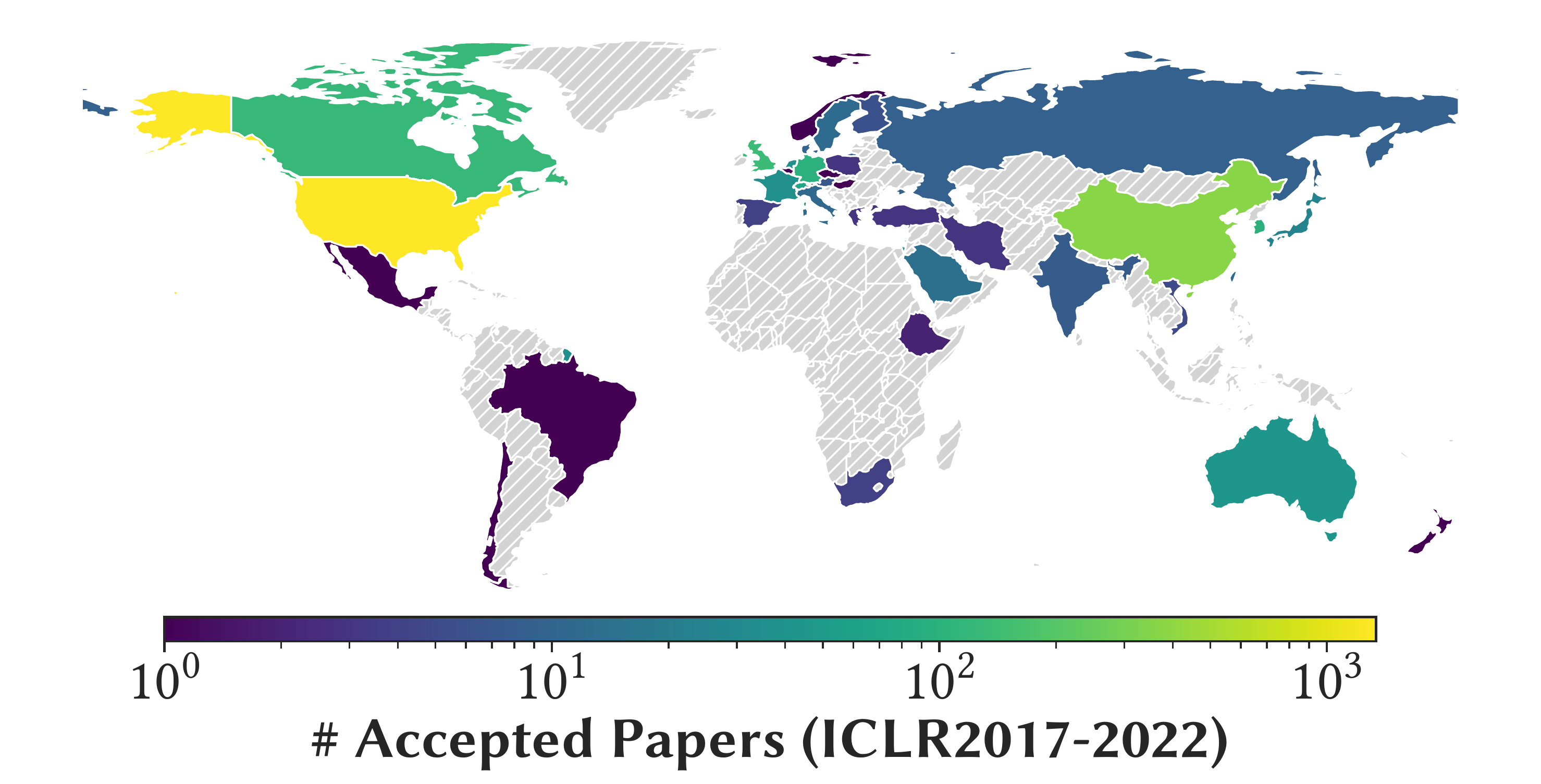}
	    \end{subfigure}%
	    \begin{subfigure}[t]{0.5\textwidth}
    	    \centering
    	    \includegraphics[width=0.9\textwidth]{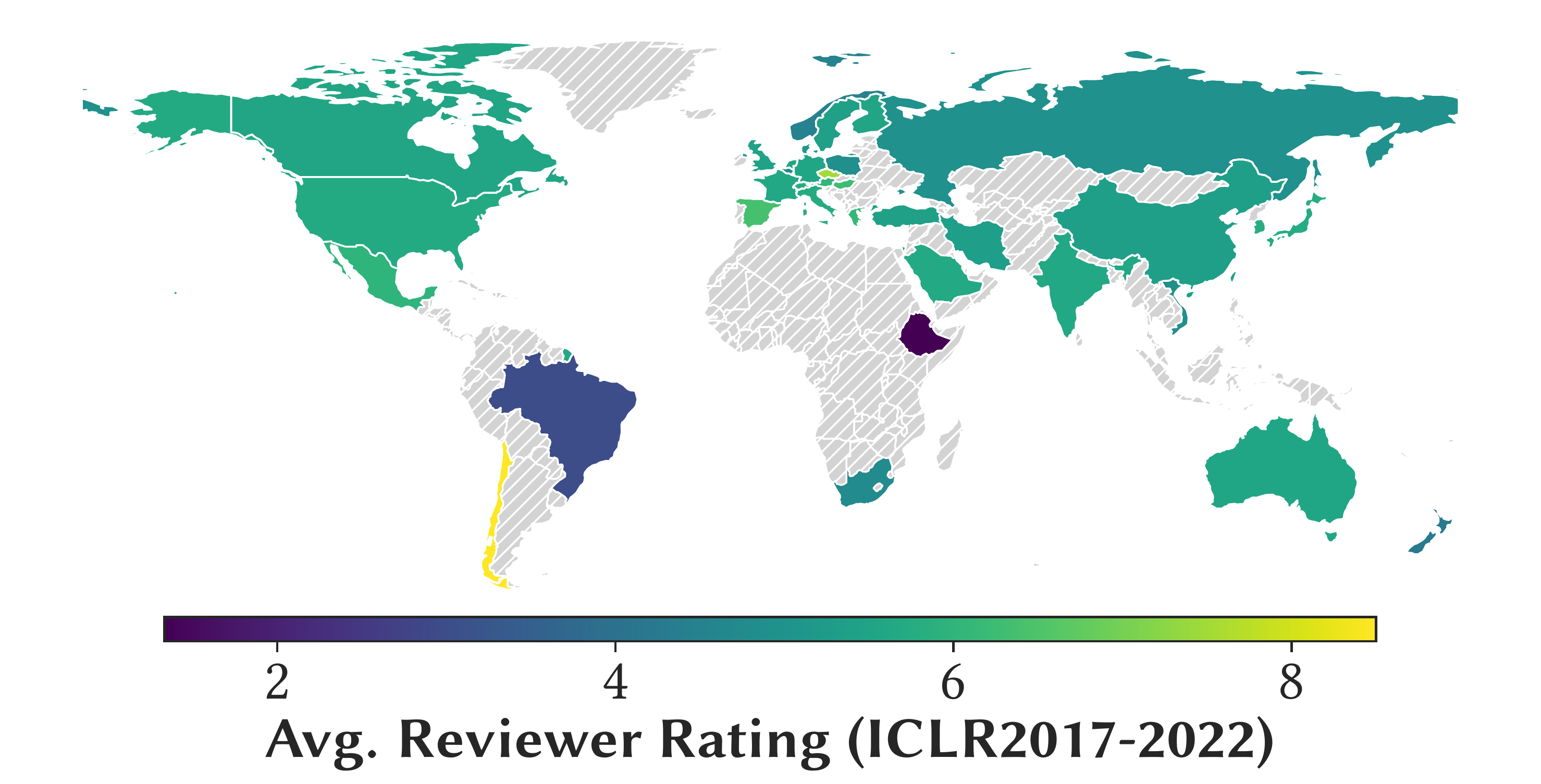}
	    \end{subfigure}
  \caption{\textbf{Geographical disparity:} does ICLR favors north American authors more?} \label{fig:teaser}
  \Description{There is a parity of submissions across different geographies.}
\end{teaserfigure}


\maketitle

\newcommand{\dpv}{\textsf{DP}}
\newcommand{\eov}{\textsf{EO}}
\newcommand{\aucv}{\textsf{AUC}}

\begin{table*}[t]
    \scriptsize
    \centering
    \resizebox{\textwidth}{!}{%
	\begin{tabular}{lclccllllll}
    \toprule
    \, & Recency    & Venue & \# Papers  & \# Reviews & \stackanchor{Author}{Features} & \stackanchor{Review}{Features} & \stackanchor{Textual}{Features} & \stackanchor{Predictive}{Model} & \stackanchor{Generation}{Model} & \stackanchor{Fairness}{Analysis} \\
    \midrule
    \citet{kang2018peerread} & 2017 & ICLR, ARR, arXiv & 586 & 1618 & \ding{55} & \ding{51}& \ding{51}  &\ding{51} &\ding{55} &\ding{55}\\
    \citet{tran2021an}       &2020 & ICLR & 5569 & - & \ding{51} & \ding{51} & \ding{55} &\ding{51} &\ding{55}  &\ding{55}\\
    \citet{dycke2022yyy}     & 2021 & ARR & 567 & 1053 & - & - & - & - &-&- \\
    This Paper                & 2022  & ICLR & 10289 & 36453 & \ding{51} & \ding{51} &\ding{51} &\ding{51} &\ding{51} &\ding{51}\\
    \bottomrule
    \end{tabular}
    }
    \caption{\textbf{Comparison with different related works.} Here author/review features
    refer to those plain features such as numerical ratings, author citation counts
    that do not need to be extracted, whereas textual features refer to those
    high-level features such as the tone/emotion of the texts.}
    \label{tab:relatedwork}
\end{table*}
\section{Introduction} \label{sec:introduction}

\begin{figure}
        \centering
    	    \includegraphics[width=\linewidth]{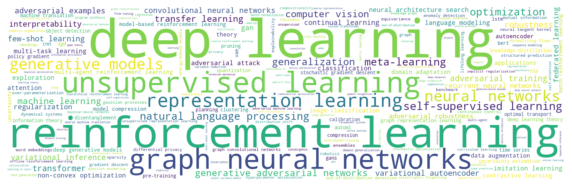}
	    \caption{\textbf{Word cloud of most frequent keywords.}} \label{fig:teaser:wordcloud}
    \end{figure}

    The current scientific development relies heavily on
    the peer review mechanism for authors to publish and share their research.
    With the sheer number of submissions, many venues have experienced
    an extremely large demands for reviewers, and many authors have complained
    they have received unfair or bogus reviews. Indeed, the famous NeurIPS
    experiments~\cite{nips2014}\footnote{Both in 2014 \cite{nips2014} and 2021, see \url{https://blog.neurips.cc/2021/12/08/the-neurips-2021-consistency-experiment/}.}  assigned a different set of reviewers to the same
    submissions, and found the reviewer ratings are sometimes strikingly different.
    In parallel, algorithmic fairness has
    attracted attention of practitioners of various
    domains to analyze the intrinsic bias of either the
    dataset or the model trained on the dataset. It is thus
    natural to wonder, whether one could use lessons from
    fairness to analyze the peer review process?
    For example, in \Cref{fig:teaser}, we note north
    America (NA) has the most accepted papers of all year, and
    in \Cref{fig:teaser:wordcloud}, we note ``reinforcement learning'' and ``graph neural networks'' are much more popular,
    does this mean ICLR favors authors in the US and/or working on those popular topics?
    In order to study questions of this flavor, a comprehensive dataset, proper fairness
    formulations, and principled application of language modeling
    techniques are needed.
    

    As the first step, we assemble a database consisting of submissions to the International Conference on Learning Representations (ICLR) that is most comprehensive and up-to-date. Our database contains submissions,
    reviews, author responses, together with author/institution profiles and extracted
    high-level features such as submission complexity, review sentiment, etc.
We decompose the peer review process into two stages and study them separately. The first stage is the \emph{review
    stage} where reviewers assign scores to submissions; and the second being the \emph{decision stage}
    where area chairs or meta reviewers provide recommendation for the submission
    based on all information available.
    In the decision process, we use predictive models as surrogates
    to study fairness disparities.
    We find that simple linear models
    can perform well, and the use of language models features allow the model to reduce its
    fairness violations. 

    Our database also motivates important new tasks relevant to the 
    NLP community: in the review stage, as one way to decipher its
    underlying mechanism, can we build models that generate reviews and score the submissions? These tasks are themselves important
    for building automatic reviewing systems, and we find it to be
    very challenging, even using the current state-of-the-art language model. We analyze the challenges and discuss desiderata of such tasks.
    
    \paragraph{Contributions.} 
    (1) We collect and maintain a comprehensive
    peer review dataset.
    (2)  We initiate the study of fairness disparities
            in peer-review with the help of language model, and demonstrate that
            the importance of using textual features in investigating disparities
            in peer review.
    (3) We demonstrate how the dataset can be used as a benchmark
    for automatic machine review systems.

    \paragraph{Summary of findings.}
    (1) Through association analysis, we observe disparities of several sensitive attributes. However, we do not have compelling evidence that those differences are significant.
    (2) The inclusion of high-level textual features from large LMs helps in both increasing predictive power as well as reducing fairness disparities.
    (3) The review process is harder to study compared with the decision process. Current state-of-the-art pre-trained large LMs have difficulty
    in completing tasks defined therein.

\section{Background}
\subsection{Problem Formulation}

    We investigate several commonly used fairness notion, including \emph{demographic disparity} (DP), \emph{equalized odds difference} (EO),
    and \emph{AUC difference} (AUC).
    For a given \emph{sensitive attribute} $A$ taking values from $\calA$, these measures
    are
    defined as
    \ba
    \resizebox{6.8cm}{!}{$\begin{cases}\dpv = \underset{a \ne a' \in\calA}{\max} \abs{\Pr(\hat{y}=1|A=a)-\Pr(\hat{y}=1|A=a')},\\
    \eov=\underset{a \ne a' \in\calA}{\max} \abs{\Pr(\hat{y}=1|A=a,y=1)-\Pr(\hat{y}=1|A=a',y=1)},\\
    \aucv=\underset{a \ne a' \in\calA}{\max} \abs{ AUC(\hat{y}, y | A=a)-AUC(\hat{y}, y | A=a')},
    \end{cases}$}
    \ea
    where $\hat{y}\in\{0,1\}$ is the binary label (e.g., either {\tt accept} or
    {\tt reject}),  $A$ is the sensitive attribute,
    and $AUC(\hat{y}, y|A=a)$ computes the area under curve (AUC) of the receiver operating characteristic (ROC) curve of the subgroup $A=a$.
    This formulation allows
    us to assess disparities in both data itself (by setting $\hat{y}$ to be the true label)
    as well as in predictive models (by setting $\hat{y}$ as
    the predictors).
    Although our database allows investigations on various
    disparities, we zoom in the following three that we think might be
    the most crucial.
    \textbf{(i) Geographical disparity:} $\calA$ is the set
        of the (dominant) geography of the authors in a submission.
    \textbf{(ii) Gender disparity:} $\calA$ represents the gender of the authors of a submission (mode, or whether a certain gender is present).
    \textbf{(iii) Prestige disparity:} $\calA$ represents the
        prestige of the authors of a submission (citation count, institution ranking).

    A popular criticism of peer review is that the reviewer rating
    is somewhat ``random'' as demonstrated by the NeurIPS experiments.
    To assess such randomness at different peer review stages
    quantitatively, we assess the goodness-of-fit when fitting
    predictive models such as logistic regression on the dataset.
    We found that in general, simple models can fit the decision stage
    well but even more complicated models fail to capture
    the essence of the reviewing stage. This suggests that the randomness
    sentiment
    many authors have echoed might be mainly from the reviewing stage.

    This observation naturally motivate us to set foot in evaluating
    existing large language models (or their fine-tuned variants)
    on their ability to \emph{approximate} human-reviewers.
    This is by definition a very challenging problem and we observe
    that more efforts need to be taken towards automatic machine reviewing
    systems.



\subsection{Related Work} \label{sec:related}

\begin{figure*}[t]
	    \centering
	    \includegraphics[width=\linewidth]{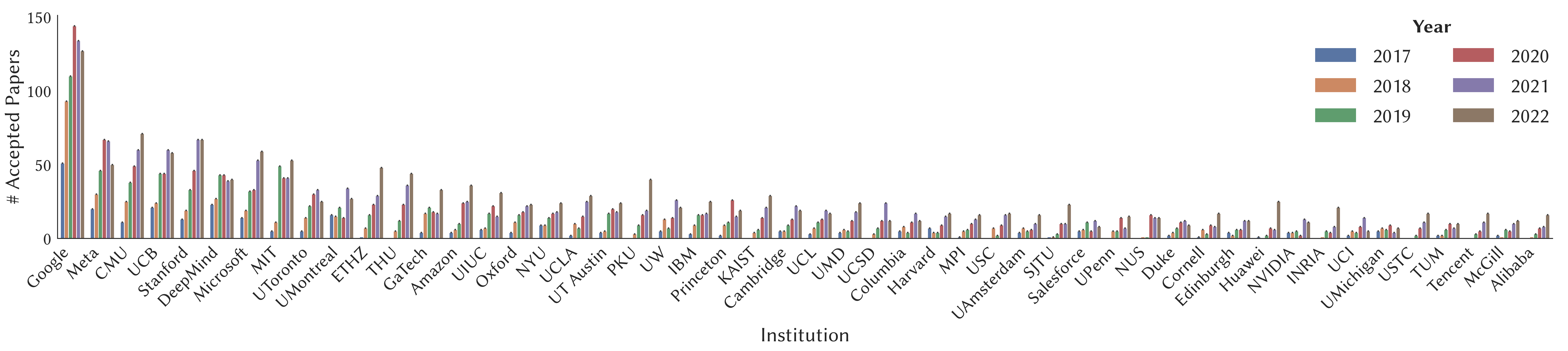}
	    \caption{Number of accepted submissions per year per institution (top $50$ ranked within all years are shown).} \label{fig:iclrranking}
\end{figure*}
\paragraph{Fairness disparities.} In modern data science, it is increasingly important for models to be non-discriminatory or fair with respect to some sensitive attributes (e.g., race or gender). Many fairness notions have been proposed to regularize models to mitigate the bias of models, both from individual level \cite{dwork2012fairness,burhanpurkar2021scaffolding} and group level \cite{hardt2016equality,certify}. Among them, demographic parity \cite{certify} and equalized odds \cite{hardt2016equality} are arguably two of the most popular fairness notions. Demographic parity states that the proportion of each segment of a protected group (e.g. gender) should receive the positive outcome (e.g. making a loan) at equal rates.
On the other hand, a classifier satisfies equalized odds if the subjects in the protected and unprotected groups have equal true positive rate and equal false positive rate. Although being of great academic and social interest, there are very few studies on
fairness disparities and equalized odds in the peer review mechanism, especially when textual features are present.

\paragraph{Peer review analysis}

As an important social network, the peer review network has attracted more and more attentions in the community.
For example, \citet{kang2018peerread} proposed a high-quality dataset contains paper from ACL, NeurIPS (formerly NIPS), and ICLR. \citet{DBLP:conf/sigir/PlankD19} create a large-scale dataset that covers over 3,000 papers from machine learning conferences.
\citet{DBLP:conf/naacl/0023EKGM19} create a dataset that focuses on investigating the effect of rebuttals in NLP conferences.
On top of these datasets, many recent research has been proposed to investigate the bias inside the review process~\cite{DBLP:conf/nips/StelmakhSS19,DBLP:conf/aaai/ManzoorS21}, argument mining~\cite{DBLP:conf/naacl/HuaNBW19}, automatically review generation~\cite{DBLP:journals/corr/abs-2102-00176}, improved review process~\cite{DBLP:conf/emnlp/RogersA20,DBLP:conf/nips/JecmenZLSCF20}, and review explanation~\cite{DBLP:conf/inlg/WangZHKJR20}.
Nonetheless, none of the existing work has investigated
the fairness violation or worked through the lens of large language models.
We summarize key differences between the most relevant datasets and ours
in \Cref{tab:relatedwork}. Notably, to the best of our knowledge, our dataset is the most comprehensive one, backed
up by various off-the-shelf features extracted from large language models
for downstream analysis.


\section{Database Construction}

\subsection{ICLR Data}
    We use the \texttt{OpenReview}\footnote{\url{https://api.openreview.net/api/}} \texttt{Python API}\footnote{\url{https://pypi.org/project/openreview-py/}} to crawl conference data
    from OpenReview, which include
    submissions, author profiles, reviews, rebuttals, and
    decisions in ICLR 2017-2022.
    We are able to obtain in total 10289 submissions,
    21808 distinct authors, 36453 reviews, 68721
    author responses, and 4436 public comments.
    The crawling process is done in Feb 2022 after
    ICLR2022 announced its decisions.
    We exclude desk-rejected or author-withdrawn submissions
    from the dataset.
    We tabulate
    per-year counts in Table~\ref{tab:data:summary}.
    Note that from table we note the natural distributional
    differences across gender groups of authors.
    We next briefly describe the schema for each data entity
    while the full Entity-Relation (ER) diagram and the covariate table are given
    in the Appendix.

\begin{table}
    \centering
    \resizebox{0.8\columnwidth}{!}{%
	\begin{tabular}{lllllll}
    \toprule
    & 2017 & 2018 & 2019 & 2020 & 2021 & 2022 \\
    \midrule
    \textbf{Submissions} & 490 & 911 & 1419 & 2213 & 2595 & 2670 \\
    \, Oral      & 15  &  23 & 24  & 48   & 53   & 54   \\
    \, Spotlight & 0   &  0  &  0  & 108  & 114  & 176  \\
    \, Poster    & 183 & 313 & 478 & 531  & 693  & 865  \\
    \, Workshop\textsuperscript{*}  & 47  & 89  & 0   & 0    & 0    & 0    \\
    \, Reject    & 245 & 486 & 917 & 1526 & 1735 & 1574 \\
    \textbf{Author} & 1416 & 2703 & 4286 & 6807 & 7968 & 8654 \\
    \, Female       & 81  & 162  & 298  & 503  & 529  & 770 \\
    \, Male         & 769 & 1565 & 2527 & 3951 & 3992 & 5524 \\
    \, Non-Binary   & 1   & 2    & 2    & 2    & 3    & 6\\
    \, Unspecified  & 565 & 974  & 1458 & 2351 & 2125 & 2354 \\
    \textbf{Review} & 1489 & 2748 & 4332 & 6721 & 10026 & 10401 \\
    \textbf{Response} & 2811 & 4404 & 9504 & 11350 & 18896 & 21756 \\
    \textbf{Comment} & 750 & 1002 & 1354 & 816 & 376 & 133 \\
    \bottomrule
    \end{tabular}
    }
    \caption{Summary of the dataset. In 2017-2018,
    submissions to the main conference may be invited to workshop tracks. Note that we exclude desk-rejected/withdrawn papers.}
    \label{tab:data:summary}
\end{table}

    \paragraph{Submission.} Each submission entity contains
    a paper number (unique within the same conference),
    a title, an abstract, a link to its pdf file, an
    one-sentence summary ({\tt tldr}), a list
    of self-provided paper keywords, and a list of
    author identifiers referring to the authors in
    author table.

    \paragraph{Author.} Author entities contain the 
    author names, their emails (only domain is
    visible), and optionally, self-reported gender,
    homepage, Google Scholar, DBLP, LinkedIn, Semantic Scholar, Wikipedia, and ORCID.
    In addition, authors can optionally report
    their current and past affiliations with corresponding
    positions.

    \paragraph{Review.} Although the specific
    review format changes each year, there is generally
    a textual review, a numerical rating (usually in the range
    of $1$ to $10$),
    and a numerical confidence score (usually in the range of $1$ to $5$).
    In certain years,
    there are more specific scores such as
    technical soundness, novelty, etc. 
    In total, there are $35717$ reviews corresponding
    to 10289 submission with an average number
    of reviews per submission being $3.47$.

    \paragraph{Author responses, public comments, and decisions.}
    Author response, public comment, and decision entities include
    a title, a comment, a forum field that refers  to the
    submission it points to, and a ``reply-to''
    field that points to its parent node in a discussion
    thread.

\begin{figure*}[t]
    \centering
    \begin{subfigure}[t]{0.2\textwidth}
	    \centering
	    \includegraphics[width=\linewidth]{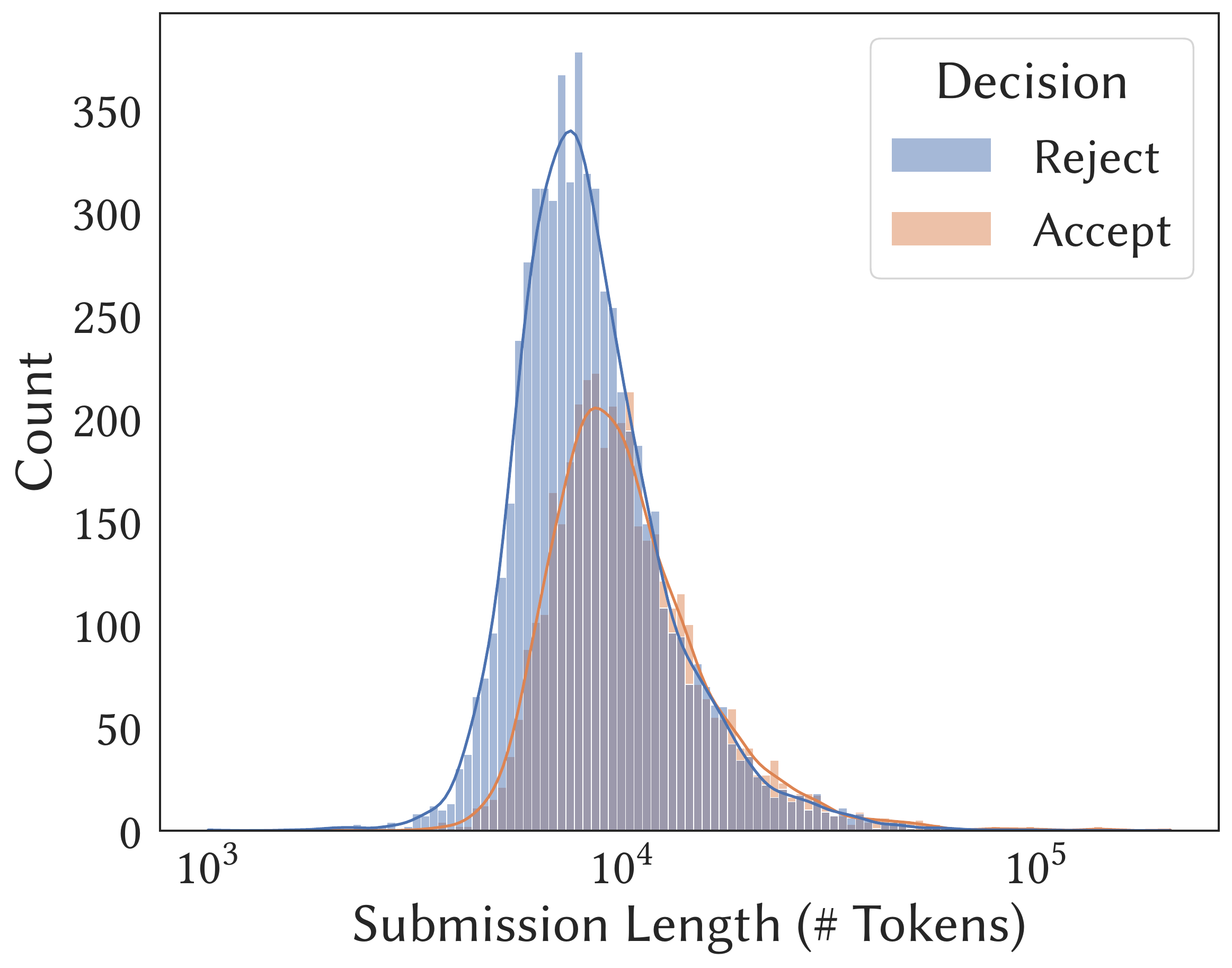}
	    \caption{\tiny Submission length.}  \label{fig:hist:sublen}
	\end{subfigure}%
	\begin{subfigure}[t]{0.2\textwidth}
	    \centering
	    \includegraphics[width=\linewidth]{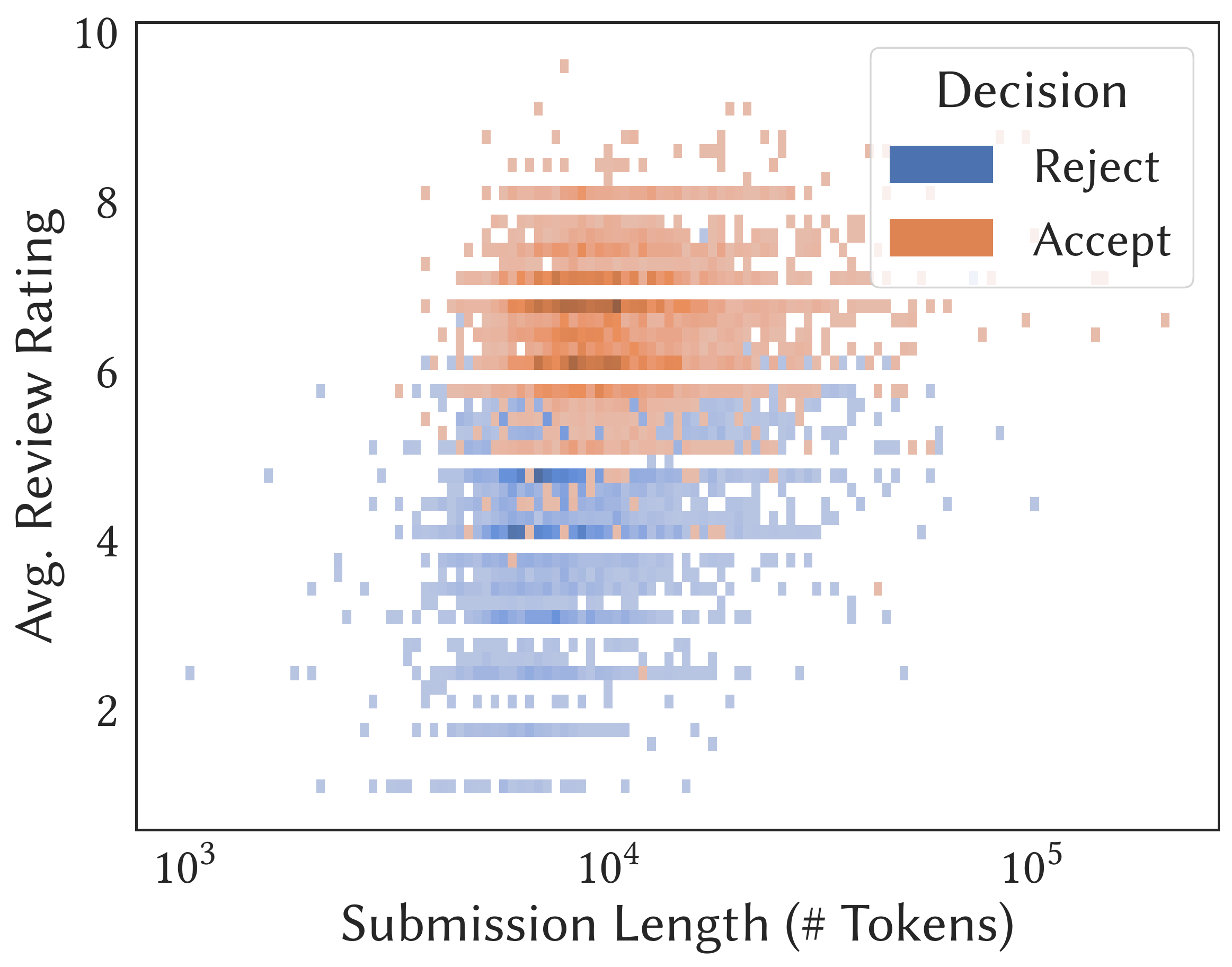}
	    \caption{\tiny Submission rating/length.} \label{fig:hist:subrat}
	\end{subfigure}%
	\begin{subfigure}[t]{0.2\textwidth}
	    \centering
	    \includegraphics[width=\linewidth]{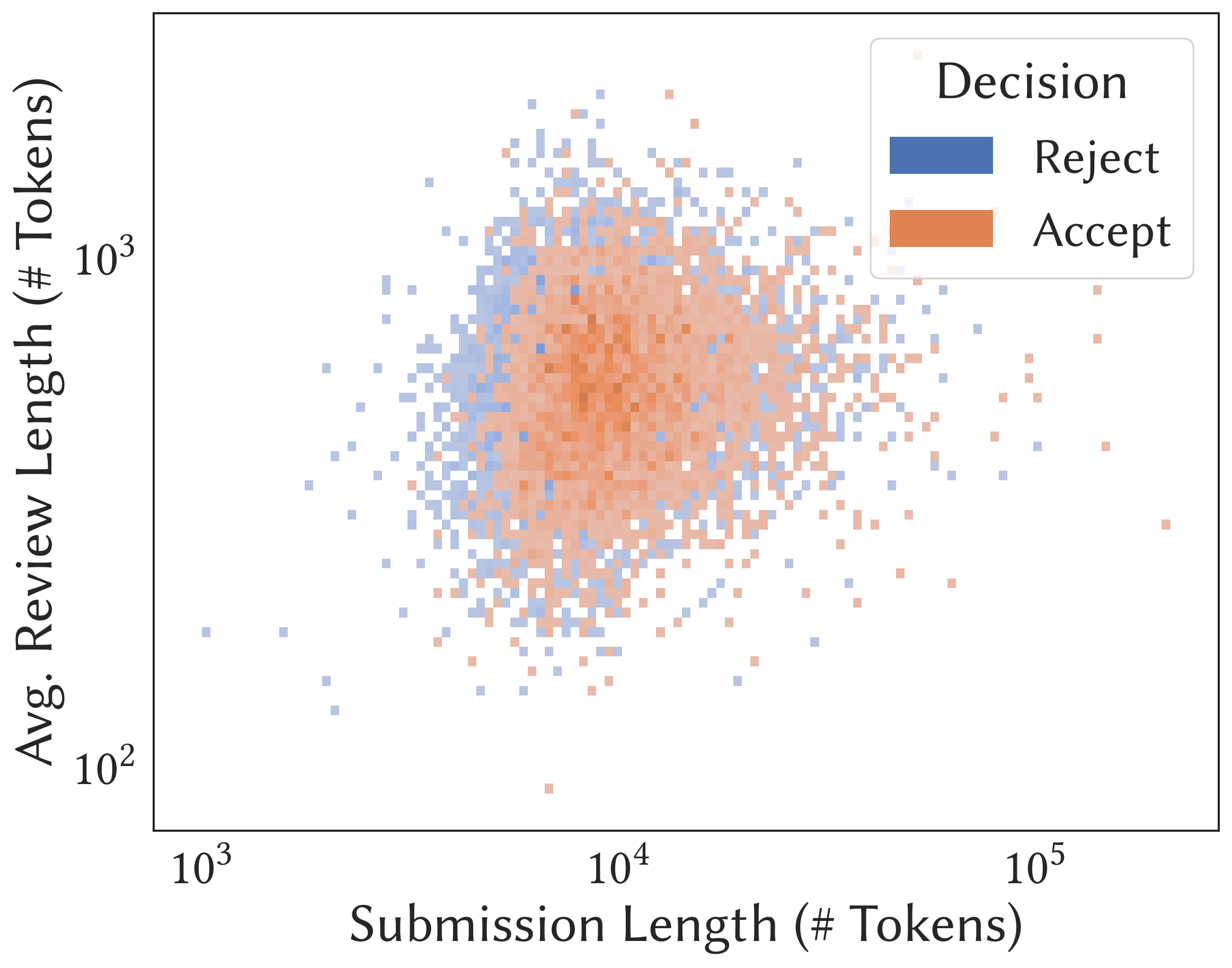}
	    \caption{\tiny Review/submission lengths.} \label{fig:hist:subrev}
	\end{subfigure}%
\begin{subfigure}[t]{0.2\textwidth}
	    \centering
	    \includegraphics[width=\linewidth]{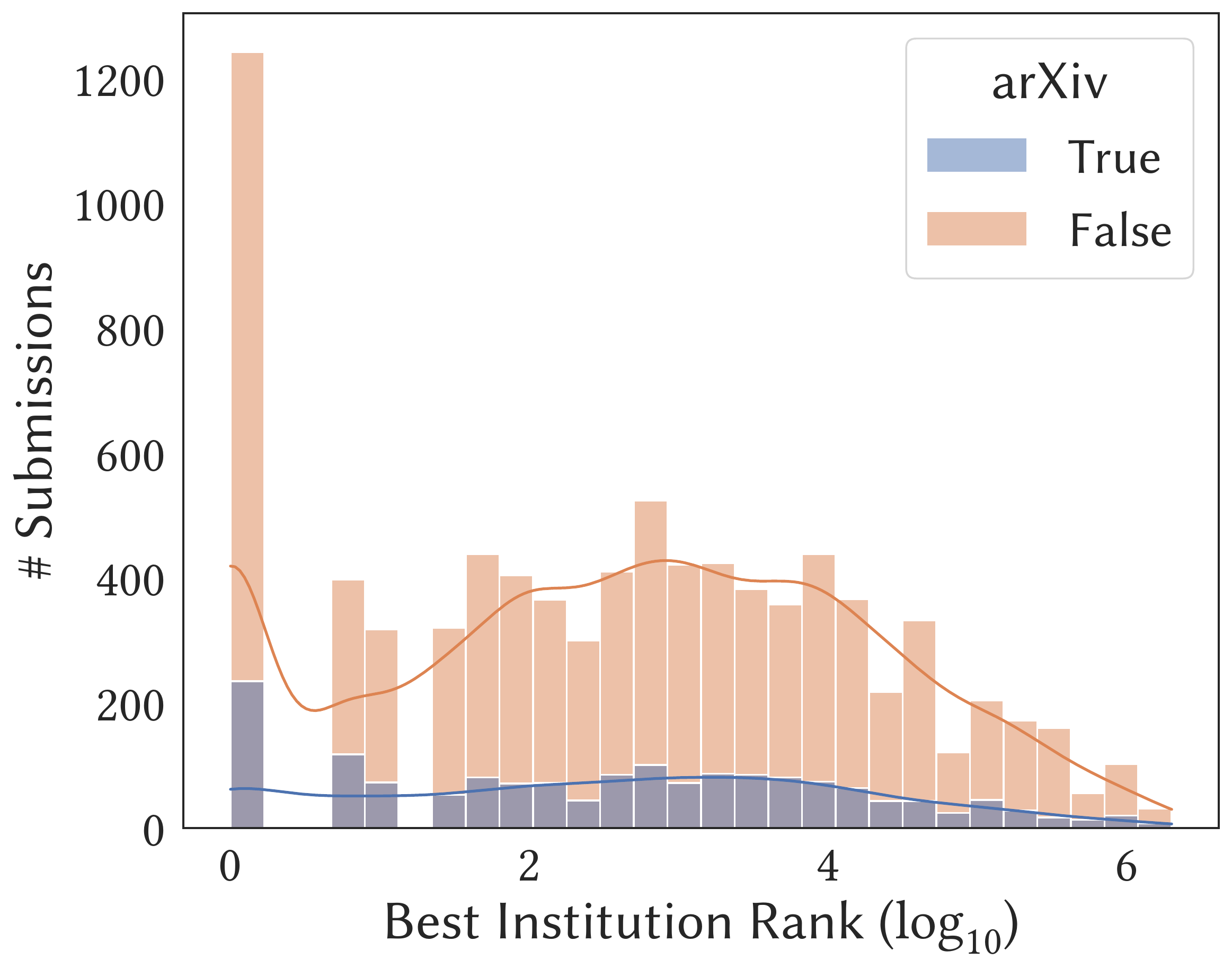}
	    \caption{\tiny Institution rank and arXiv.} \label{fig:hist:arxiv}
	\end{subfigure}%
     \begin{subfigure}[t]{0.2\textwidth}
	    \centering
	    \includegraphics[width=\linewidth]{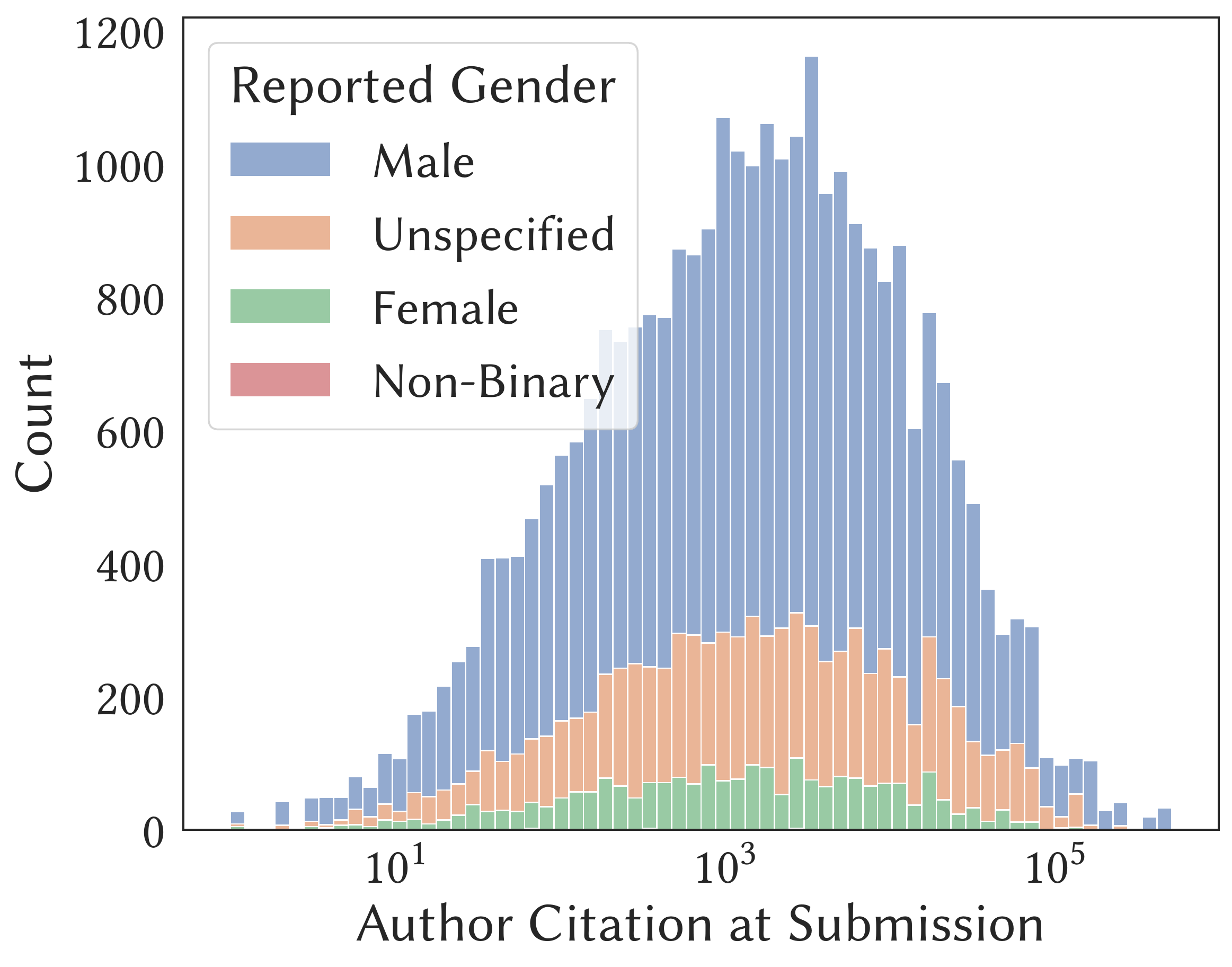}
	    \caption{\tiny  Author citation.} \label{fig:hist:citation}
	\end{subfigure}\\

	\begin{subfigure}[t]{0.2\textwidth}
	    \centering
	    \includegraphics[width=\linewidth]{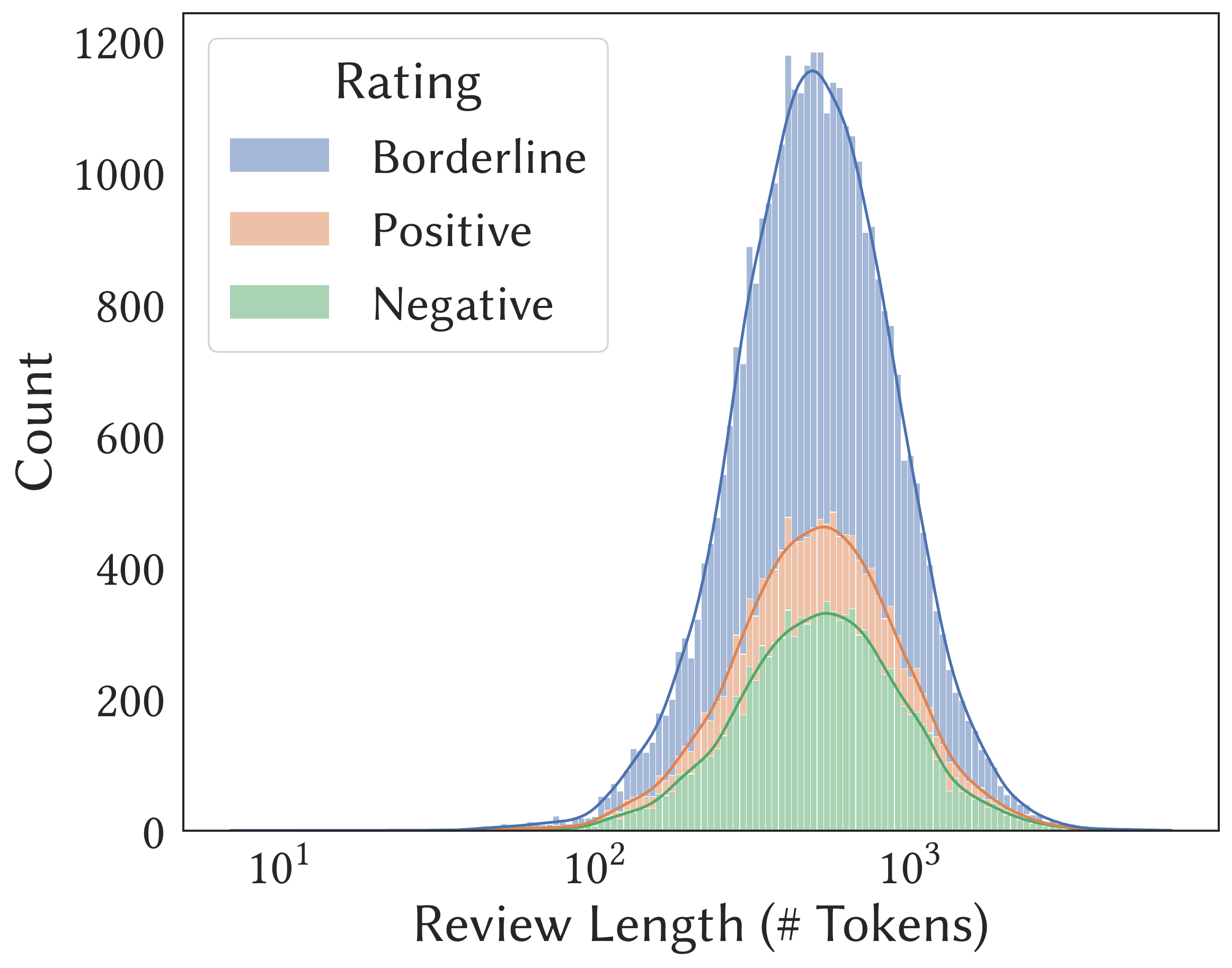}
	    \caption{\tiny Review rating/length.} \label{fig:hist:revlen}
	\end{subfigure}%
	\begin{subfigure}[t]{0.2\textwidth}
	    \centering
	    \includegraphics[width=\linewidth]{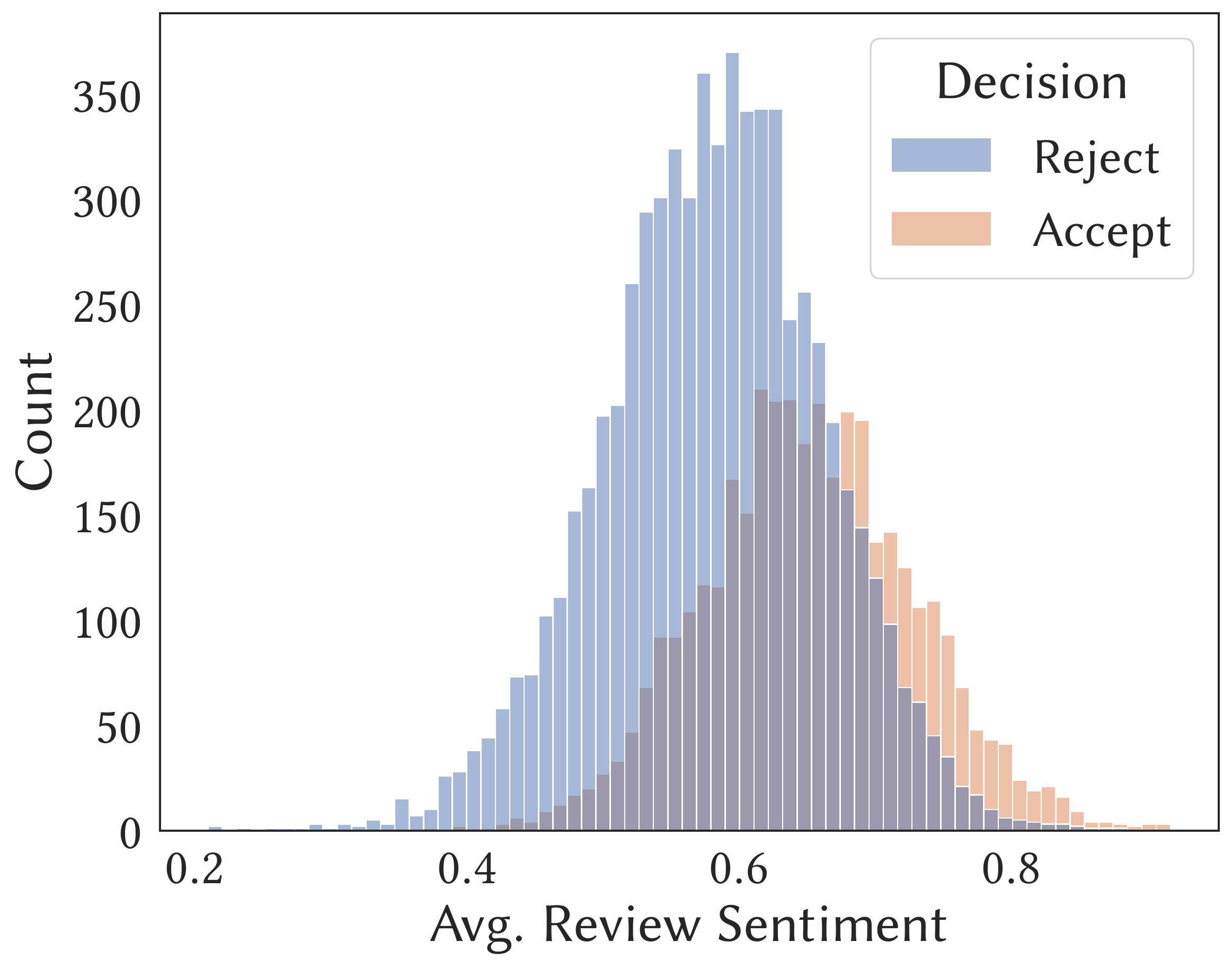}
	    \caption{\tiny Review sentiment/decision} \label{fig:hist:revdec}
	\end{subfigure}%
	\begin{subfigure}[t]{0.2\textwidth}
	    \centering
	    \includegraphics[width=\linewidth]{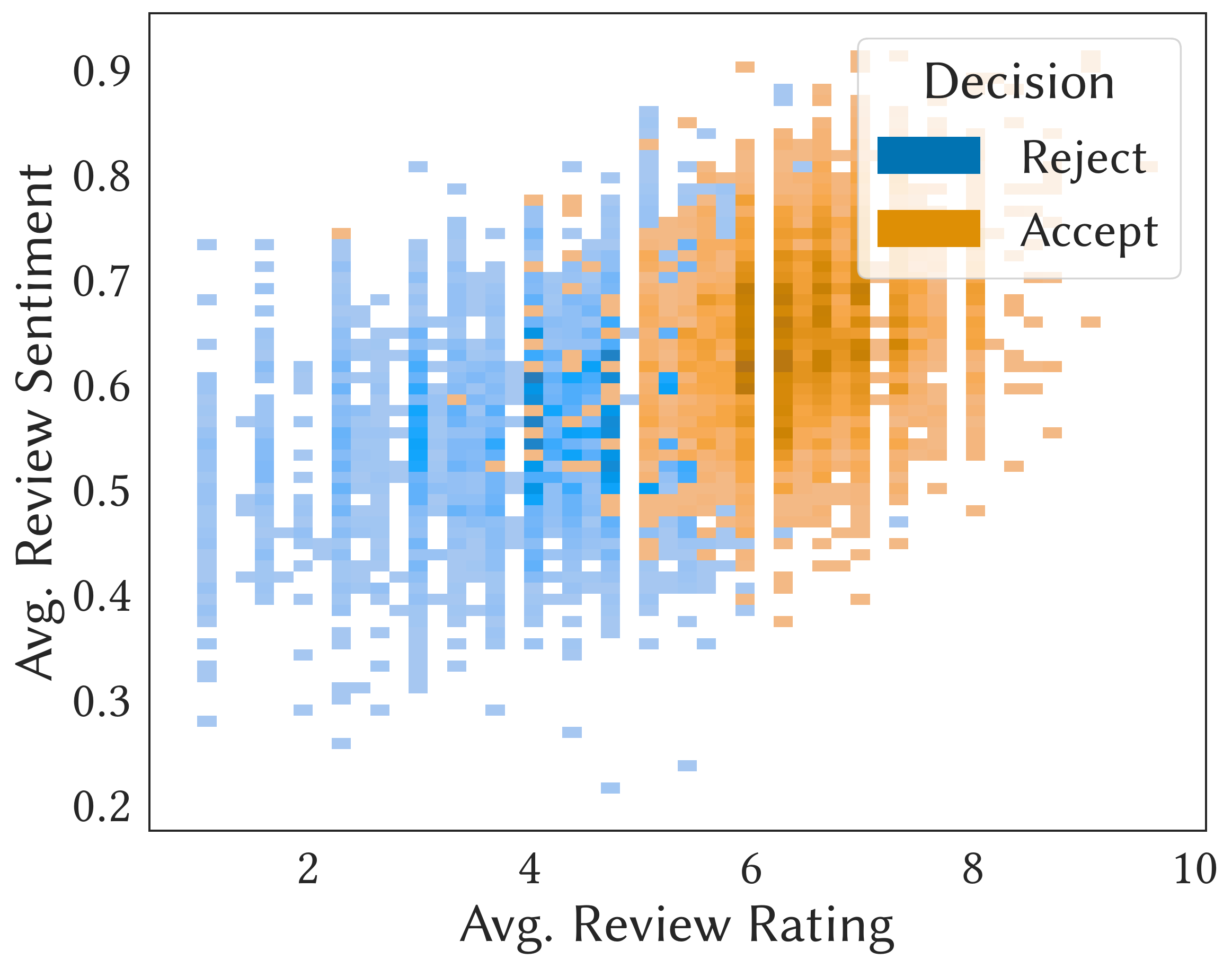}
	    \caption{\tiny Review sentiment/rating.} \label{fig:hist:revsen}
	\end{subfigure}%
    \begin{subfigure}[t]{0.2\textwidth}
	    \centering
	    \includegraphics[width=\linewidth]{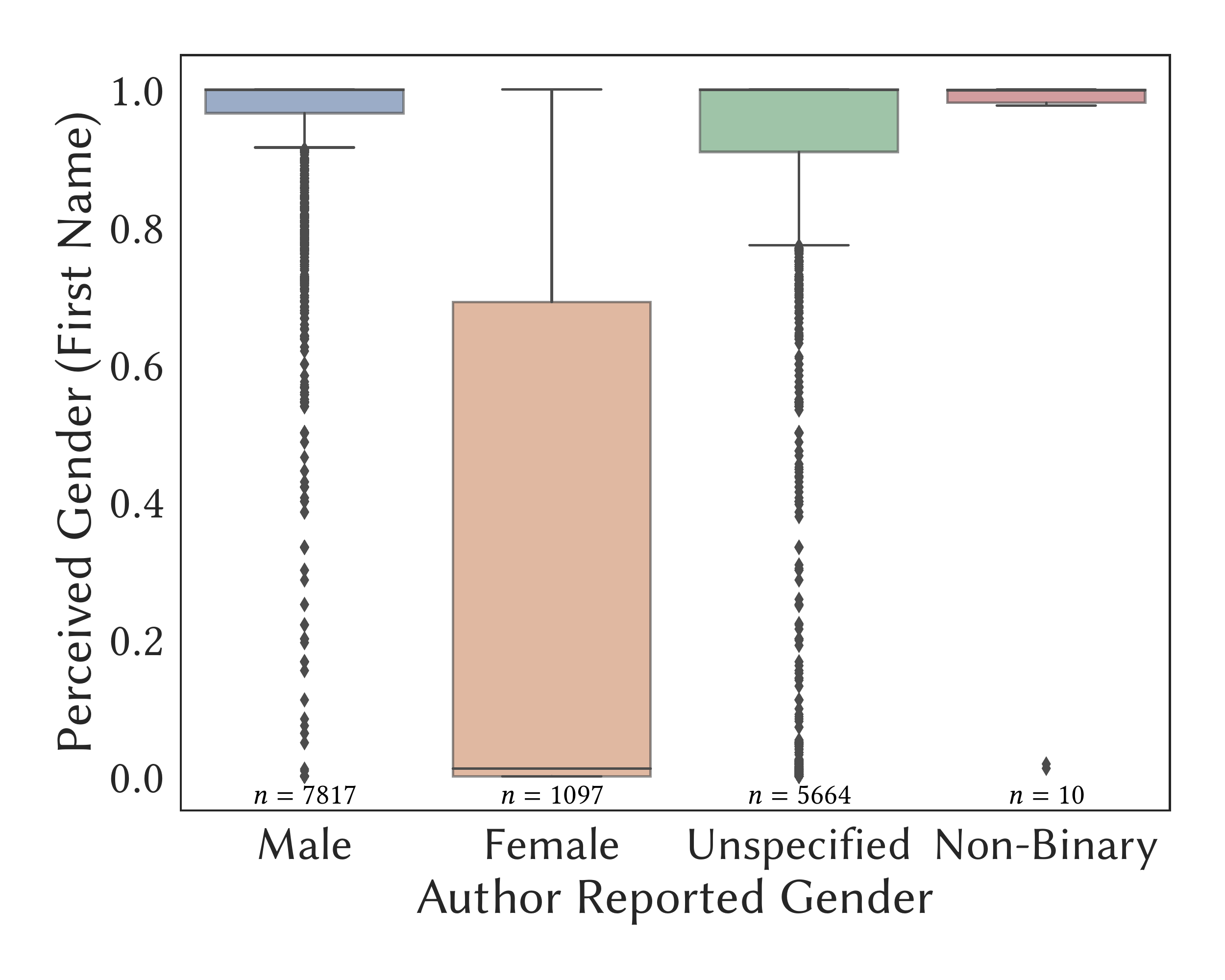}
	    \caption{\tiny Author reported gender.} \label{fig:hist:repgen}
	\end{subfigure}%
	\begin{subfigure}[t]{0.2\textwidth}
	    \centering
	    \includegraphics[width=\linewidth]{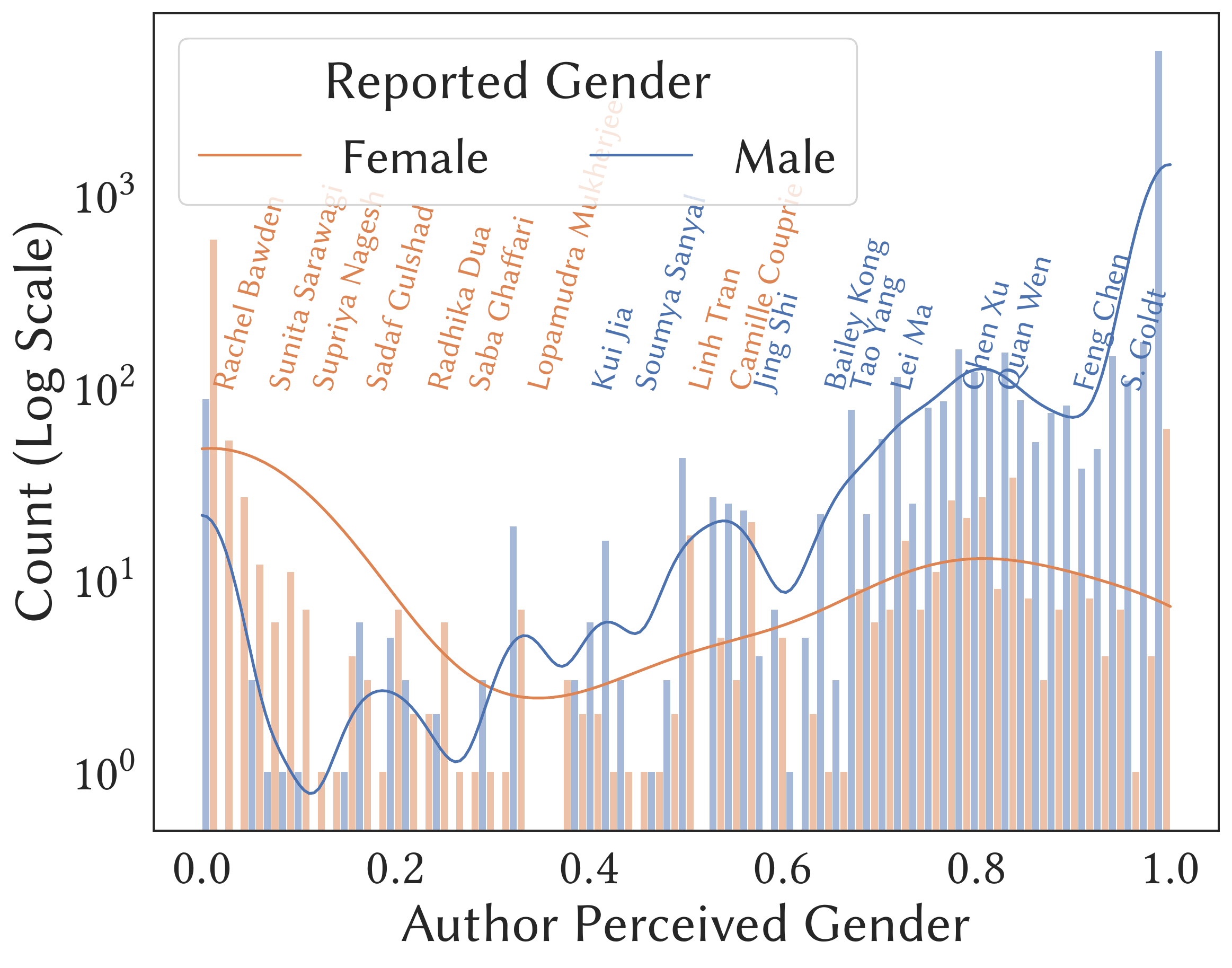}
	    \caption{\tiny Author perceived gender.}\label{fig:hist:pergen}
	\end{subfigure}\\

    \begin{subfigure}[t]{0.2\textwidth}
	    \centering
	    \includegraphics[width=\linewidth]{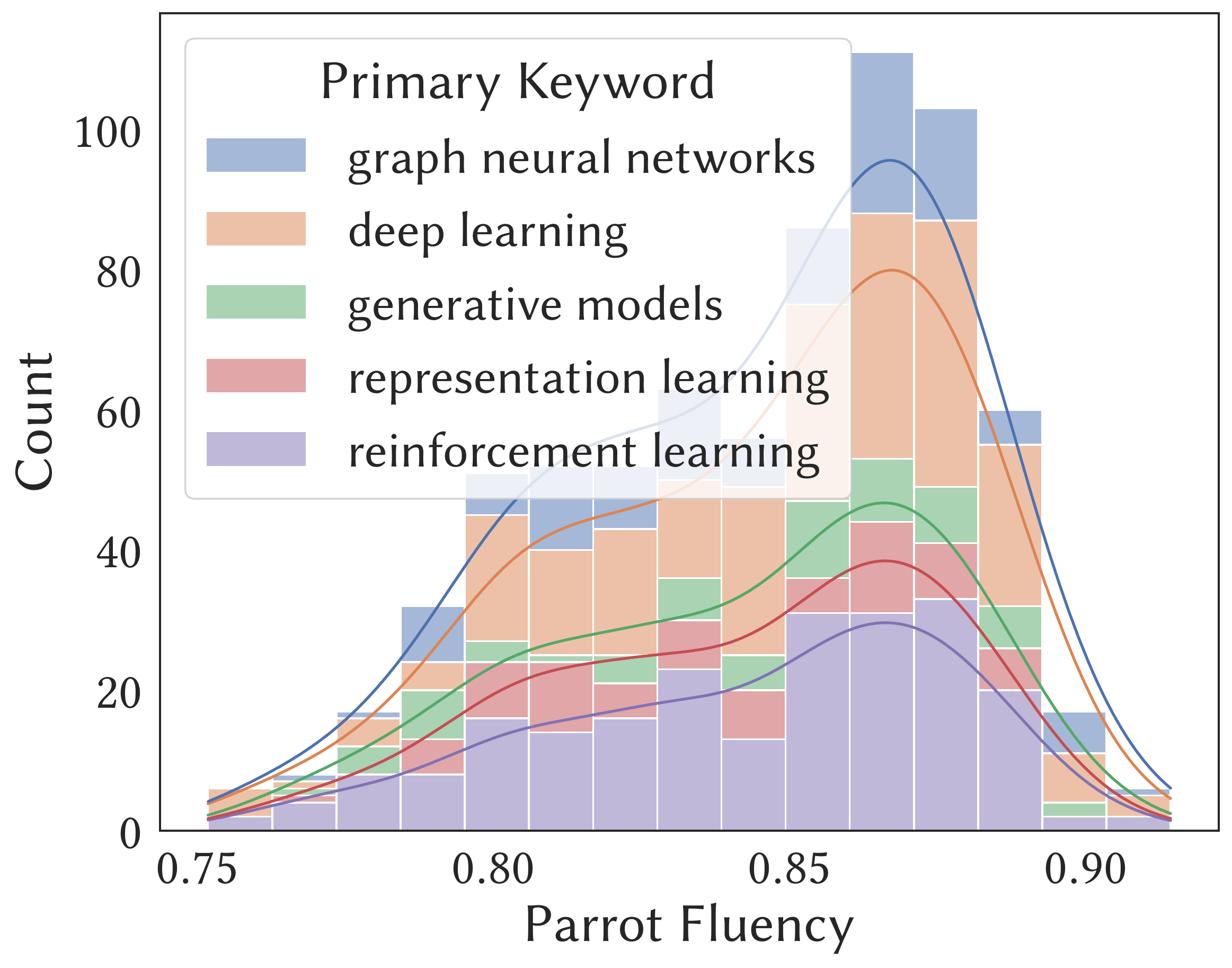}
	    \caption{\tiny Submission fluency.} \label{fig:hist:subflu}
	\end{subfigure}%
	\begin{subfigure}[t]{0.2\textwidth}
	    \centering
	    \includegraphics[width=\linewidth]{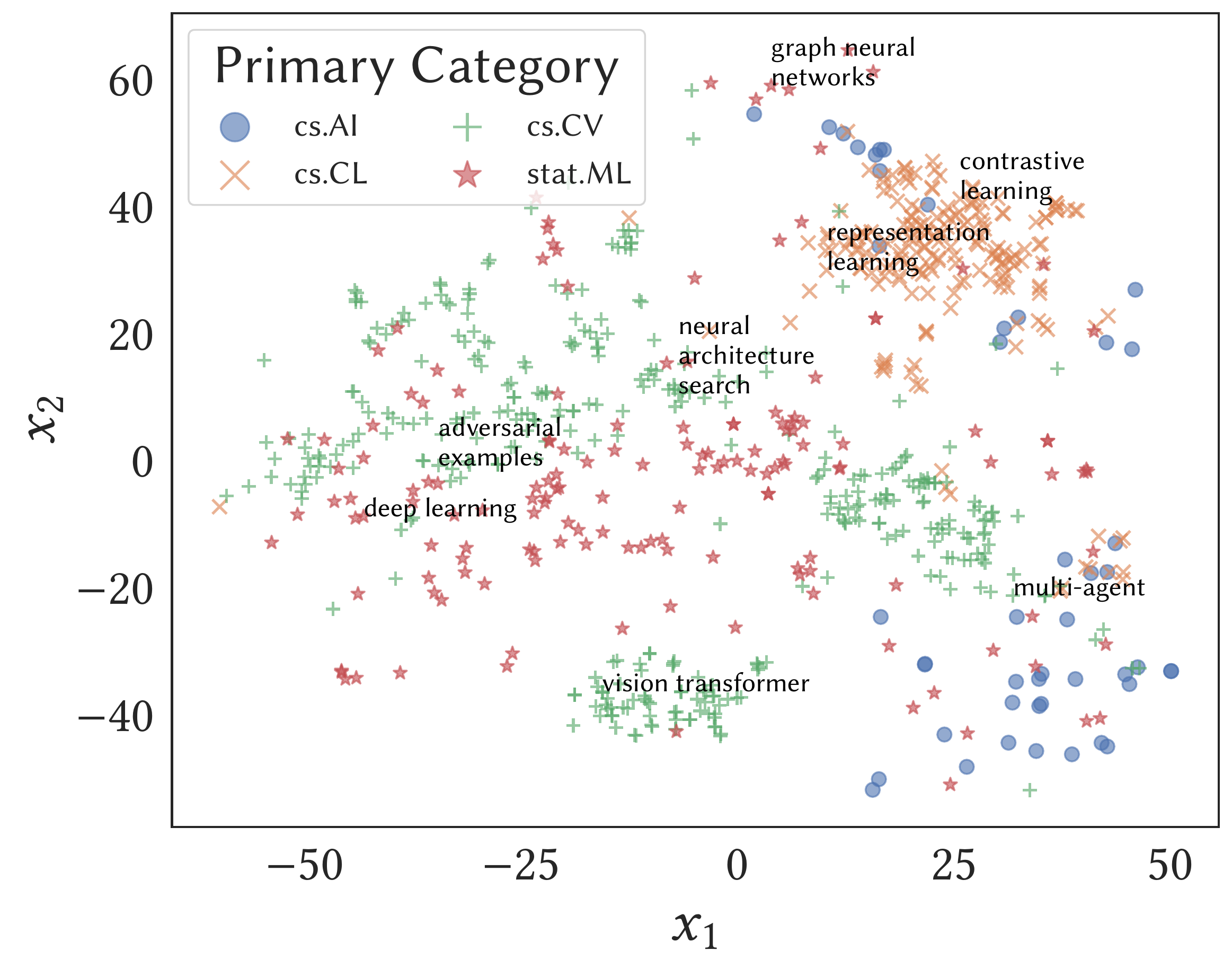}
	    \caption{\tiny Specter embedding.} \label{fig:hist:subspe}
	\end{subfigure}%
    \begin{subfigure}[t]{0.2\textwidth}
	    \centering
	    \includegraphics[width=\linewidth]{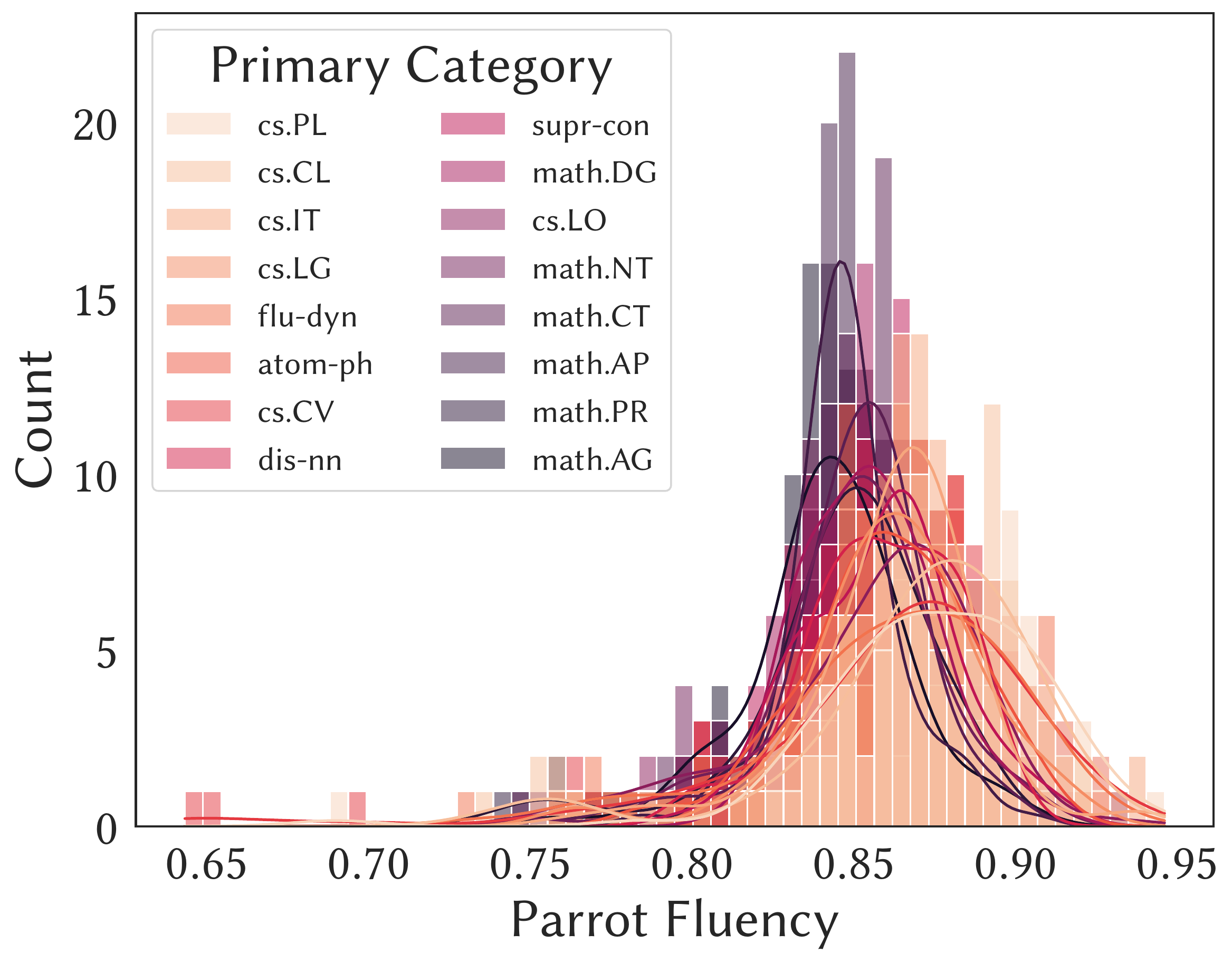}
	    \caption{\tiny Histogram of fluency score.} 
     \label{fig:sanity:hist}
	\end{subfigure}~
    \begin{subfigure}[t]{0.2\textwidth}
	    \centering
	    \includegraphics[width=\linewidth]{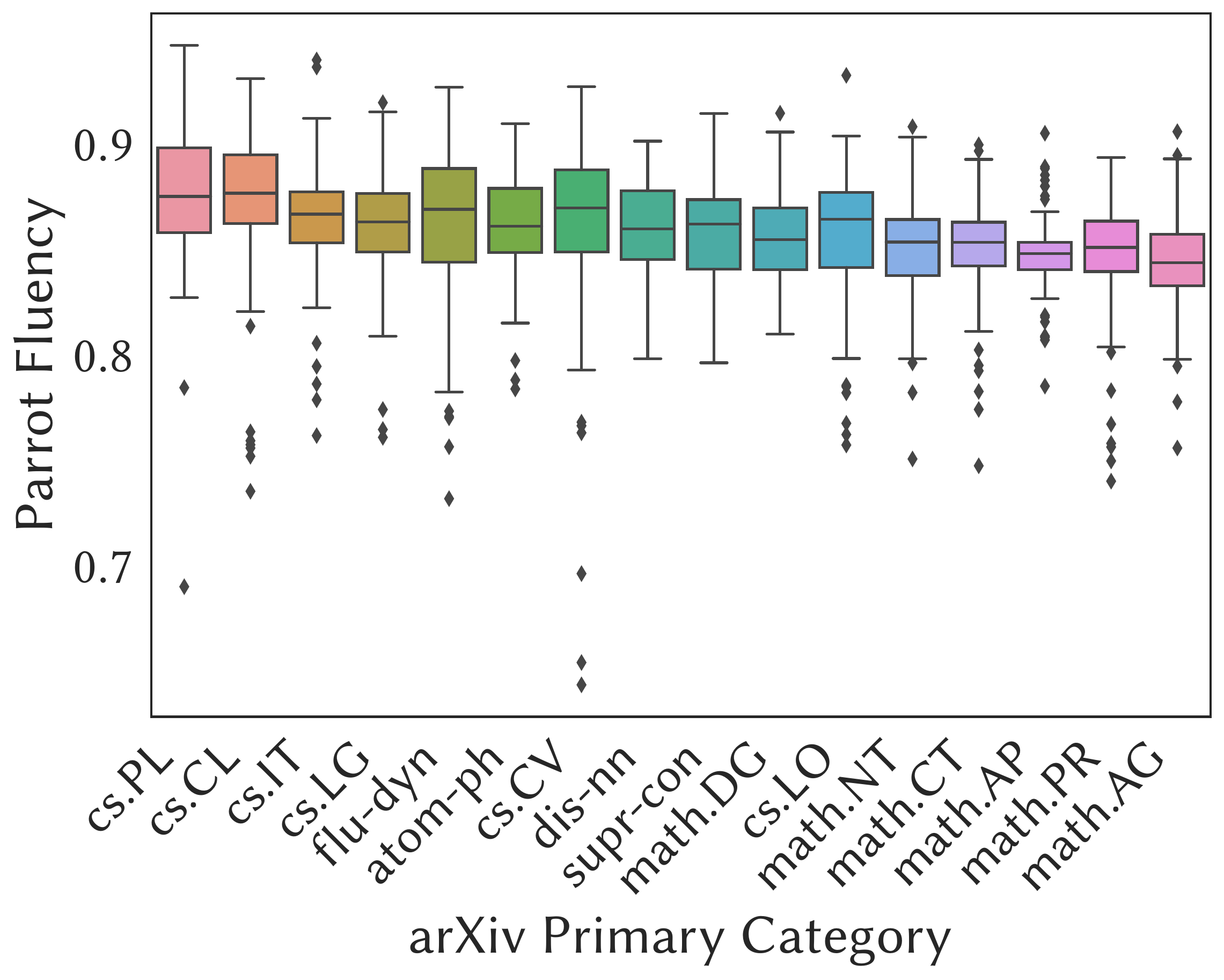}
	    \caption{\tiny Box plot of fluency score.}
        \label{fig:sanity:box}
	\end{subfigure}~
    \begin{subfigure}[t]{0.2\textwidth}
        \centering
        \includegraphics[width=\linewidth]{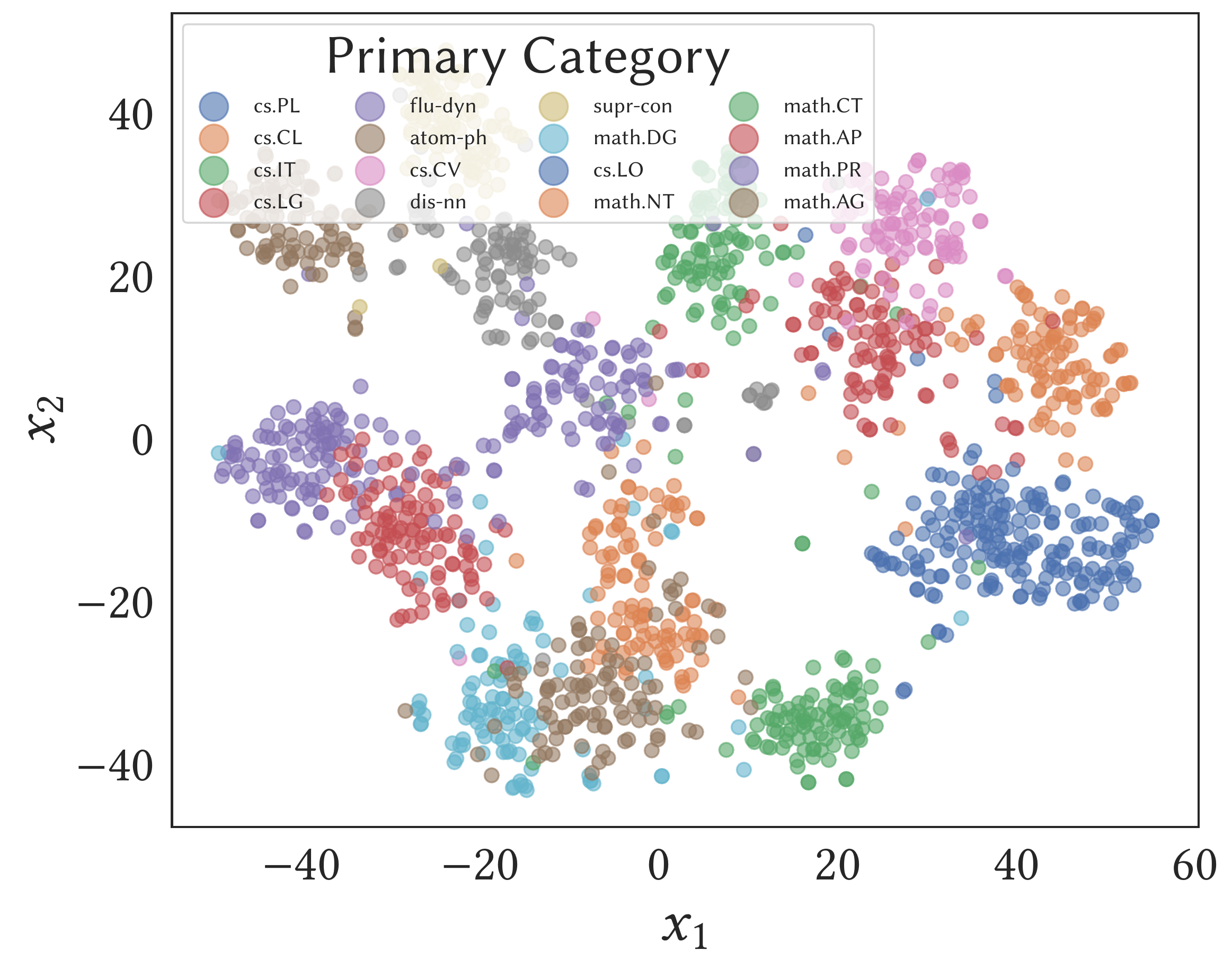}
    	\caption{Specter embedding.} 
     \label{fig:sanity:specter}
    \end{subfigure}%

        \caption{\textbf{Exploratory data analysis}. Roughly speaking, the first row focuses on the
        submissions, the second row on reviews and authors, and the third row on high-level features extracted from LMs. (m)-(o) are done on a control dataset we randomly sampled from arXiv to illustrate those high-level features work as expected.
        We provide
        more details and discussions of the plots in \Cref{sec:eda}.}
        \label{fig:hist}
\end{figure*}

\subsection{Academic Profile Data}
    To complement the author information, 
    we obtain academic profiles using the scholarly API\footnote{\url{https://pypi.org/project/scholarly}} for Google Scholar\footnote{\url{https://scholar.google.com}},
    which include the citation counts (as of Feb 2022),
    five-year citation counts, $h$-index, five-year
    $h$-index, $i10$-index, five-year $i10$-index,
    and per-year citation counts, $h$-index, and
    $i10$-index.
    Among all 21808 distinct authors, we are
    able to find 21031 associated Google Scholar profiles.
    There are 4180 authors report their Google Scholar
    profile in their OpenReview profile. For other authors,
    we concatenate their name with their latest institution (if reported) to form the search string,
    and select the most relevant query result from
    Google Scholar, which results in 3566 matches.
    If no results are found, we remove the institution
    string, search again, and select the most relevant query result. We are able to find another 4180
    matches.

\subsection{Institution Ranking}
    We obtain institutional rankings from
    CSRankings\footnote{\url{https://csrankings.org/}},
    a metric-based ranking of $582$ computer science institutions based on per-year citations counts of
    publications in various venues that are divided
    into $4$ areas (Artificial Intelligence, Systems,
    Theory, and Interdisciplinary Areas) and $26$
    sub-areas. We match the reported institution
    from author profiles with those in the CSRanking
    by thresholding the normalized Levenshtein distance
    at $0.8$ between
    institution names (both are uncased). We find
    $852$ matches out of $4745$ unique institutions.

    Nonetheless, since the ranking from CSRanking is the aggregated
    ranking weighting all sub-domain of computer science
    equally, it might not be the most representative; furthermore,
    the institutions available in the CSRanking only cover about
    $1/6$  institutions authors reported in OpenReview.
    To this end, we also consider a data-driven
    ranking that ranks each institute at a particular year
    by the total
    number of accepted papers in all previous years,
    which we refer to as the ``ICLR ranking.'' We
    show the top-$50$ institutions with highest
    ranks as of $200$, grouped by their per-year acceptance
    counts in \Cref{fig:iclrranking}.

\subsection{arXiv Matching}
    A crucial component in the analysis of reviewing stage
    bias is
    whether a submission is put onto the arXiv before the
    review releasing date. This provides us with a proxy to assess
    whether there might be a possible mechanism for the reviewers
    to gain information on authors' identities.
    We use the python wrapper for the arXiv API\footnote{\url{https://pypi.org/project/arxiv/}}
    to obtain the five most relevant results based
    on the title of the submission for subsequent filtering.
    Similar to \citet{kang2018peerread},
    we compute Jacard similarity
    and normalized  Levenshtein similarity between
    authors; in additional, we also use the Specter
    model\footnote{\url{https://huggingface.co/allenai/specter}} developed by \citet{specter2020cohan}
    to compute the cosine-similarity of the title-abstract
    embedding. We fine-tune the filtering threshold
    to be $0.5$ for all similarity measures
    such that the number of the matched papers are
    approximately the same as in \citet{tran2021an}.

\section{Language Model Enhancements}
    In this section, we describe how we augment the dataset
    by extracting various high-level features, mostly
    with the help of large language models.

    \subsection{Submission Features}
        We process all submissions in {\tt pdf} format in the dataset,
        with a total of 36GB to extract
        full texts.
        \paragraph{Summary statistics.}
         We use Grobid\footnote{\url{https://github.com/kermitt2/grobid}} to extract bibliographical data and obtain sections/figures/tables with their corresponding headings. We are able to obtain the counts of figures, tables, and sections
        in this way.

        \paragraph{Keywords.}
        Authors can optionally provide a list of keywords alongside
        with their submission. There are $9537$ submissions that provide a total of $36004$ keywords and among them on average
        $3.84$ keywords are provided per submission.
        On top of the raw keywords provided by authors,
        we also cluster keywords together by thresholding the Levenshtein
        distance~\cite{levenshtein1966binary}. Note that processed keyword clusters may correspond to
        different but relevant research topics (e.g., convex optimization vs.~non-convex optimization).

        \paragraph{Textual complexity.}
        An essential problem is to quantify the
        mathematical complexity of a paper.
        A ``superficial'' way of doing this is
        to assess how well the texts of the paper
        aligns with English grammar since equations and
        mathematical symbols  usually violate it.
        We used a pre-trained RoBERTa model used for
        assessing text fluency called
        Parrot\footnote{\url{ https://huggingface.co/prithivida/parrot_paraphraser_on_T5}}for this purpose. The fluency score ranges from $0$ to $1$, with
        a higher score represents less complicated texts.
        As a sanity check, we compute this fluency score on $100$
        randomly drawn papers from
        $16$ different arXiv categories spanning pure mathematics, physics,
        and various domains of computer science.
        In \Cref{fig:sanity:hist} and \Cref{fig:sanity:box}, we plot the histogram of complexity scores
        against $16$ categories. We note that the bulk of the scores
        concentrate around $0.85$ and the distributional difference
        is aligned with intuition (harder subjects such as algebraic geometry
        have a generally lower complexity score).

        \paragraph{Specter embedding.}
         The Specter embedding provides us a means
        of clustering submissions into different
        cluster through spectral clustering \cite{specter2020cohan}.
        We show the $t$-SNE \cite{van2008visualizing} plot in \Cref{fig:sanity:specter}, where each color/marker
        corresponds to a different arXiv primary
        category; we also show altogether the
        primary category of a few random data points.
        We observe the clusters are interpretable:
        ``language modelling'' and ``contrastive
        learning'' are far away from each other
        while ``deep learning'' are prevalent in
        many clusters. On a practical note, this embedding can
        be used to assess the relevance among submissions.
        
    \subsection{Review Features}
        \paragraph{Tone and sentiment.}
            Aside from integer-valued rating
            and confidence (and sometimes
            more aspects such as technical soundness),
            the tone or sentiment of the review
            may also affect the decision stage.
            We use the RoBERTa model trained on
            Twitter sentiment\footnote{\url{https://huggingface.co/cardiffnlp/twitter-roberta-base-sentiment}} to extract a sentiment score
            ranges between $0$ and $1$
            for each review ($1$ signifies most
            positive). Although this model
            was trained on Twitter, we found the
            sentiments it generates are highly
            correlated with the review rating,
            indicating a good representation, as
            shown in \Cref{fig:hist:revsen,fig:hist:revdec}.

\begin{figure*}
    \centering
	\begin{subfigure}[t]{0.2\textwidth}
	    \centering
	    \includegraphics[width=\linewidth]{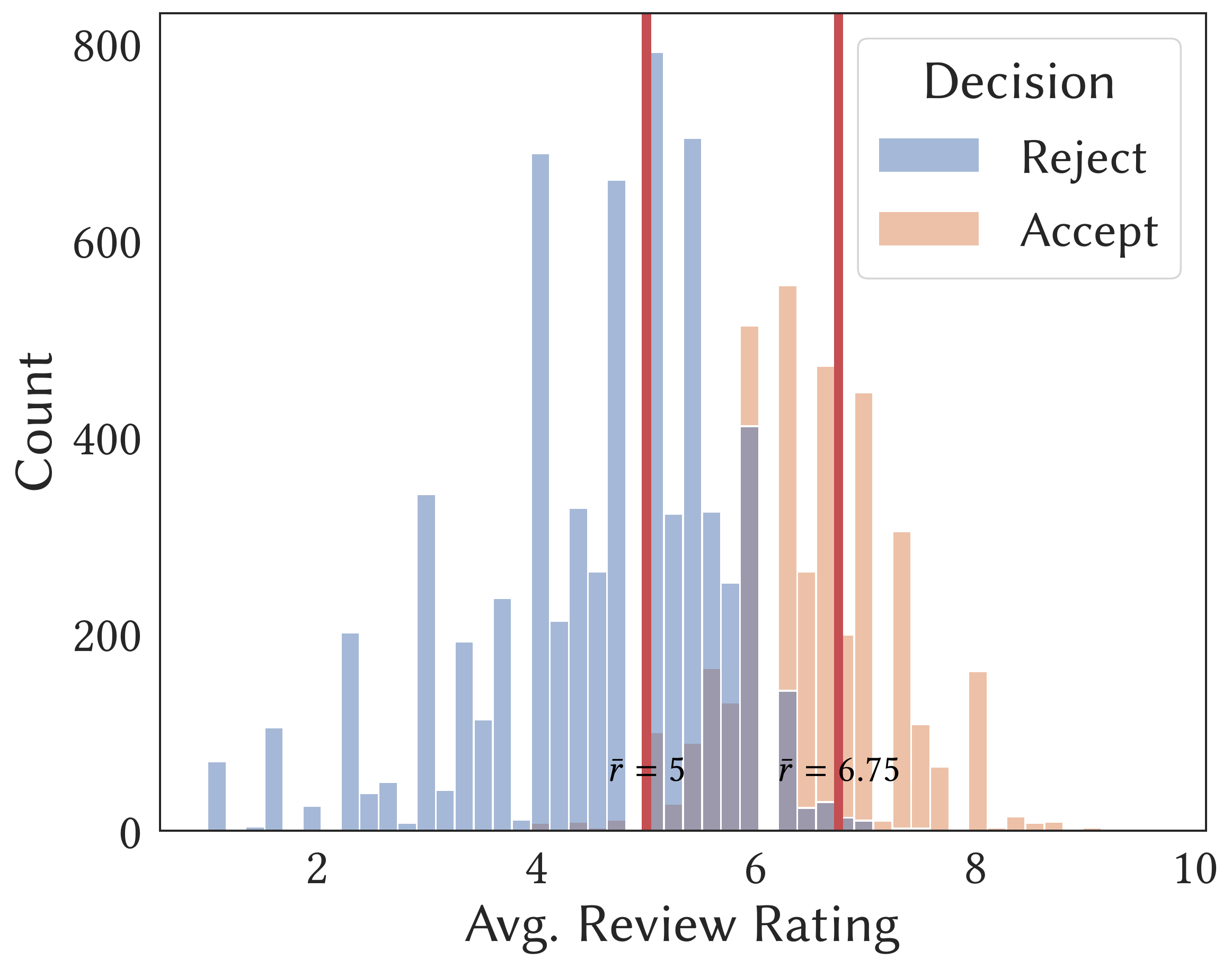}
	    \caption{\tiny Avg.~rating and borderline scores.} \label{fig:mar:hist}
	\end{subfigure}~
	\begin{subfigure}[t]{0.2\textwidth}
	    \centering
	    \includegraphics[width=\linewidth]{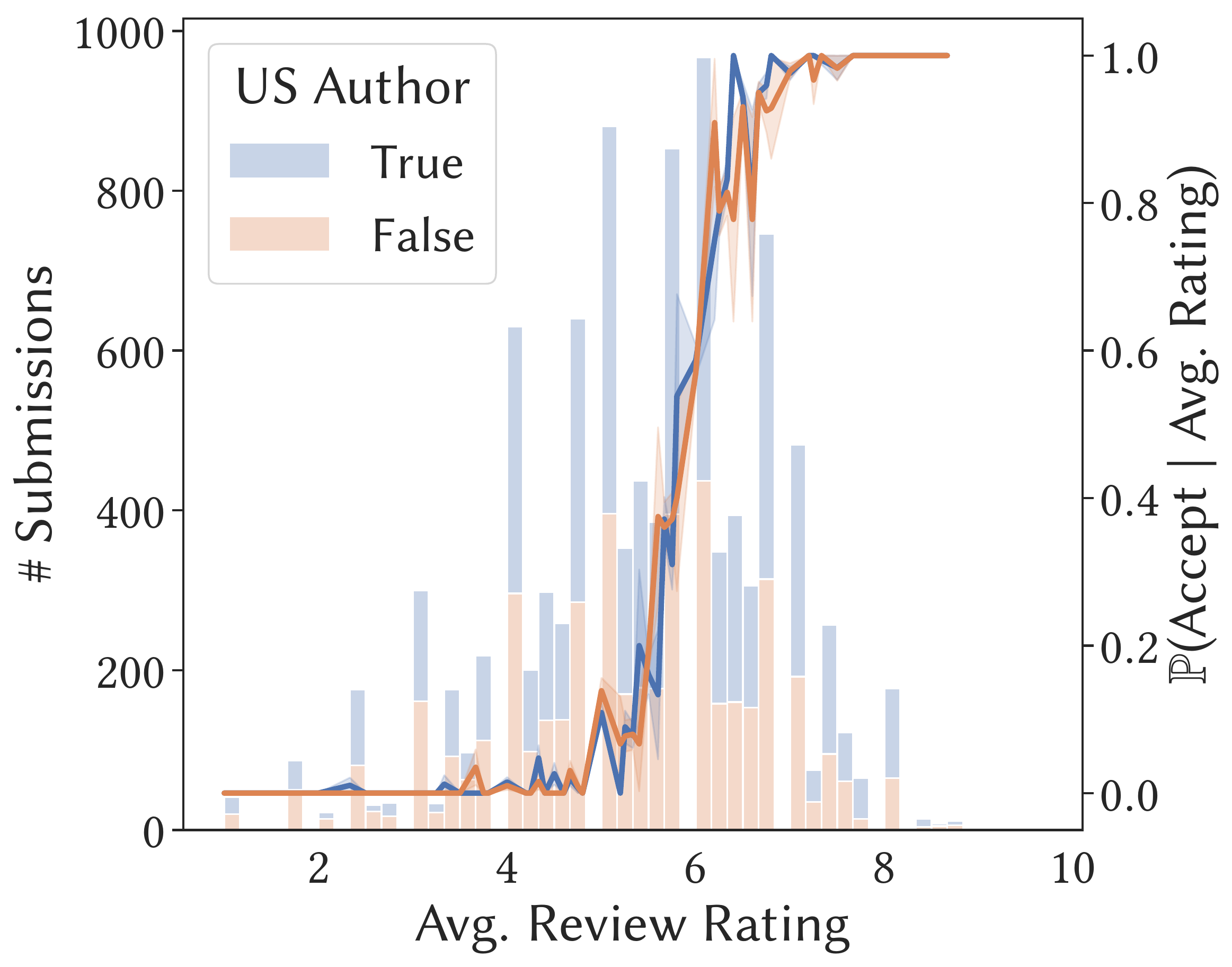}
	    \caption{\tiny Geographical disparity.} \label{fig:mar:na}
	\end{subfigure}~
    \begin{subfigure}[t]{0.2\textwidth}
	    \centering
	    \includegraphics[width=\linewidth]{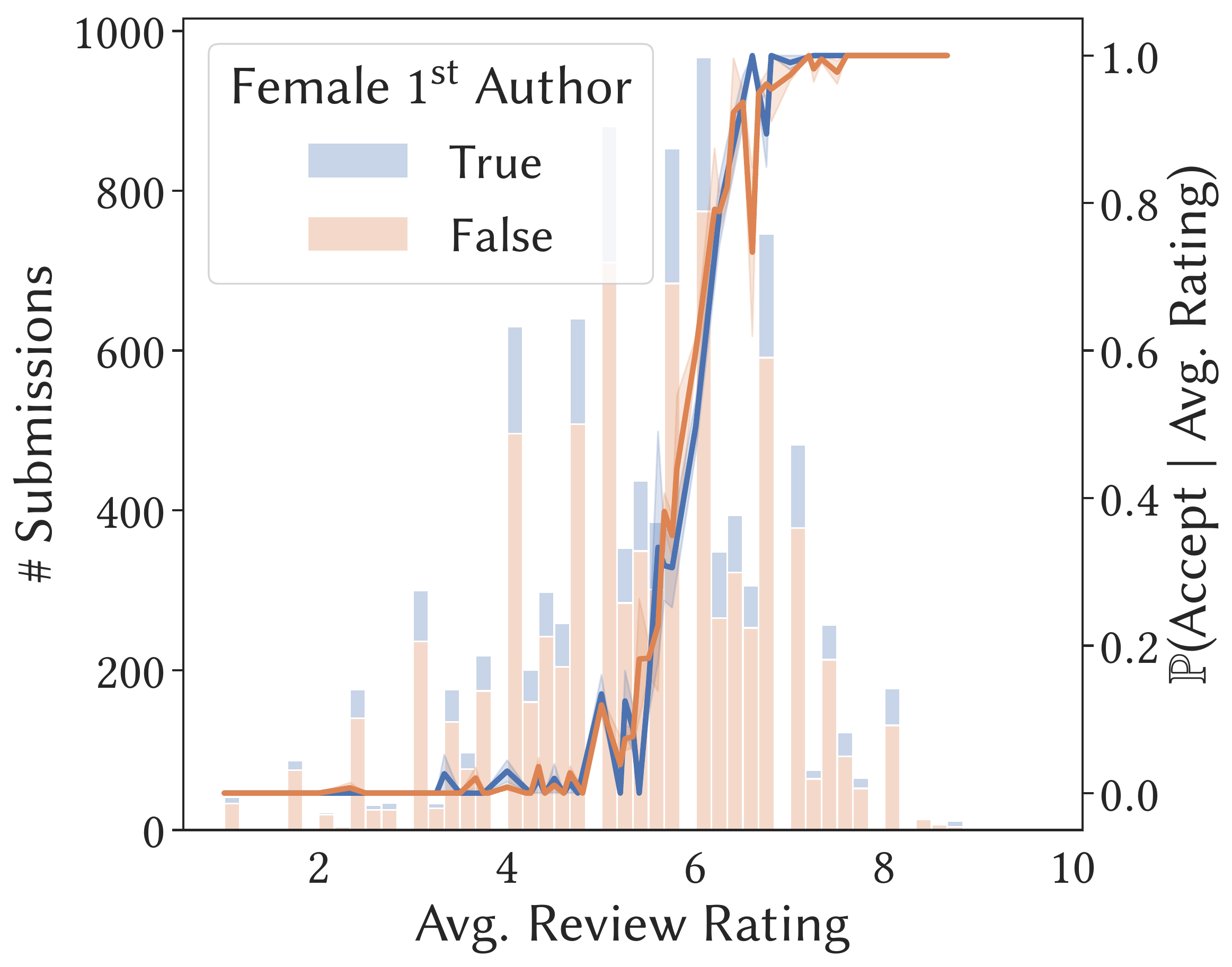}
	    \caption{\tiny Gender disparity.} \label{fig:mar:gender}
	\end{subfigure}~
    \begin{subfigure}[t]{0.2\textwidth}
	    \centering
	    \includegraphics[width=\linewidth]{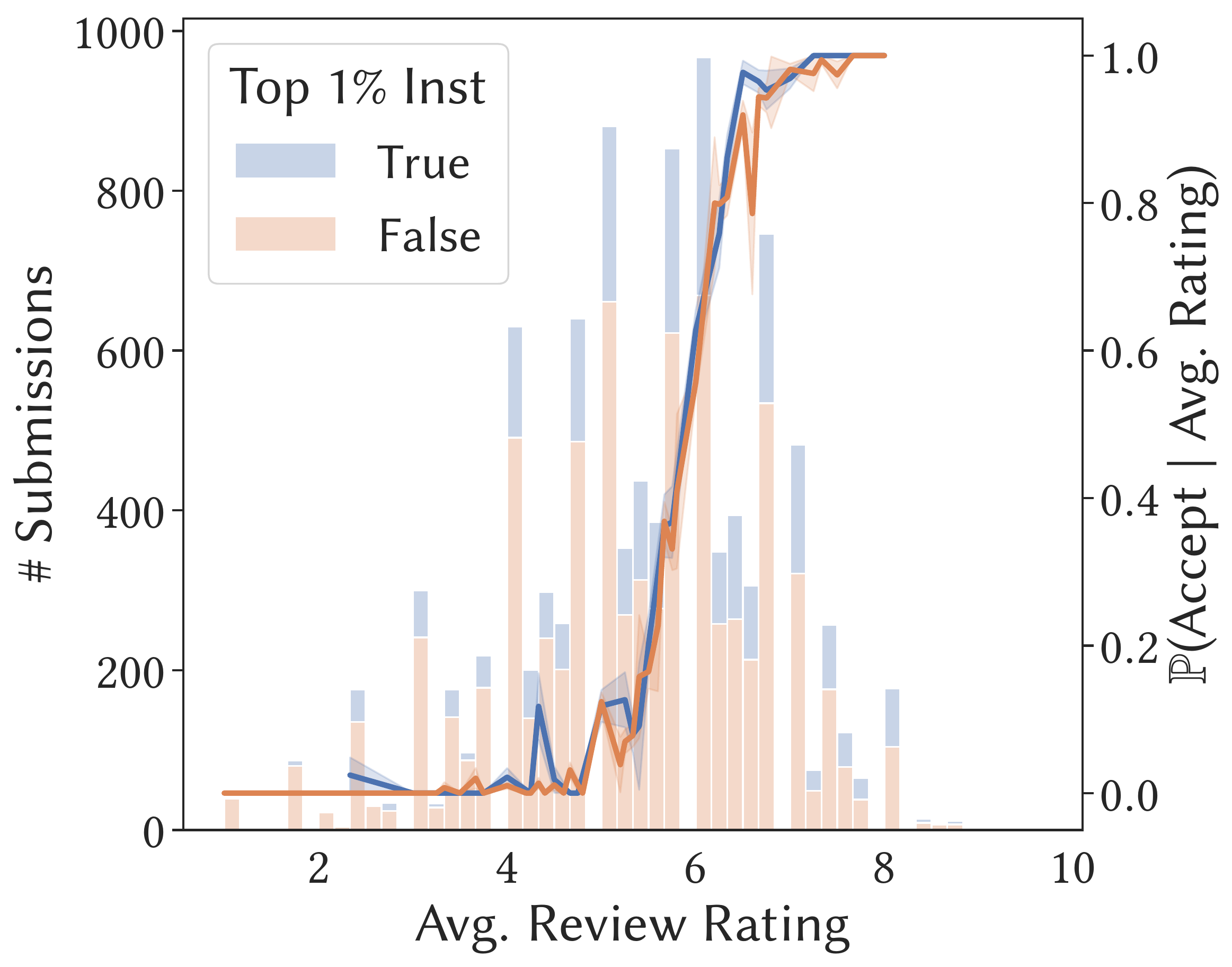}
	    \caption{\tiny Author prestige disparity.} \label{fig:mar:author}
	\end{subfigure}~
    \begin{subfigure}[t]{0.2\textwidth}
	    \centering
	    \includegraphics[width=\linewidth]{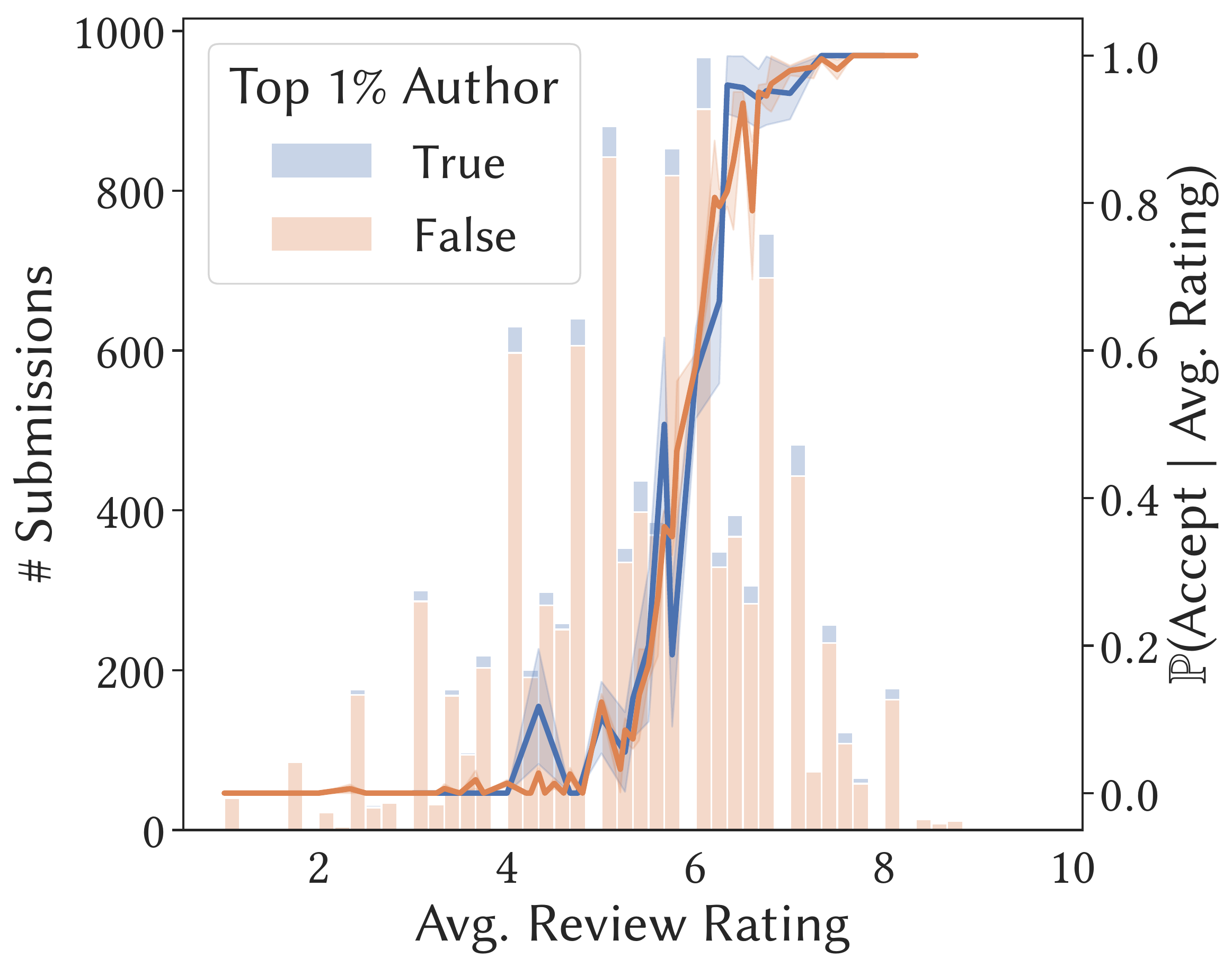}
	    \caption{\tiny Institution prestige disparity.} \label{fig:mar:inst}
	\end{subfigure}%
        \caption{\textbf{Acceptance probability conditioning on average review rating}. (a) Histogram of average review ratings of accepted/rejected papers with borderline range marked. (b)-(e) Empirical acceptance
        probability across groups. The horizontal axis signifies average rating, the left vertical axis corresponds to the histogram of samples from two groups and the right vertical axis corresponds to the probability. We observe that at certain ratings, there appears to be a statistically significant difference between two groups.}
        \label{fig:mar}
\end{figure*}

\subsection{Author Features}
        \paragraph{Reported and perceived gender.}
            Although each author may optionally report their
            gender information in their OpenReview profiles,
            this reported gender may not be the same as
            the \emph{perceived} gender
            We use the first name gender
            dictionary appraoch appeared in
            \cite{tsai2016cross}, which
            assigns a ``male'' score ranging from $0$
            to $1$ to each first name according to its
            frequency on Wikipedia.

        \paragraph{Geographical information.}
            We use the domain name from the email addresses
            of author profile to identify the geographical
            information of each author. Since the author affiliation might change over time, if the author
            provides its affiliation history, we use the
            email record at the year of the conference submission
            to identify the author's geography, and thus
            the same author's geography might change
            for different submissions.


\section{Investigating Fairness Disparities}
    In this section, we investigate fairness
    disparities in the decision process.
    For the ease of exposition,
    we dichotomize the sensitive attributes as
    (i) whether a paper's majority of authors are from
    north America;
    (ii) whether a paper's leading author is Female;
    (iii) whether a paper's most highly-cited authors falls
    in the top $1\%$ authors.
    We will first perform exploratory data analysis to explore
    potential fairness disparities in the data. We then zoom in
    several sensitive attributes of interest and study their marginal
    effects on the submission acceptance. To imitate the decision process, we fit predictive models given various features
    to predict acceptance probability. If the model captures this
    process well, we may use it as a surrogate to study the existence
    of fairness disparities and trace their roots.
    

\subsection{Exploratory Data Analysis} \label{sec:eda}

We first explore the dataset among
various dimensions and provide intuitive insights
into the dataset \cite{tukey1977exploratory}. We select several features of interest
and plot their relationships in \Cref{fig:hist}.

\paragraph{Submission.}
In \Cref{fig:hist:sublen,fig:hist:subrat,fig:hist:subrev}, we plot the histograms
of the length of all submissions (counting number of tokens using
the Longformer tokenizer developed \citet{Beltagy2020Longformer}), across different decisions,
average review rating and average review length. The rejected papers as a whole
are generally slightly shorter, while the review lengths do not differ much
for submissions of various lengths.
The derived features for submissions, fluency scores and Specter embeddings
are plotted in \Cref{fig:hist:subflu,fig:hist:subspe} across different
submission keywords. We observe that the paper complexity varies little
across the five most common keywords. In \Cref{fig:hist:subspe},
we select a subset of submissions that have a arXiv counterpart,
and visualize the $t$-SNE plot of their
Specter embeddings. We mark the top keywords of a random subset of the samples
and observe the similarities of submissions defined by primary categories or keywords are
well captured. Finally, in \Cref{fig:hist:arxiv}, we plot the histogram of the
best institution rank of a paper, grouped by whether the submissions was on arXiv
prior to review release.
\paragraph{Reviews}
We observe that the review rating type (negative, borderline, or above) does
not have a significance difference across reviews of various lengths (\Cref{fig:hist:revlen}). The extracted review sentiment feature, however,
has a distributional difference across accept/reject groups (\Cref{fig:hist:revdec}),
and is correlated with average review rating (\Cref{fig:hist:revsen}), which
aligns with our intuition.
\paragraph{Author.} 
As shown in \Cref{tab:data:summary}, the distribution across gender groups in 
all authors differ significantly, which is further quantified in \Cref{fig:hist:repgen}.
Together with \Cref{fig:hist:pergen}, which shows the histogram of perceived gender
across reported gender groups with random examples shown, we observe the perceived
gender is in general aligned with reported gender, and is more preferable
in downstream analyses due to its higher coverage. Finally in \Cref{fig:hist:citation},
we note the distribution of citation counts of authors in different
gender groups are similar.

\subsection{Marginal Disparity Analysis}
\begin{figure}[t]
\centering
	\begin{subfigure}[t]{0.45\columnwidth}
	    \centering
	    \includegraphics[width=\linewidth]{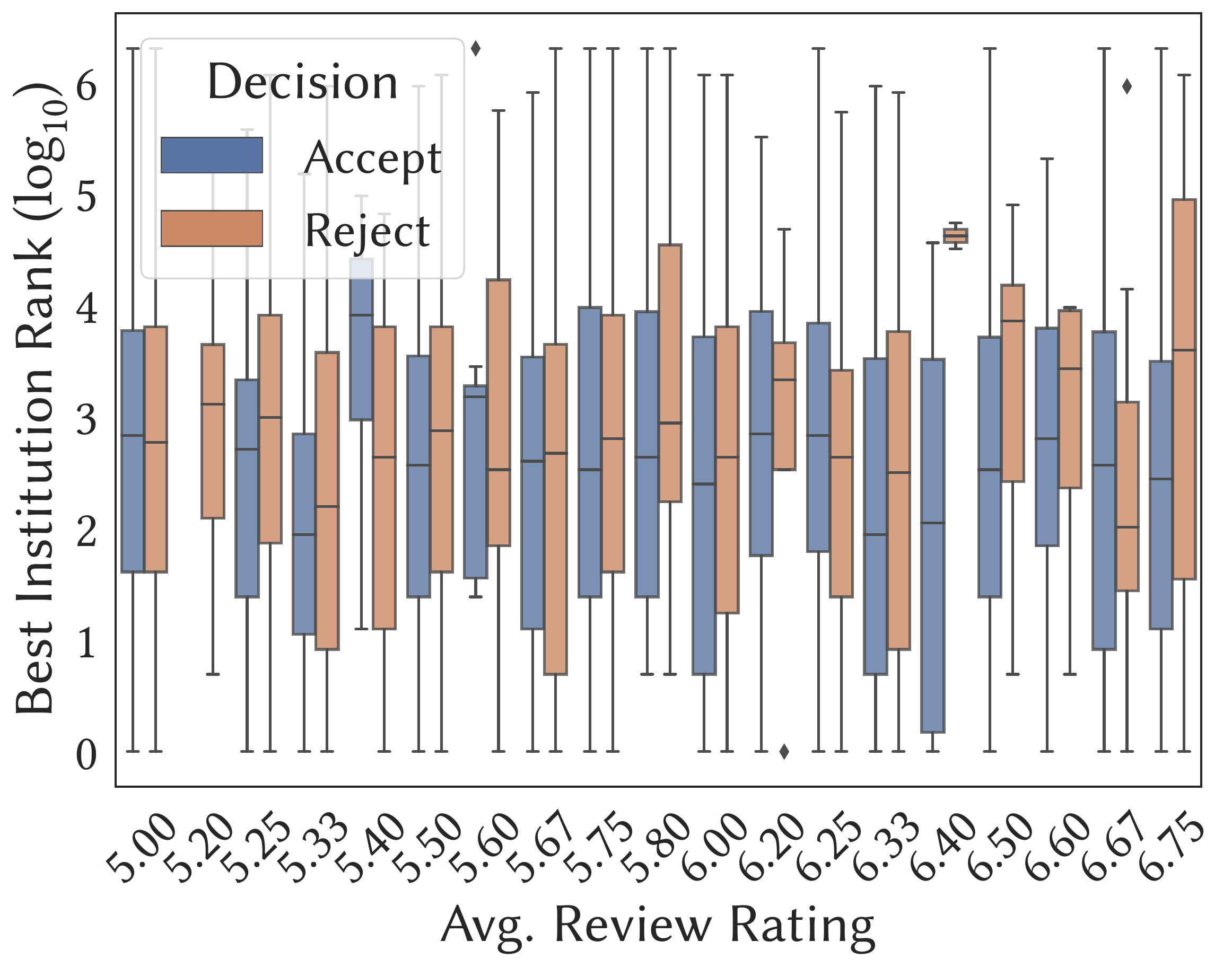}
	    \caption{\tiny Best author institution.} 
     \label{fig:box:inst}
	\end{subfigure}\hspace{1em}
    \begin{subfigure}[t]{0.45\columnwidth}
	    \centering
	    \includegraphics[width=\linewidth]{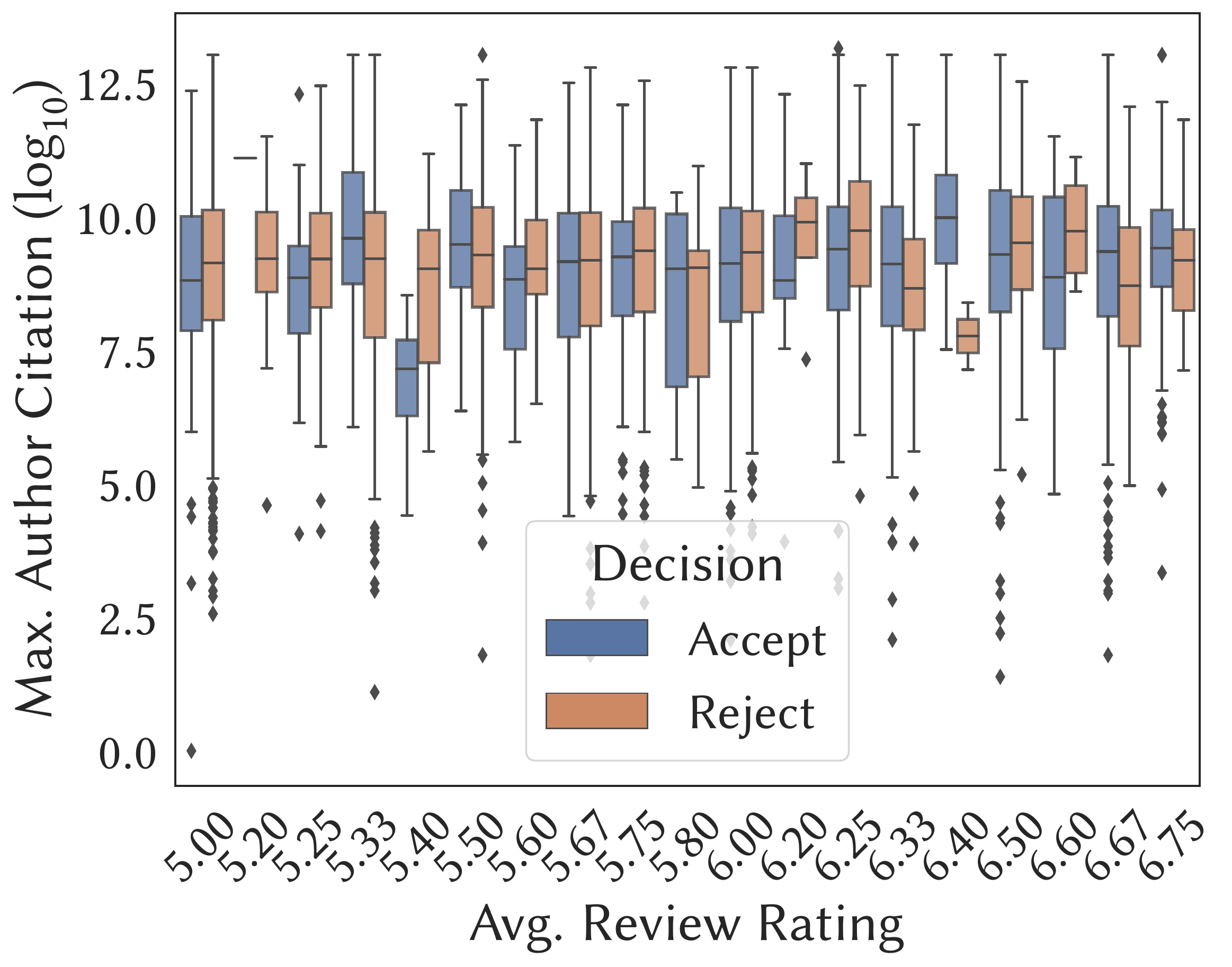}
	    \caption{\tiny Best author citation.}
        \label{fig:box:author}
	\end{subfigure}%
    \caption{\textbf{Decision for papers of various prestige among
    borderline scores.} Note that although samples overlap, their means
    can differ significantly at certain score levels.} \label{fig:box}
    
\end{figure}

We first study fairness disparities based on the \emph{marginals}
of various sensitive attributes while averaging out all other factors given reviewer ratings.
In this setup, we estimate the (empirical) marginal probability
of a acceptance at each average rating level and construct confidence
bands based on sample variances.
We summarize the results in \Cref{fig:mar,fig:box}.
We note that the empirical acceptance probabilities
are visually similar across four sensitive attributes considered. Nonetheless, we note at certain level of average rating (mostly borderline ratings between $5$ to $7$), some sensitive
attributes exhibit statistically significant difference (e.g., in \Cref{fig:mar:na} around rating
$6.25$). In \Cref{fig:box}, we zoom in to the borderline papers for the prestige disparity (measured by author institution ranking and citation count at the year of submission). We found that although the ranges of accepted/rejected
papers overlap, their means can be strikingly different.
Marginal analysis provides us with rough idea
of when disparities may occur but falls short of
providing explanations, which will be the focus of the next subsection.
    
\subsection{Joint Disparity Analysis}
\begin{figure}[t]
\centering
	\begin{subfigure}[t]{0.5\columnwidth}
	    \centering
	    \includegraphics[width=\linewidth]{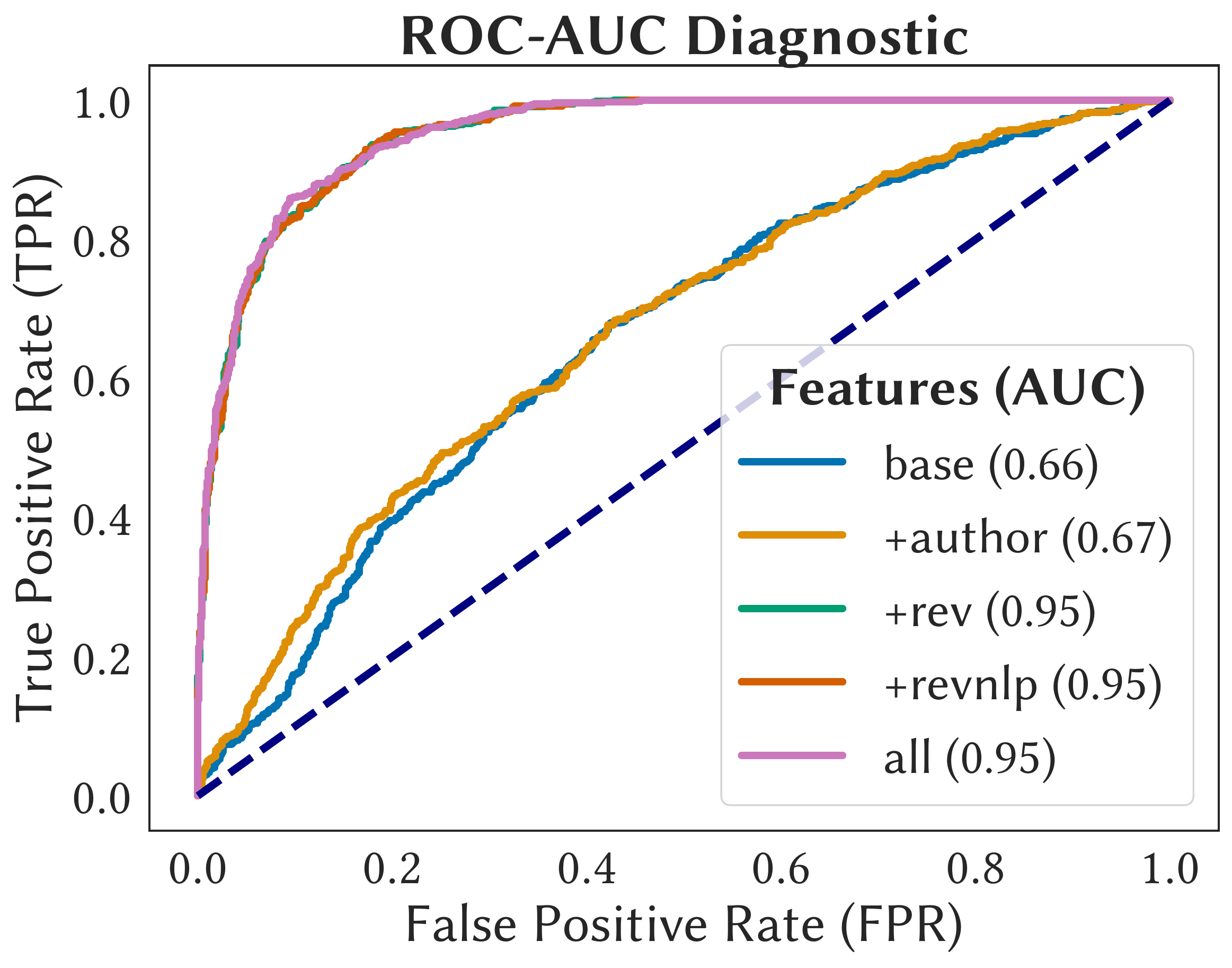}
	    \caption{\tiny ROC curve.} 
     \label{fig:glm:roc}
	\end{subfigure}\hfill
    \begin{subfigure}[t]{0.5\columnwidth}
	    \centering
	    \includegraphics[width=\linewidth]{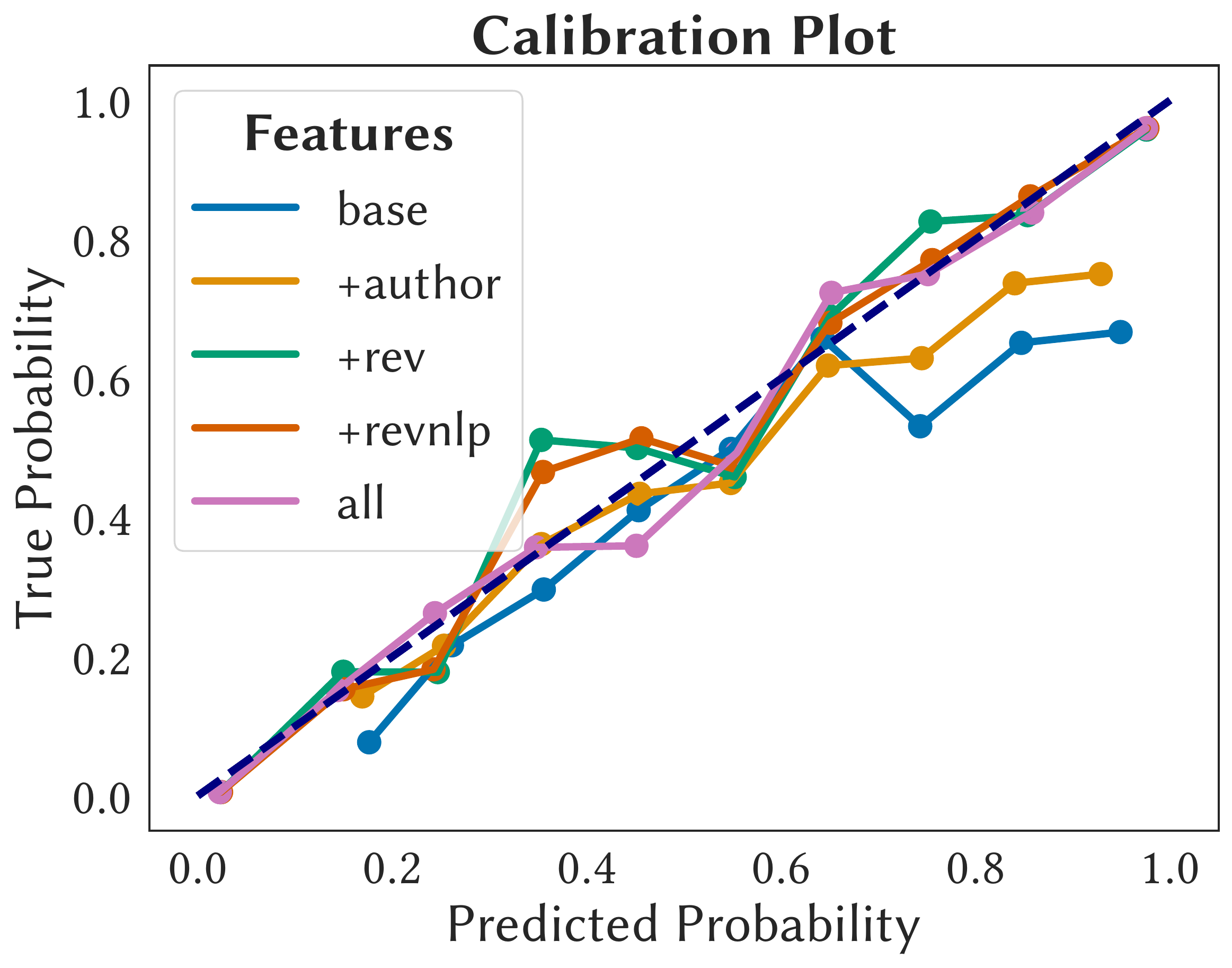}
	    \caption{\tiny Calibration curve.}
        \label{fig:glm:cal}
	\end{subfigure}%
    \caption{\textbf{Diagnosis plots for decision-stage predictive models.} The ROC curve together with AUC captures the discrimination power of classifiers and the calibration curve assesses its bias. We note that with review features included, simple models such as logistic regression captures the decision stage well.}\label{fig:glm}
    
\end{figure}
Although the estimations of the marginal probability
and their confidence bands are non-parametric thus not relying on
any modelling assumptions, it is inconvenient to condition on multiple
factors simultaneously. In this subsection, we aim at studying fairness violations in the decision stage
using predictive models, which also brings the additional benefits of
selecting the most important features for the prediction.

\begin{figure*}[t]
    \centering
    \begin{subfigure}[t]{0.2\textwidth}
	    \centering
	    \includegraphics[width=\linewidth]{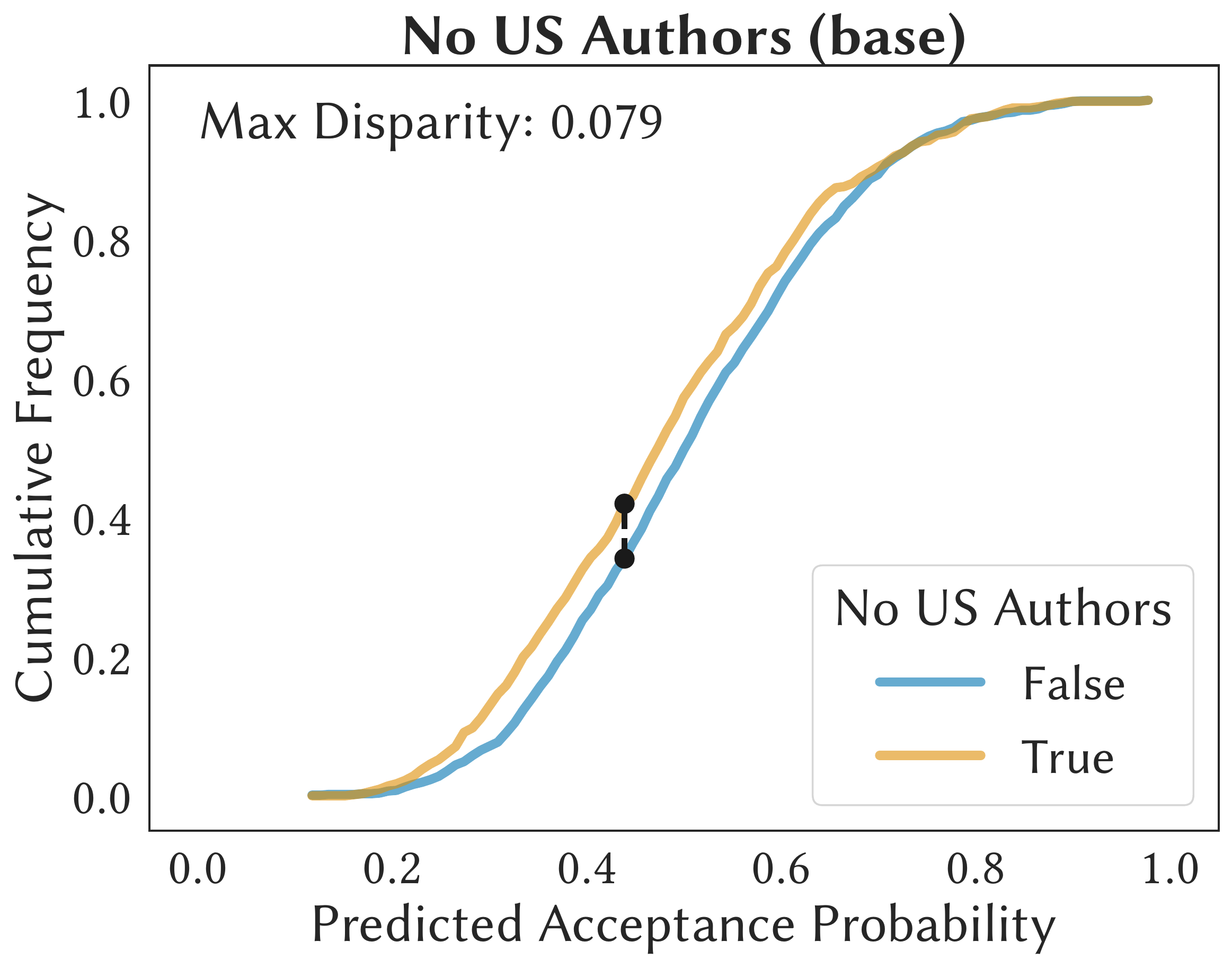}
	    \caption{\tiny US Authors (base).}  \label{fig:dis:nous:b}
	\end{subfigure}%
	\begin{subfigure}[t]{0.2\textwidth}
	    \centering
	    \includegraphics[width=\linewidth]{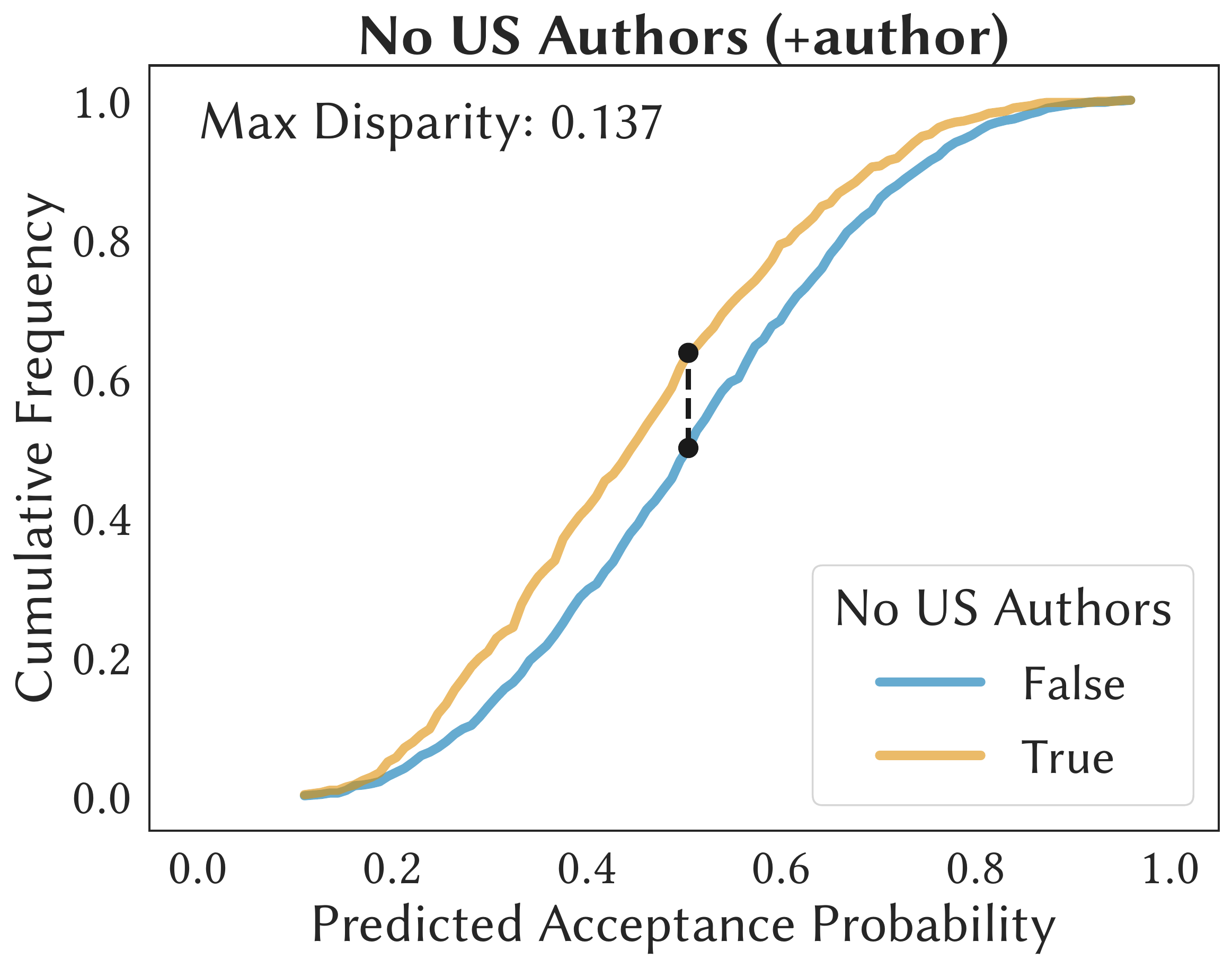}
        \caption{\tiny US Authors (+author).}  \label{fig:dis:nous:a}
	\end{subfigure}%
	\begin{subfigure}[t]{0.2\textwidth}
	    \centering
	    \includegraphics[width=\linewidth]{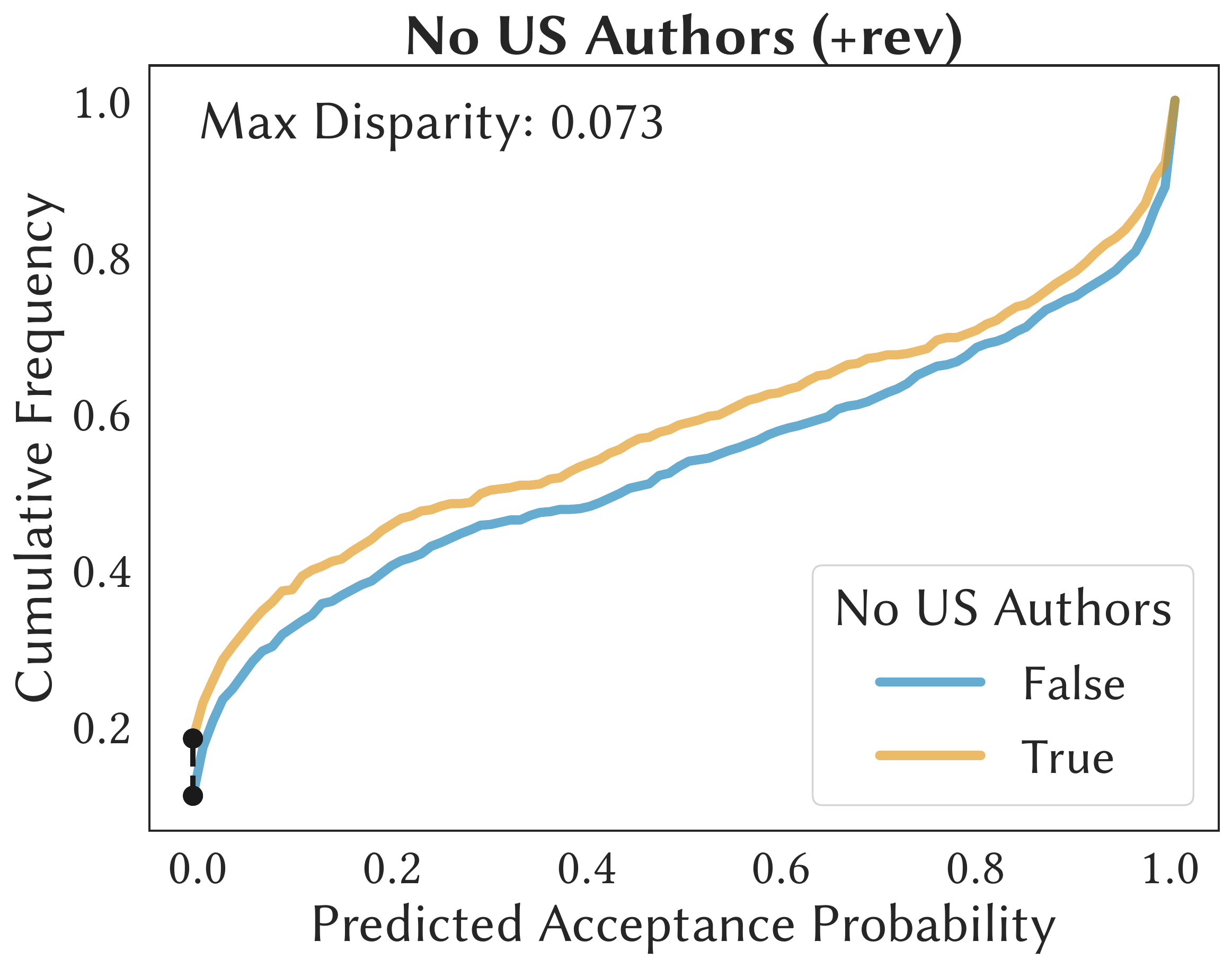}
\caption{\tiny US Authors (+rev).}  \label{fig:dis:nous:r}
	\end{subfigure}%
\begin{subfigure}[t]{0.2\textwidth}
	    \centering
	    \includegraphics[width=\linewidth]{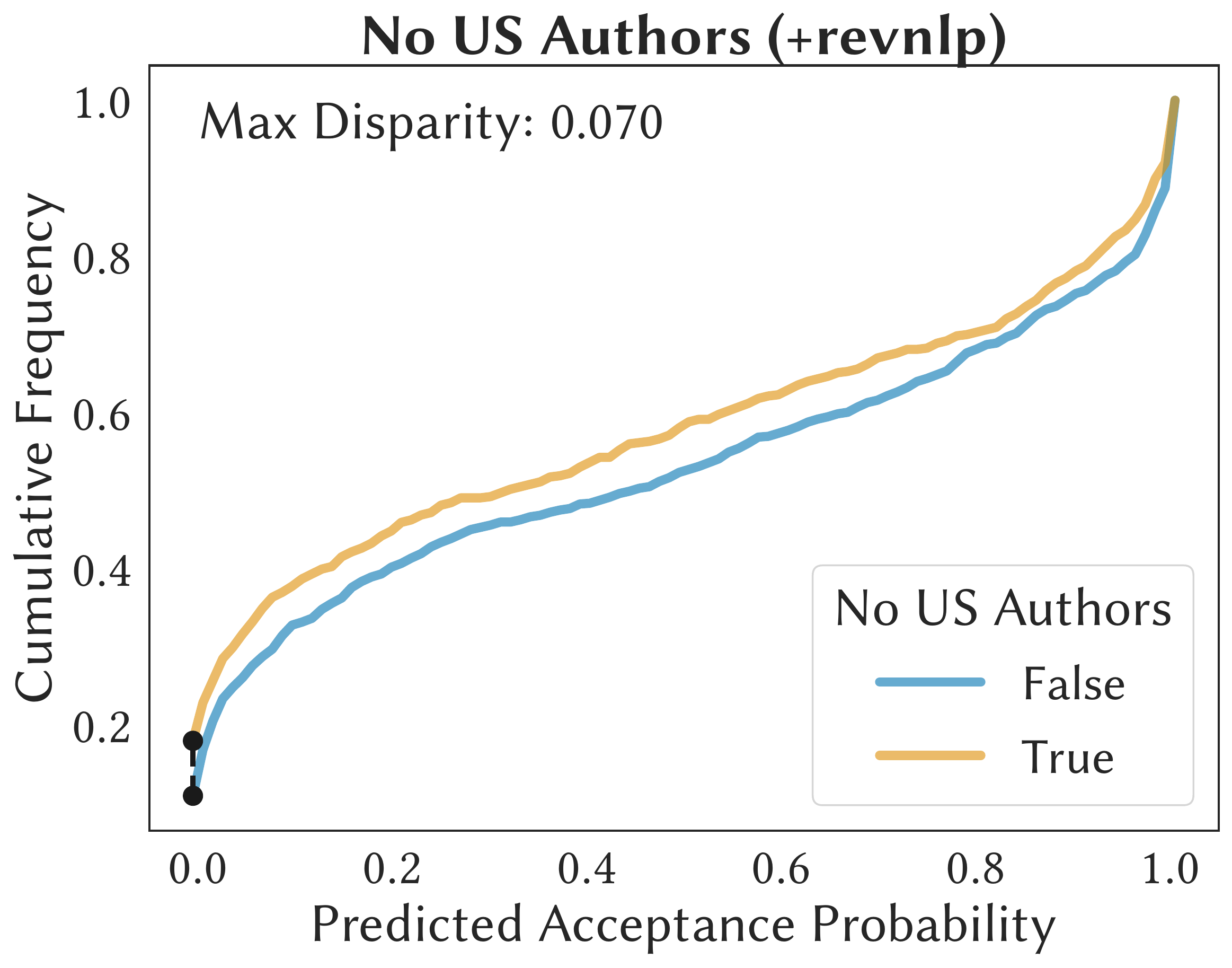}
\caption{\tiny US Authors (+revnlp).}  \label{fig:dis:nous:rn}
	\end{subfigure}%
     \begin{subfigure}[t]{0.2\textwidth}
	    \centering
	    \includegraphics[width=\linewidth]{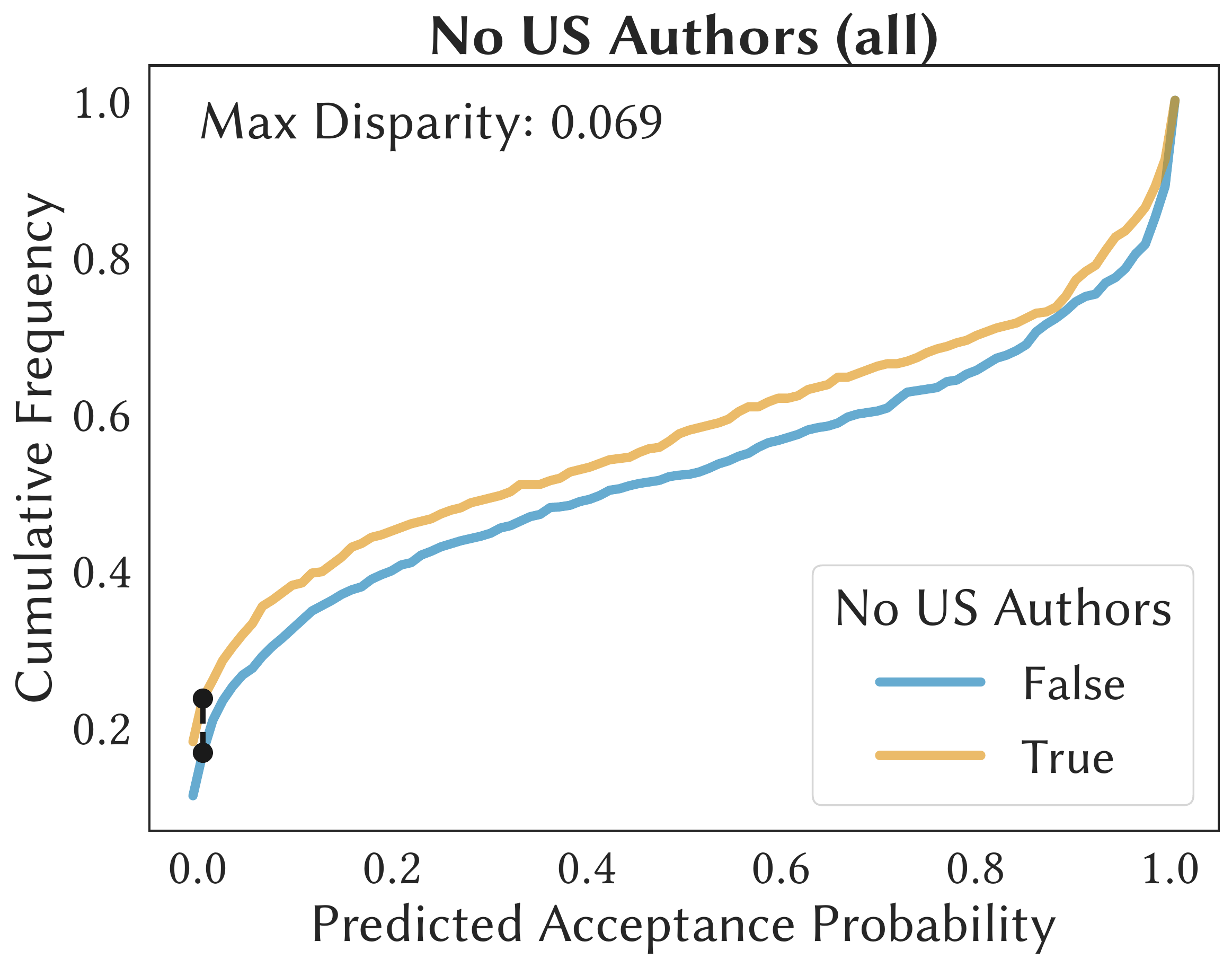}
\caption{\tiny US Authors (all).}  \label{fig:dis:nous:all}
	\end{subfigure}\\

    \begin{subfigure}[t]{0.2\textwidth}
	    \centering
	    \includegraphics[width=\linewidth]{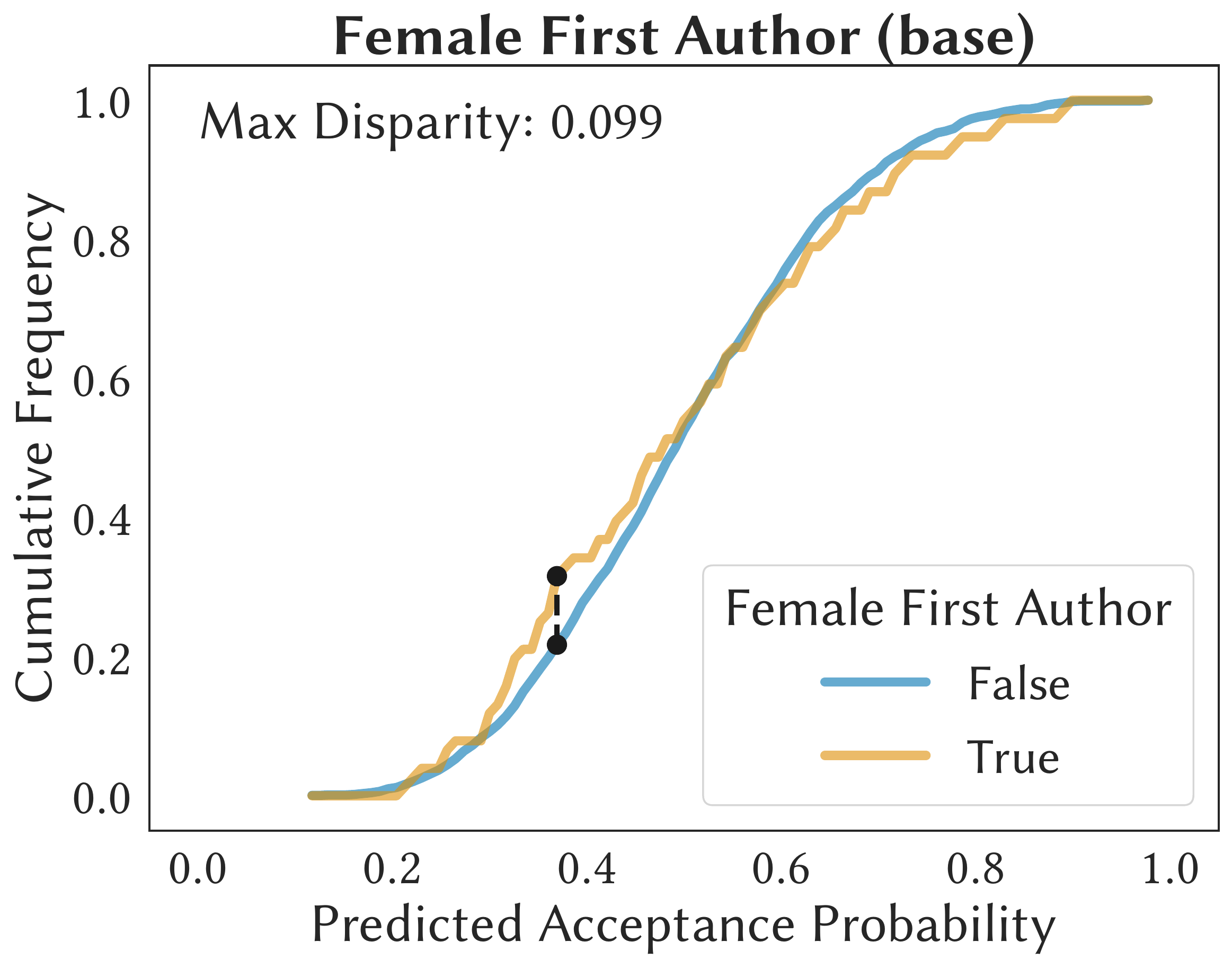}
	    \caption{\tiny Female $1^{\text{st}}$ Author (base).}  \label{fig:dis:fstf:b}
	\end{subfigure}%
	\begin{subfigure}[t]{0.2\textwidth}
	    \centering
	    \includegraphics[width=\linewidth]{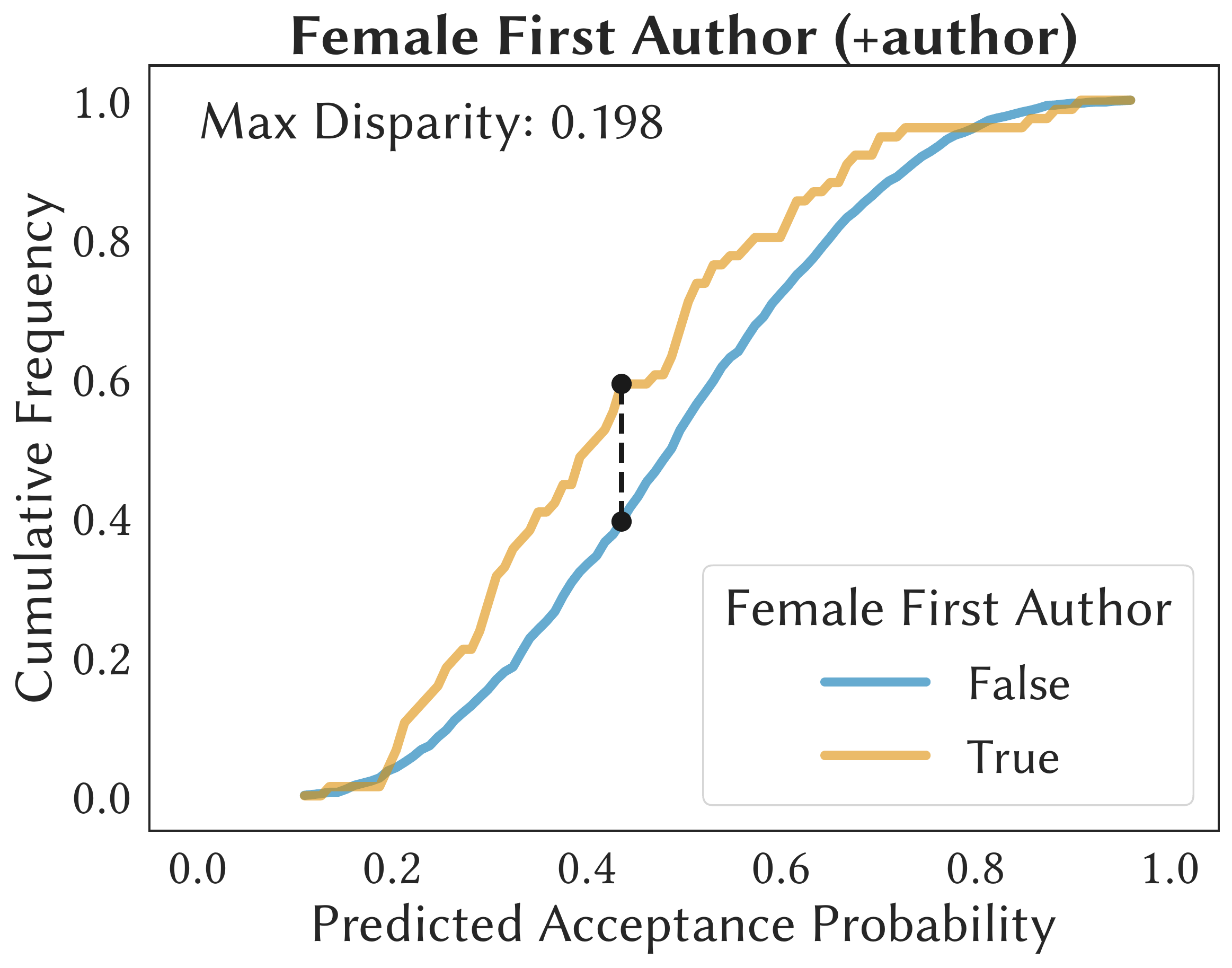}
        \caption{\tiny Female $1^{\text{st}}$ Author (+author).}  \label{fig:dis:fstf:a}
	\end{subfigure}%
	\begin{subfigure}[t]{0.2\textwidth}
	    \centering
	    \includegraphics[width=\linewidth]{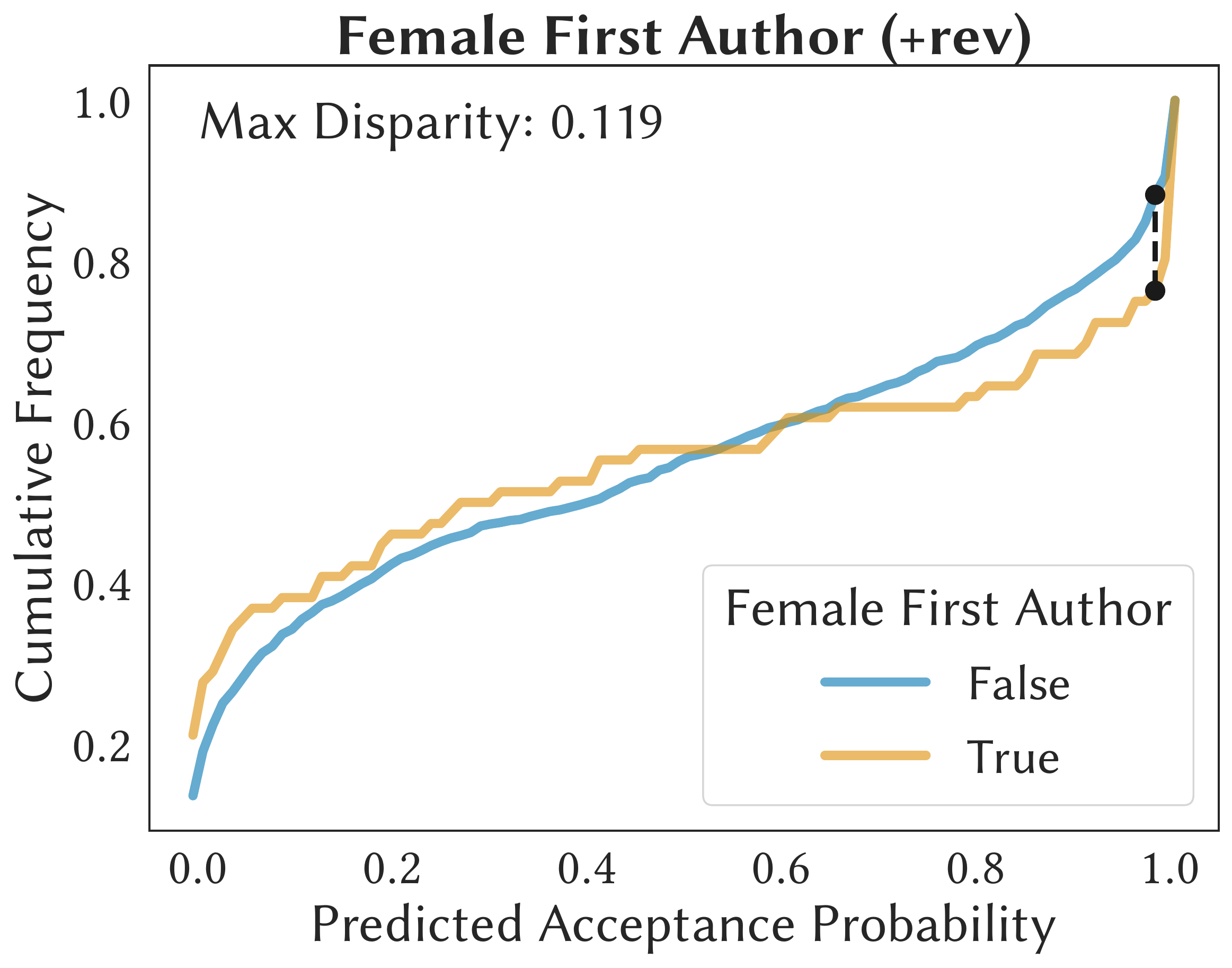}
\caption{\tiny Female $1^{\text{st}}$ Author (+rev).}  \label{fig:dis:fstf:r}
	\end{subfigure}%
\begin{subfigure}[t]{0.2\textwidth}
	    \centering
	    \includegraphics[width=\linewidth]{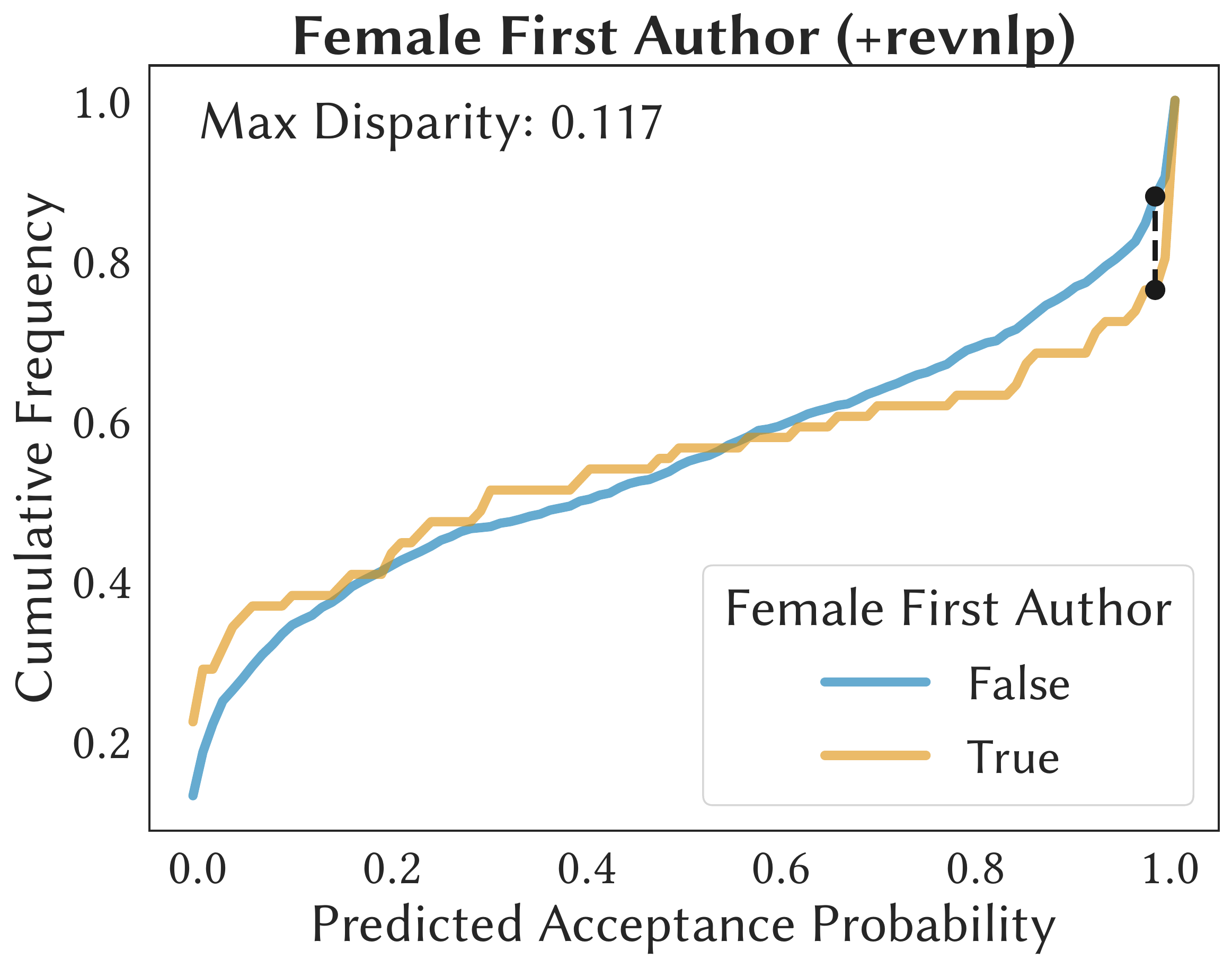}
\caption{\tiny Female $1^{\text{st}}$ Author (+revnlp).}  \label{fig:dis:fstf:rn}
	\end{subfigure}%
     \begin{subfigure}[t]{0.2\textwidth}
	    \centering
	    \includegraphics[width=\linewidth]{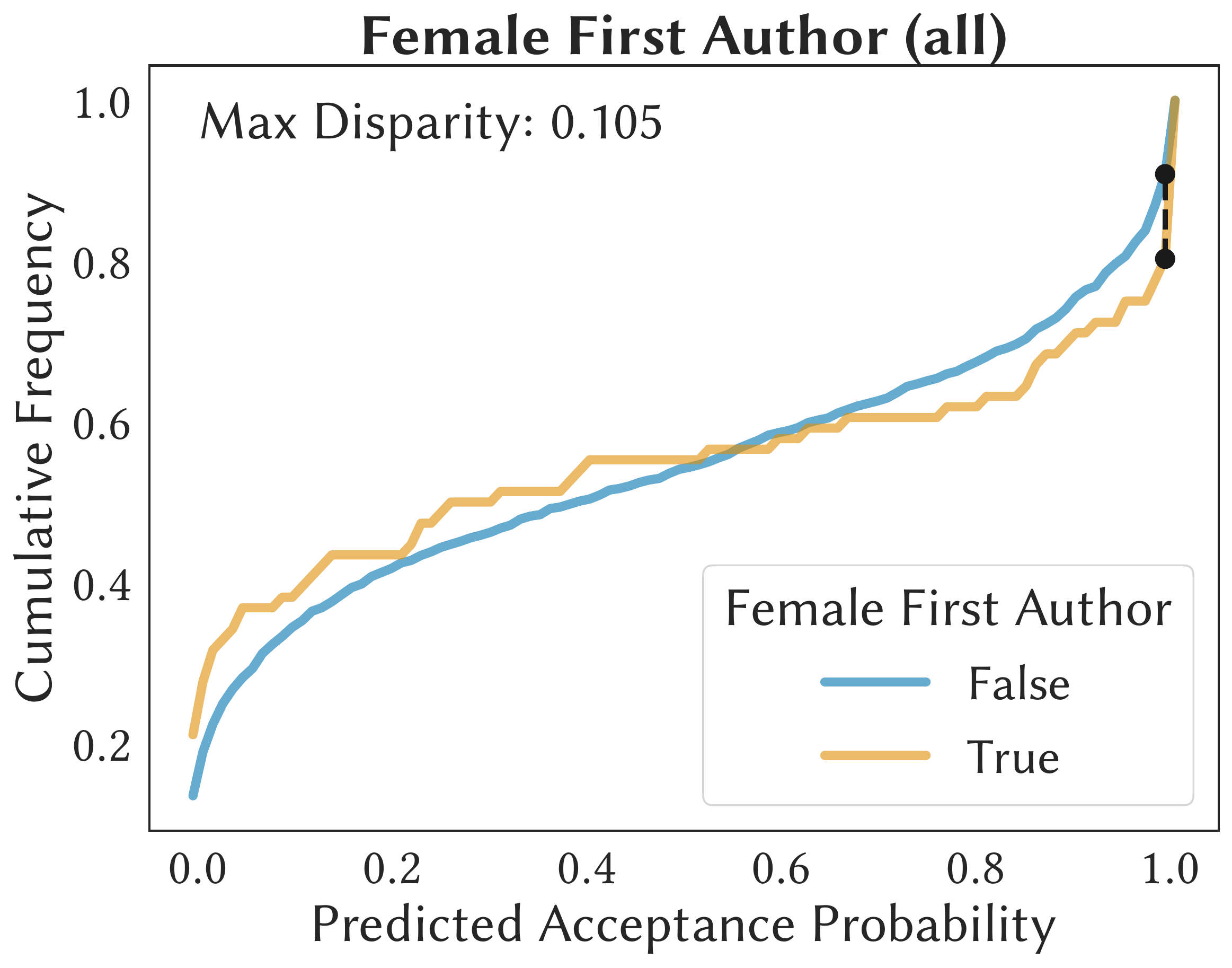}
\caption{\tiny Female $1^{\text{st}}$ Author (all).}  \label{fig:dis:fstf:all}
	\end{subfigure}\\

    \begin{subfigure}[t]{0.2\textwidth}
	    \centering
	    \includegraphics[width=\linewidth]{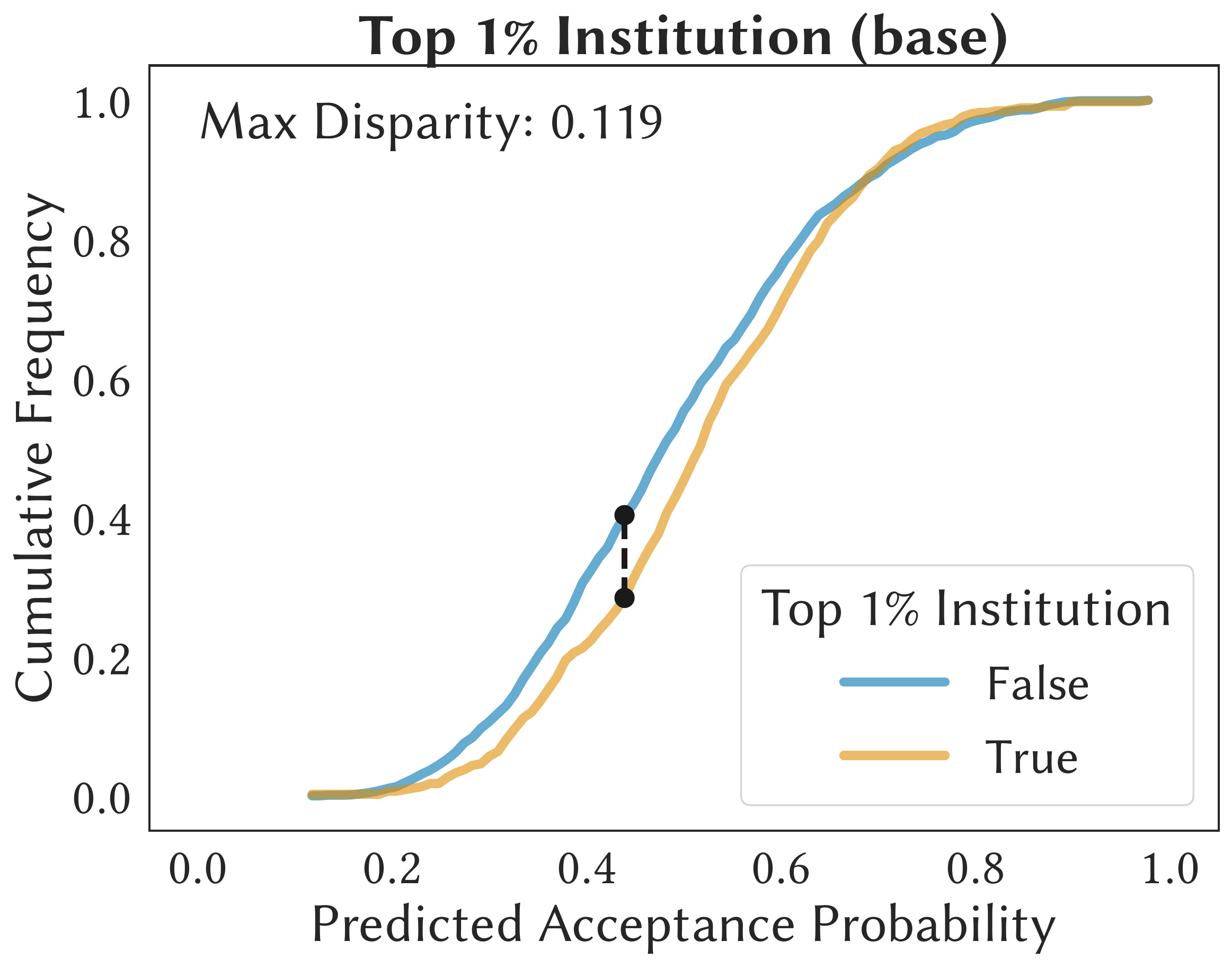}
	    \caption{\tiny Top Inst Author (base).}  \label{fig:dis:tinst:b}
	\end{subfigure}%
	\begin{subfigure}[t]{0.2\textwidth}
	    \centering
	    \includegraphics[width=\linewidth]{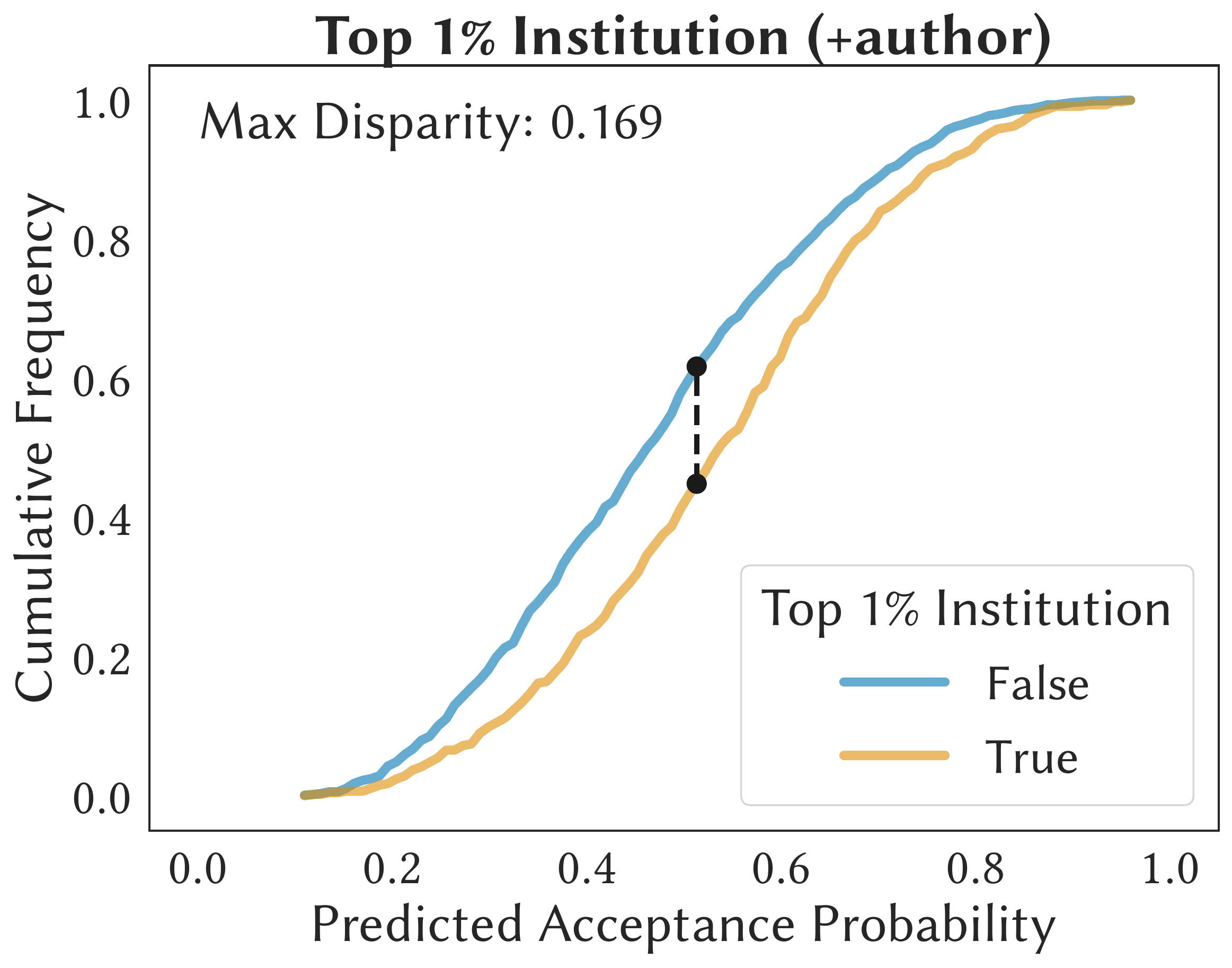}
        \caption{\tiny Top Inst Author (+author).}  \label{fig:dis:tinst:a}
	\end{subfigure}%
	\begin{subfigure}[t]{0.2\textwidth}
	    \centering
	    \includegraphics[width=\linewidth]{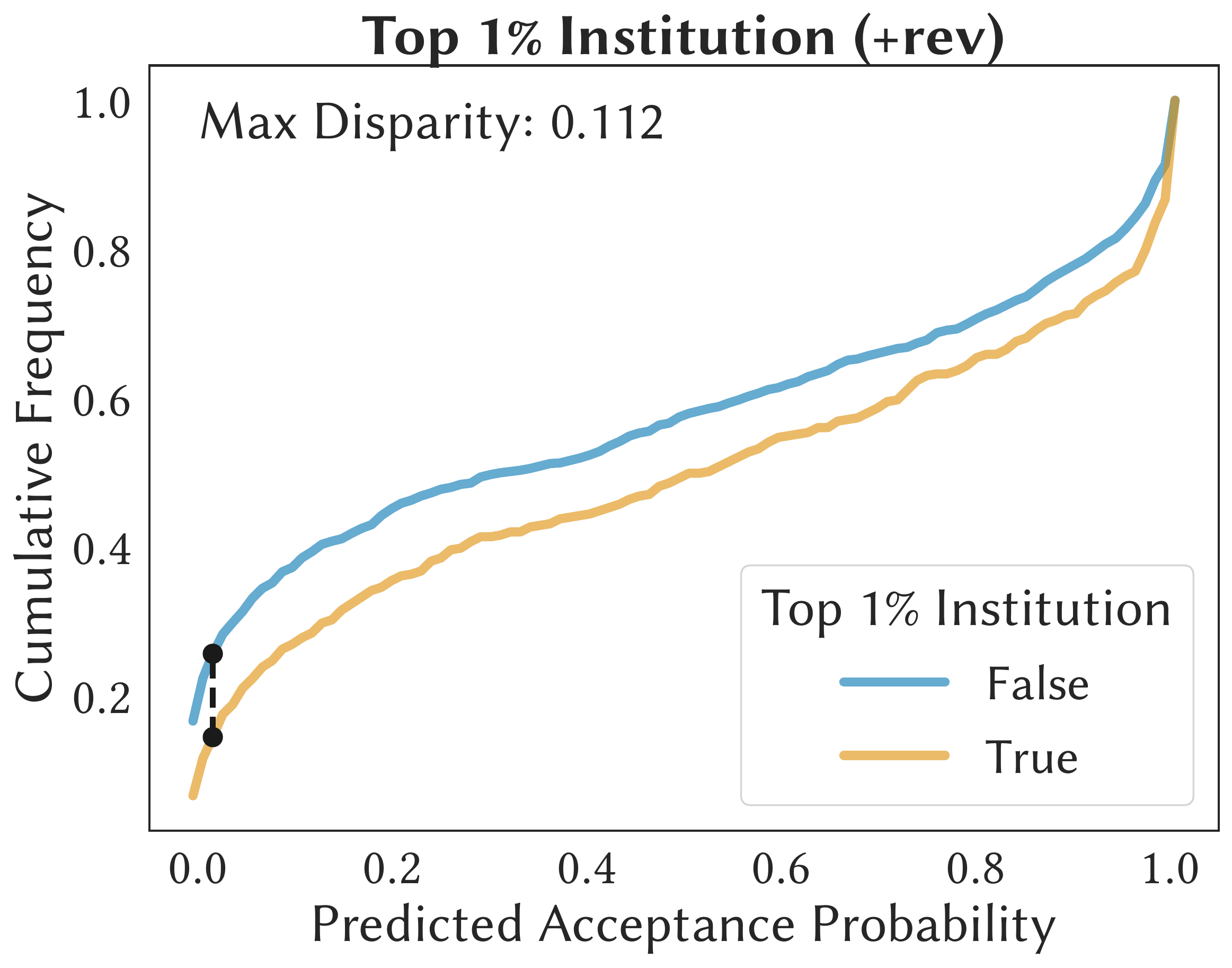}
\caption{\tiny Top Inst Author (+rev).}  \label{fig:dis:tinst:r}
	\end{subfigure}%
\begin{subfigure}[t]{0.2\textwidth}
	    \centering
	    \includegraphics[width=\linewidth]{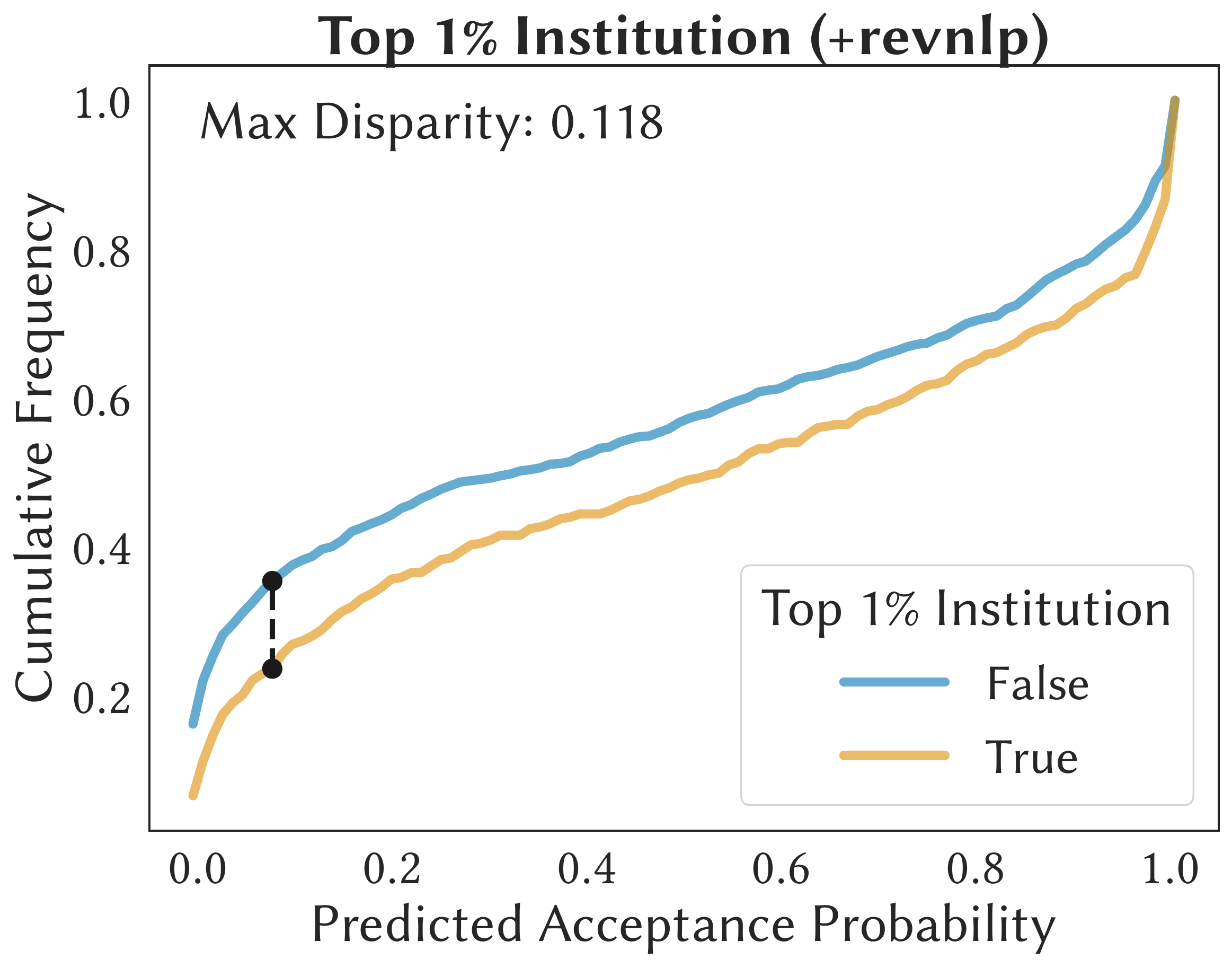}
\caption{\tiny Top Inst Author (+revnlp).}  \label{fig:dis:tinst:rn}
	\end{subfigure}%
     \begin{subfigure}[t]{0.2\textwidth}
	    \centering
	    \includegraphics[width=\linewidth]{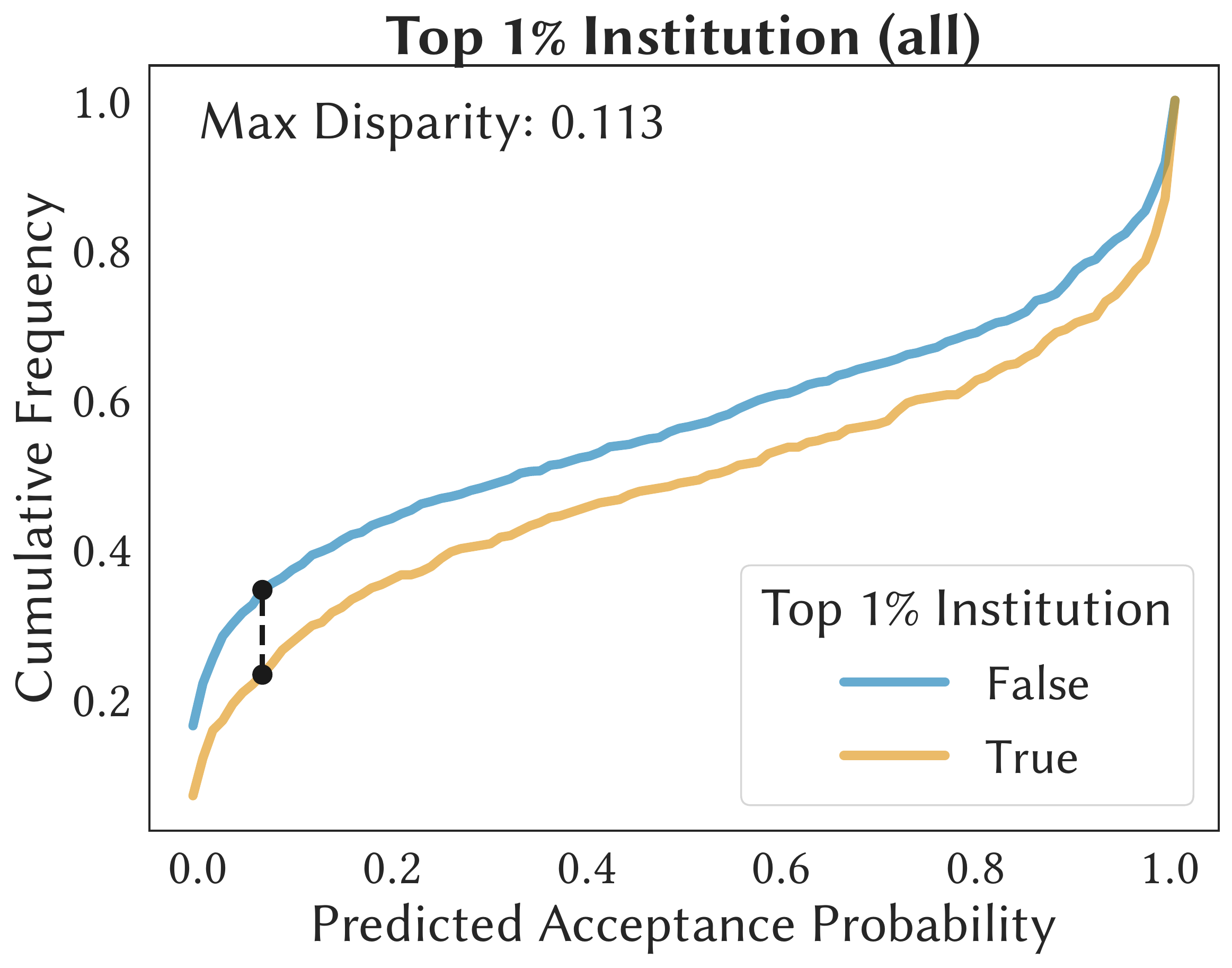}
\caption{\tiny Top Inst Author (all).}  \label{fig:dis:tinst:all}
	\end{subfigure}\\

        \caption{\textbf{Model disparity when including different sets of features.} Here the disparity is measured by the maximum
        disparity of the cumulative probability function of predicted acceptance probabilities over sensitive groups. We observe that in general, the inclusion of high-level textual features to the model helps to reduce this type of fairness disparity.}
        \label{fig:dis}
\end{figure*}

We use the submission data form 2017-2021 as the training set and data
in 2022 as the test set to fit a logistic regression model using the combination
of the following sets of features:
(i) {\tt base}: submission features (including
textual features); (ii) {\tt +author}:
base features plus all author features;
(iii) {\tt +rev} base features plus review ratings and confidences; (iv) {\tt +revnlp}
base features plus all review features including sentiment and length.
Specifically, for submission features, we also
assign a cluster number $k\in[20]$ to each submission
by running spectral clustering using cosine
similarity on the specter embeddings.
We show the diagnosis curves in \Cref{fig:glm}
including the receiver operating
characteristics (ROC) curve and
the calibration curves of the models
trained on different sets of features when evaluating
on the test set.  In \Cref{tab:dp}, we tabulate several
disparity measures on models trained using different sets of features and in \Cref{fig:dis} we plot the distributional disparity
across the sensitive attributes. In what follows we discuss
in greater detail the results.

\paragraph{Decision stage is less noisy.}
The ROC curve together with its area under the curve
(AUC) capture the discrimination power of classifier. Notice that after review features are
added, the ROC curve is close to the $(0,1)$ diagonal and its AUC is around $0.95$.
In \Cref{fig:glm:cal}, we plot the calibration curve with true probabilities
and predicted probabilities. A well-calibrated
model should be centered around the main diagonal.
We found that although the model is simple, after
including review features, it can capture the most of the variations in the dataset and generalizes well. This implies that the biases and noises in the decision process are low given the review ratings.

\paragraph{Fairness violations differ.}
Together with DP, we also compute two other commonly
used fairness metrics, equalized odds difference (EO) and AUC difference (AUC). We tabulate
the violations on both {\tt +rev} and {\tt +revnlp} (marked by +R) in \Cref{tab:dp}. We note that the violations are relatively mild (mostly less than $0.1$). Together with \Cref{fig:glm},
we see that in the decision stage, a simple model can capture both the discriminatory power as well as
fairness constraints.
Nonetheless, the level of disparity varies across
sensitive attributes: geographical disparity
appears to be  generally higher than gender disparity.

\paragraph{Large language models help explain disparities.}
In \Cref{fig:glm}, we note that the ROC behavior 
of both {\tt +rev} and {\tt +revnlp} models are similarly good but the latter generally has smaller
fairness violations (columns marked with +R
in \Cref{tab:dp}). This implies that 
    \emph{textual features produced by large LMs are helpful in assessing fairness disparities,}
and thus should be incorporated into other related analyses.
The same observation can be drawn from \Cref{fig:dis},
where we plot the cumulative probability function of the predicted
acceptance probability across different sensitive attribute groups.
A fair classifier would not distinguish the two curves and the maximum disparity between the two curves is used to measure distributional disparity. We observe that the inclusion of textual
features ({\tt +revnelp}) often helps in reducing the disparity.

\subsection{Summary}
In this section we perform fairness analysis on models
trained to mimic the decision process in peer-review. 
Although fairness disparities often exist as a result of
the underlying data distribution being imbalanced, the inclusion
of high-level textual features generated by LMs can often help
to ameliorate such parities.


\begin{table}[t]
    \scriptsize
    \centering
    \resizebox{\columnwidth}{!}{%
	\begin{tabular}{lllllll}%
    \toprule
    Disparity & DP & DP (+R) & EO & EO (+R) & AUC & AUC (+R) \\
    \midrule
    Geographical & 0.050 & 0.050 & 0.025 & 0.029 & 0.069 & 0.068 \\
    Gender & 0.022 & 0.014 & 0.108 & 0.087 & 0.121 & 0.119 \\
    Author Prestige & 0.107 & 0.101 & 0.031 & 0.024 & 0.138 & 0.138 \\
    \bottomrule
    \end{tabular}%
}
    \caption{\textbf{Fairness disparities in the decision stage prediction model.} We consider three fairness notions, DP, EO, and AUC, and compare whether adding LM features (marked by +R) from reviews reduces fairness violation. We found that most of the time, adding LM features helps reduce fairness disparities in the model.}
    \label{tab:dp}
\end{table}


\section{Towards Automatic Reviewing Systems}

 In this section, we discuss concrete
examples how our database has the potential of motivating more
challenging NLP tasks. Our focus lies in the review stage and
we aim at defining tasks that
facilitate the understanding of the review process, and more broadly, towards 
automatic machine reviewing systems.
We train all models
on NVIDIA RTX 2080 Ti GPUs with 10GB memory, except for the Longformer,
which requires 45GB memory to exploit its maximum token limit. All fine-tuning are completed within 12 hours.

\newcommand{\txtdata}[1]{{\tiny \texttt{#1}}}
\begin{table*}[t]
    \tiny\centering
	\begin{tabular}{lp{0.9\linewidth}}
    \toprule
    Submission ID & \texttt{figzpGMrdD}  \\
    LED Output (+) &%
    \begin{tabminted}{text}
     Summary:\n\nThis paper studies the continual learning problem in PLMs. The authors analyze the performance of different combinations of PLMs and CL methods on 3 benchmark datasets, including task-incremental learning and class-incremental learning.\n\nStrengths:\n\n1. The paper is well-written and easy to follow.\n\n2. The paper is well-motivated and easy to follow.\n\nWeaknesses:\n\n1. The paper does not provide a thorough analysis of the continual learning problem in PLMs.\n\n2. The paper does not provide a thorough analysis of the continual learning problem in PLMs.
     \end{tabminted}
     \\
    LED Output (-) &%
    \begin{tabminted}{text}
	This paper conducts a comparative study of the performance of different continual learning methods and CL methods. The authors conduct experiments on three continual learning settings, namely task-incremental learning, class-incremental learning, and class-incremental learning.\n\nThe paper is well written and easy to follow. However, I have several concerns:\n\n1. The paper is not well organized. For example, in section 3.1, the authors claim that "the continual learning problem with the PLM’s ability to retrain the inner representations, which leads to catastrophic forgetting". However, in section 3.2, the authors claim that
    \end{tabminted}
    \\
    Review Sample &%
    \begin{tabminted}{text}
This paper explores the continual learning performance when combining different PLMs and common continual learning methods with 3 challenging NLP classification tasks.\n\nTo benchmark these combinations the methods are evaluated in task-incremental and class-incremental learning settings over various NLP end-tasks, which covers common learning settings in continual learning and NLP...
\end{tabminted}
\\
    \bottomrule
    \end{tabular}
    \caption{\textbf{Samples of generated reviews using LED.}
    The LED model is fine-tuned on ICLR2017-2021 while the above samples
    are from ICLR2022. The symbols (+) and (-) signify whether the model is trained on positive or negative reviews. We note that the
    generated samples are on-topic (as compared with the real review sample), contains a summary of the paper, and also provides opinions, however, the model still struggles to general more consistent and congruent argument, highlighting the difficulty of this task.
    }
    \label{tab:res:review}
\end{table*}

\paragraph{Review Generation}

The first step of an automatic reviewing
system is generating reviews from
the submission. There has been a large literature
on using LMs for summarization tasks (\cite{tas2007survey});
the difference here is that in the review generation
task, the desirable output is the \emph{opinions} rather than summaries of the paper. 
To this end, we fine-tune Longformer Encoder-Decoder (LED) models for conditional generation tasks on submission and review samples. Generated reviews are tabulated in \Cref{tab:res:review}, where (+) and (-) signify whether
the model is trained on positive or negative samples. We note that the generated sample
contain not only summaries, but also opinions and comments of the paper.

\paragraph{Review Score Prediction}
    Directly predicting the quality of a submission
    is a daunting task even for human reviewers. To this end, we can construct a two-stage model to generate reviews and score the reviews.
    The predicted scores can then be used as a
    surrogate for the paper quality.
    We fine-tune RoBERTa and Longformer models
    on the 2017-2021 submissions and evaluate them on
    the 2022 submissions, which achive
    $0.881$ and $0.891$ F1 scores respectively.

\section{Discussions and Conclusions} \label{sec:conclusions}

We assemble a comprehensive data of peer-reviews
and postulate the study of fairness violations
using three commonly used fairness metrics.
We observe that in the decision
stage, such violations differ for sensitive attributes, and language model features help alleviate disparities in the decision stage models. Yet, we do not find compelling eveidence that such differences are significant. 
We also demonstrate the potential of our database to be used in benchmarking new and important NLP tasks, we provide baseline models for the example of automatic machine review tasks.


\paragraph{Limitations and Future Work.}
\textbf{(i) Move beyond association analysis:}
the fairness analysis we considered are by nature association analysis and we are not able to draw any causal conclusion. 
\textbf{(ii) Choice of statistical parity measures:} there are various type of statistical parity measures, and we believe there may be more interesting conclusions could be drawn for other measures besides those we chose.

\begin{acks}
J.Z.~and D.R.~conceived of the presented idea.
J.Z.~collected the database and performed analysis.
H.Z.~advised on natural language processing techniques
and Z.D.~advised on algorithmic fairness techniques.
H.Z.~and Z.D.~verified the methodology, results, and analysis.
All authors discussed the results and contributed to the writing and polishing of the manuscript. D.R.~supervised
and oversaw the whole project.

The authors would like to thank Bo Zhang
from the Fred Hutchinson Cancer Center
and Jennifer Sheffield from the University of Pennsylvania
for stimulating discussions and helpful suggestions.
This work is in part supported by the
\grantsponsor{DARPA}{US Defense Advanced Research Projects Agency (DARPA)}{https://www.darpa.mil/}
under Contract \grantnum{DARPA}{FA8750-19-2-1004}; 
the \grantsponsor{ODNI}{Oﬃce of the Director of National Intelligence (ODNI)}{https://www.dni.gov/}, \grantsponsor{IARPA}{Intelligence Advanced Research Projects Activity (IARPA)}{https://www.iarpa.gov/}, via IARPA Contract No.~\grantnum{IARPA}{2019-19051600006} under the BETTER Program;
\grantsponsor{ONR}{Office of Naval Research (ONR)}{https://www.nre.navy.mil/} Contract \grantnum{ONR}{N00014-19-1-2620}; 
\grantsponsor{NSF}{National Science Foundation (NSF)}{https://www.nsf.gov/} under
Contract \grantnum{NSF}{CCF-1934876}.
The views and conclusions contained herein are those of the authors and should not be interpreted as necessarily representing the oﬃcial policies, either expressed or implied, of ODNI, IARPA, the Department of Defense, or the U.S.~Government. The U.S.~Government is authorized to reproduce and distribute reprints for governmental purposes notwithstanding any copyright annotation therein.
\end{acks}

\bibliographystyle{ACM-Reference-Format}
\bibliography{bib/new.bib}


\begin{thebibliography}{24}


\ifx \showCODEN    \undefined \def \showCODEN     #1{\unskip}     \fi
\ifx \showDOI      \undefined \def \showDOI       #1{#1}\fi
\ifx \showISBNx    \undefined \def \showISBNx     #1{\unskip}     \fi
\ifx \showISBNxiii \undefined \def \showISBNxiii  #1{\unskip}     \fi
\ifx \showISSN     \undefined \def \showISSN      #1{\unskip}     \fi
\ifx \showLCCN     \undefined \def \showLCCN      #1{\unskip}     \fi
\ifx \shownote     \undefined \def \shownote      #1{#1}          \fi
\ifx \showarticletitle \undefined \def \showarticletitle #1{#1}   \fi
\ifx \showURL      \undefined \def \showURL       {\relax}        \fi
\providecommand\bibfield[2]{#2}
\providecommand\bibinfo[2]{#2}
\providecommand\natexlab[1]{#1}
\providecommand\showeprint[2][]{arXiv:#2}

\bibitem[Beltagy et~al\mbox{.}(2020)]%
        {Beltagy2020Longformer}
\bibfield{author}{\bibinfo{person}{Iz Beltagy}, \bibinfo{person}{Matthew~E.
  Peters}, {and} \bibinfo{person}{Arman Cohan}.}
  \bibinfo{year}{2020}\natexlab{}.
\newblock \showarticletitle{Longformer: The Long-Document Transformer}.
\newblock \bibinfo{journal}{\emph{arXiv:2004.05150}} (\bibinfo{year}{2020}).
\newblock


\bibitem[Burhanpurkar et~al\mbox{.}(2021)]%
        {burhanpurkar2021scaffolding}
\bibfield{author}{\bibinfo{person}{Maya Burhanpurkar}, \bibinfo{person}{Zhun
  Deng}, \bibinfo{person}{Cynthia Dwork}, {and} \bibinfo{person}{Linjun
  Zhang}.} \bibinfo{year}{2021}\natexlab{}.
\newblock \showarticletitle{Scaffolding sets}.
\newblock \bibinfo{journal}{\emph{arXiv preprint arXiv:2111.03135}}
  (\bibinfo{year}{2021}).
\newblock


\bibitem[Cohan et~al\mbox{.}(2020)]%
        {specter2020cohan}
\bibfield{author}{\bibinfo{person}{Arman Cohan}, \bibinfo{person}{Sergey
  Feldman}, \bibinfo{person}{Iz Beltagy}, \bibinfo{person}{Doug Downey}, {and}
  \bibinfo{person}{Daniel~S. Weld}.} \bibinfo{year}{2020}\natexlab{}.
\newblock \showarticletitle{{SPECTER: Document-level Representation Learning
  using Citation-informed Transformers}}. In \bibinfo{booktitle}{\emph{ACL}}.
\newblock


\bibitem[Cortes and Lawrence(2021)]%
        {nips2014}
\bibfield{author}{\bibinfo{person}{Corinna Cortes} {and}
  \bibinfo{person}{Neil~D. Lawrence}.} \bibinfo{year}{2021}\natexlab{}.
\newblock \bibinfo{title}{Inconsistency in Conference Peer Review: Revisiting
  the 2014 NeurIPS Experiment}.
\newblock
\newblock
\urldef\tempurl%
\url{https://doi.org/10.48550/ARXIV.2109.09774}
\showDOI{\tempurl}


\bibitem[Dwork et~al\mbox{.}(2012)]%
        {dwork2012fairness}
\bibfield{author}{\bibinfo{person}{Cynthia Dwork}, \bibinfo{person}{Moritz
  Hardt}, \bibinfo{person}{Toniann Pitassi}, \bibinfo{person}{Omer Reingold},
  {and} \bibinfo{person}{Richard Zemel}.} \bibinfo{year}{2012}\natexlab{}.
\newblock \showarticletitle{Fairness through awareness}. In
  \bibinfo{booktitle}{\emph{Proceedings of the 3rd innovations in theoretical
  computer science conference}}. \bibinfo{pages}{214--226}.
\newblock


\bibitem[Dycke et~al\mbox{.}(2022)]%
        {dycke2022yyy}
\bibfield{author}{\bibinfo{person}{Nils Dycke}, \bibinfo{person}{Ilia
  Kuznetsov}, {and} \bibinfo{person}{Iryna Gurevych}.}
  \bibinfo{year}{2022}\natexlab{}.
\newblock \bibinfo{title}{Yes-Yes-Yes: Donation-based Peer Reviewing Data
  Collection for ACL Rolling Review and Beyond}.
\newblock
\newblock
\showeprint[arxiv]{2201.11443}~[cs.CL]


\bibitem[Feldman et~al\mbox{.}(2015)]%
        {certify}
\bibfield{author}{\bibinfo{person}{Michael Feldman}, \bibinfo{person}{Sorelle~A
  Friedler}, \bibinfo{person}{John Moeller}, \bibinfo{person}{Carlos
  Scheidegger}, {and} \bibinfo{person}{Suresh Venkatasubramanian}.}
  \bibinfo{year}{2015}\natexlab{}.
\newblock \showarticletitle{Certifying and removing disparate impact}. In
  \bibinfo{booktitle}{\emph{proceedings of the 21th ACM SIGKDD international
  conference on knowledge discovery and data mining}}.
  \bibinfo{pages}{259--268}.
\newblock


\bibitem[Gao et~al\mbox{.}(2019)]%
        {DBLP:conf/naacl/0023EKGM19}
\bibfield{author}{\bibinfo{person}{Yang Gao}, \bibinfo{person}{Steffen Eger},
  \bibinfo{person}{Ilia Kuznetsov}, \bibinfo{person}{Iryna Gurevych}, {and}
  \bibinfo{person}{Yusuke Miyao}.} \bibinfo{year}{2019}\natexlab{}.
\newblock \showarticletitle{Does My Rebuttal Matter? Insights from a Major
  {NLP} Conference}. In \bibinfo{booktitle}{\emph{Proceedings of the 2019
  Conference of the North American Chapter of the Association for Computational
  Linguistics: Human Language Technologies, {NAACL-HLT} 2019, Minneapolis, MN,
  USA, June 2-7, 2019, Volume 1 (Long and Short Papers)}},
  \bibfield{editor}{\bibinfo{person}{Jill Burstein}, \bibinfo{person}{Christy
  Doran}, {and} \bibinfo{person}{Thamar Solorio}} (Eds.).
  \bibinfo{publisher}{Association for Computational Linguistics},
  \bibinfo{pages}{1274--1290}.
\newblock
\urldef\tempurl%
\url{https://doi.org/10.18653/v1/n19-1129}
\showURL{%
\tempurl}


\bibitem[Hardt et~al\mbox{.}(2016)]%
        {hardt2016equality}
\bibfield{author}{\bibinfo{person}{Moritz Hardt}, \bibinfo{person}{Eric Price},
  {and} \bibinfo{person}{Nati Srebro}.} \bibinfo{year}{2016}\natexlab{}.
\newblock \showarticletitle{Equality of opportunity in supervised learning}.
\newblock \bibinfo{journal}{\emph{Advances in neural information processing
  systems}}  \bibinfo{volume}{29} (\bibinfo{year}{2016}).
\newblock


\bibitem[Hua et~al\mbox{.}(2019)]%
        {DBLP:conf/naacl/HuaNBW19}
\bibfield{author}{\bibinfo{person}{Xinyu Hua}, \bibinfo{person}{Mitko Nikolov},
  \bibinfo{person}{Nikhil Badugu}, {and} \bibinfo{person}{Lu Wang}.}
  \bibinfo{year}{2019}\natexlab{}.
\newblock \showarticletitle{Argument Mining for Understanding Peer Reviews}. In
  \bibinfo{booktitle}{\emph{Proceedings of the 2019 Conference of the North
  American Chapter of the Association for Computational Linguistics: Human
  Language Technologies, {NAACL-HLT} 2019, Minneapolis, MN, USA, June 2-7,
  2019, Volume 1 (Long and Short Papers)}},
  \bibfield{editor}{\bibinfo{person}{Jill Burstein}, \bibinfo{person}{Christy
  Doran}, {and} \bibinfo{person}{Thamar Solorio}} (Eds.).
  \bibinfo{publisher}{Association for Computational Linguistics},
  \bibinfo{pages}{2131--2137}.
\newblock
\urldef\tempurl%
\url{https://doi.org/10.18653/v1/n19-1219}
\showURL{%
\tempurl}


\bibitem[Jecmen et~al\mbox{.}(2020)]%
        {DBLP:conf/nips/JecmenZLSCF20}
\bibfield{author}{\bibinfo{person}{Steven Jecmen}, \bibinfo{person}{Hanrui
  Zhang}, \bibinfo{person}{Ryan Liu}, \bibinfo{person}{Nihar~B. Shah},
  \bibinfo{person}{Vincent Conitzer}, {and} \bibinfo{person}{Fei Fang}.}
  \bibinfo{year}{2020}\natexlab{}.
\newblock \showarticletitle{Mitigating Manipulation in Peer Review via
  Randomized Reviewer Assignments}. In \bibinfo{booktitle}{\emph{Advances in
  Neural Information Processing Systems 33: Annual Conference on Neural
  Information Processing Systems 2020, NeurIPS 2020, December 6-12, 2020,
  virtual}}, \bibfield{editor}{\bibinfo{person}{Hugo Larochelle},
  \bibinfo{person}{Marc'Aurelio Ranzato}, \bibinfo{person}{Raia Hadsell},
  \bibinfo{person}{Maria{-}Florina Balcan}, {and} \bibinfo{person}{Hsuan{-}Tien
  Lin}} (Eds.).
\newblock
\urldef\tempurl%
\url{https://proceedings.neurips.cc/paper/2020/hash/93fb39474c51b8a82a68413e2a5ae17a-Abstract.html}
\showURL{%
\tempurl}


\bibitem[Kang et~al\mbox{.}(2018)]%
        {kang2018peerread}
\bibfield{author}{\bibinfo{person}{Dongyeop Kang}, \bibinfo{person}{Waleed
  Ammar}, \bibinfo{person}{Bhavana Dalvi}, \bibinfo{person}{Madeleine van
  Zuylen}, \bibinfo{person}{Sebastian Kohlmeier}, \bibinfo{person}{Eduard
  Hovy}, {and} \bibinfo{person}{Roy Schwartz}.}
  \bibinfo{year}{2018}\natexlab{}.
\newblock \showarticletitle{A Dataset of Peer Reviews ({P}eer{R}ead):
  Collection, Insights and {NLP} Applications}. In
  \bibinfo{booktitle}{\emph{Proceedings of the 2018 Conference of the North
  {A}merican Chapter of the Association for Computational Linguistics: Human
  Language Technologies, Volume 1 (Long Papers)}}.
  \bibinfo{publisher}{Association for Computational Linguistics},
  \bibinfo{address}{New Orleans, Louisiana}, \bibinfo{pages}{1647--1661}.
\newblock
\urldef\tempurl%
\url{https://doi.org/10.18653/v1/N18-1149}
\showDOI{\tempurl}


\bibitem[Levenshtein et~al\mbox{.}(1966)]%
        {levenshtein1966binary}
\bibfield{author}{\bibinfo{person}{Vladimir~I Levenshtein} {et~al\mbox{.}}}
  \bibinfo{year}{1966}\natexlab{}.
\newblock \showarticletitle{Binary codes capable of correcting deletions,
  insertions, and reversals}. In \bibinfo{booktitle}{\emph{Soviet physics
  doklady}}, Vol.~\bibinfo{volume}{10}. Soviet Union,
  \bibinfo{pages}{707--710}.
\newblock


\bibitem[Manzoor and Shah(2021)]%
        {DBLP:conf/aaai/ManzoorS21}
\bibfield{author}{\bibinfo{person}{Emaad~A. Manzoor} {and}
  \bibinfo{person}{Nihar~B. Shah}.} \bibinfo{year}{2021}\natexlab{}.
\newblock \showarticletitle{Uncovering Latent Biases in Text: Method and
  Application to Peer Review}. In \bibinfo{booktitle}{\emph{Thirty-Fifth {AAAI}
  Conference on Artificial Intelligence, {AAAI} 2021, Thirty-Third Conference
  on Innovative Applications of Artificial Intelligence, {IAAI} 2021, The
  Eleventh Symposium on Educational Advances in Artificial Intelligence, {EAAI}
  2021, Virtual Event, February 2-9, 2021}}. \bibinfo{publisher}{{AAAI} Press},
  \bibinfo{pages}{4767--4775}.
\newblock
\urldef\tempurl%
\url{https://ojs.aaai.org/index.php/AAAI/article/view/16608}
\showURL{%
\tempurl}


\bibitem[Plank and van Dalen(2019)]%
        {DBLP:conf/sigir/PlankD19}
\bibfield{author}{\bibinfo{person}{Barbara Plank} {and}
  \bibinfo{person}{Reinard van Dalen}.} \bibinfo{year}{2019}\natexlab{}.
\newblock \showarticletitle{CiteTracked: {A} Longitudinal Dataset of Peer
  Reviews and Citations}. In \bibinfo{booktitle}{\emph{Proceedings of the 4th
  Joint Workshop on Bibliometric-enhanced Information Retrieval and Natural
  Language Processing for Digital Libraries {(BIRNDL} 2019) co-located with the
  42nd International {ACM} {SIGIR} Conference on Research and Development in
  Information Retrieval {(SIGIR} 2019), Paris, France, July 25, 2019}}
  \emph{(\bibinfo{series}{{CEUR} Workshop Proceedings},
  Vol.~\bibinfo{volume}{2414})}, \bibfield{editor}{\bibinfo{person}{Muthu~Kumar
  Chandrasekaran} {and} \bibinfo{person}{Philipp Mayr}} (Eds.).
  \bibinfo{publisher}{CEUR-WS.org}, \bibinfo{pages}{116--122}.
\newblock
\urldef\tempurl%
\url{http://ceur-ws.org/Vol-2414/paper12.pdf}
\showURL{%
\tempurl}


\bibitem[Rogers and Augenstein(2020)]%
        {DBLP:conf/emnlp/RogersA20}
\bibfield{author}{\bibinfo{person}{Anna Rogers} {and} \bibinfo{person}{Isabelle
  Augenstein}.} \bibinfo{year}{2020}\natexlab{}.
\newblock \showarticletitle{What Can We Do to Improve Peer Review in NLP?}. In
  \bibinfo{booktitle}{\emph{Findings of the Association for Computational
  Linguistics: {EMNLP} 2020, Online Event, 16-20 November 2020}}
  \emph{(\bibinfo{series}{Findings of {ACL}}, Vol.~\bibinfo{volume}{{EMNLP}
  2020})}, \bibfield{editor}{\bibinfo{person}{Trevor Cohn},
  \bibinfo{person}{Yulan He}, {and} \bibinfo{person}{Yang Liu}} (Eds.).
  \bibinfo{publisher}{Association for Computational Linguistics},
  \bibinfo{pages}{1256--1262}.
\newblock
\urldef\tempurl%
\url{https://doi.org/10.18653/v1/2020.findings-emnlp.112}
\showURL{%
\tempurl}


\bibitem[Stelmakh et~al\mbox{.}(2019)]%
        {DBLP:conf/nips/StelmakhSS19}
\bibfield{author}{\bibinfo{person}{Ivan Stelmakh}, \bibinfo{person}{Nihar~B.
  Shah}, {and} \bibinfo{person}{Aarti Singh}.} \bibinfo{year}{2019}\natexlab{}.
\newblock \showarticletitle{On Testing for Biases in Peer Review}. In
  \bibinfo{booktitle}{\emph{Advances in Neural Information Processing Systems
  32: Annual Conference on Neural Information Processing Systems 2019, NeurIPS
  2019, December 8-14, 2019, Vancouver, BC, Canada}},
  \bibfield{editor}{\bibinfo{person}{Hanna~M. Wallach}, \bibinfo{person}{Hugo
  Larochelle}, \bibinfo{person}{Alina Beygelzimer}, \bibinfo{person}{Florence
  d'Alch{\'{e}}{-}Buc}, \bibinfo{person}{Emily~B. Fox}, {and}
  \bibinfo{person}{Roman Garnett}} (Eds.). \bibinfo{pages}{5287--5297}.
\newblock
\urldef\tempurl%
\url{https://proceedings.neurips.cc/paper/2019/hash/d3d80b656929a5bc0fa34381bf42fbdd-Abstract.html}
\showURL{%
\tempurl}


\bibitem[Tas and Kiyani(2007)]%
        {tas2007survey}
\bibfield{author}{\bibinfo{person}{Oguzhan Tas} {and} \bibinfo{person}{Farzad
  Kiyani}.} \bibinfo{year}{2007}\natexlab{}.
\newblock \showarticletitle{A survey automatic text summarization}.
\newblock \bibinfo{journal}{\emph{PressAcademia Procedia}} \bibinfo{volume}{5},
  \bibinfo{number}{1} (\bibinfo{year}{2007}), \bibinfo{pages}{205--213}.
\newblock


\bibitem[Tran et~al\mbox{.}(2021)]%
        {tran2021an}
\bibfield{author}{\bibinfo{person}{David Tran}, \bibinfo{person}{Alexander~V
  Valtchanov}, \bibinfo{person}{Keshav~R Ganapathy}, \bibinfo{person}{Raymond
  Feng}, \bibinfo{person}{Eric~Victor Slud}, \bibinfo{person}{Micah Goldblum},
  {and} \bibinfo{person}{Tom Goldstein}.} \bibinfo{year}{2021}\natexlab{}.
\newblock \bibinfo{title}{An Open Review of OpenReview: A Critical Analysis of
  the Machine Learning Conference Review Process}.
\newblock
\newblock
\urldef\tempurl%
\url{https://openreview.net/forum?id=Cn706AbJaKW}
\showURL{%
\tempurl}


\bibitem[Tsai et~al\mbox{.}(2016)]%
        {tsai2016cross}
\bibfield{author}{\bibinfo{person}{Chen-Tse Tsai}, \bibinfo{person}{Stephen
  Mayhew}, {and} \bibinfo{person}{Dan Roth}.} \bibinfo{year}{2016}\natexlab{}.
\newblock \showarticletitle{Cross-Lingual Named Entity Recognition via
  Wikification}. In \bibinfo{booktitle}{\emph{Proceedings of The 20th {SIGNLL}
  Conference on Computational Natural Language Learning}}.
  \bibinfo{publisher}{Association for Computational Linguistics},
  \bibinfo{address}{Berlin, Germany}, \bibinfo{pages}{219--228}.
\newblock
\urldef\tempurl%
\url{https://doi.org/10.18653/v1/K16-1022}
\showDOI{\tempurl}


\bibitem[Tukey et~al\mbox{.}(1977)]%
        {tukey1977exploratory}
\bibfield{author}{\bibinfo{person}{John~W Tukey} {et~al\mbox{.}}}
  \bibinfo{year}{1977}\natexlab{}.
\newblock \bibinfo{booktitle}{\emph{Exploratory data analysis}}.
  Vol.~\bibinfo{volume}{2}.
\newblock \bibinfo{publisher}{Reading, MA}.
\newblock


\bibitem[Van~der Maaten and Hinton(2008)]%
        {van2008visualizing}
\bibfield{author}{\bibinfo{person}{Laurens Van~der Maaten} {and}
  \bibinfo{person}{Geoffrey Hinton}.} \bibinfo{year}{2008}\natexlab{}.
\newblock \showarticletitle{Visualizing data using t-SNE.}
\newblock \bibinfo{journal}{\emph{Journal of machine learning research}}
  \bibinfo{volume}{9}, \bibinfo{number}{11} (\bibinfo{year}{2008}).
\newblock


\bibitem[Wang et~al\mbox{.}(2020)]%
        {DBLP:conf/inlg/WangZHKJR20}
\bibfield{author}{\bibinfo{person}{Qingyun Wang}, \bibinfo{person}{Qi Zeng},
  \bibinfo{person}{Lifu Huang}, \bibinfo{person}{Kevin Knight},
  \bibinfo{person}{Heng Ji}, {and} \bibinfo{person}{Nazneen~Fatema Rajani}.}
  \bibinfo{year}{2020}\natexlab{}.
\newblock \showarticletitle{ReviewRobot: Explainable Paper Review Generation
  based on Knowledge Synthesis}. In \bibinfo{booktitle}{\emph{Proceedings of
  the 13th International Conference on Natural Language Generation, {INLG}
  2020, Dublin, Ireland, December 15-18, 2020}},
  \bibfield{editor}{\bibinfo{person}{Brian Davis}, \bibinfo{person}{Yvette
  Graham}, \bibinfo{person}{John~D. Kelleher}, {and} \bibinfo{person}{Yaji
  Sripada}} (Eds.). \bibinfo{publisher}{Association for Computational
  Linguistics}, \bibinfo{pages}{384--397}.
\newblock
\urldef\tempurl%
\url{https://aclanthology.org/2020.inlg-1.44/}
\showURL{%
\tempurl}


\bibitem[Yuan et~al\mbox{.}(2021)]%
        {DBLP:journals/corr/abs-2102-00176}
\bibfield{author}{\bibinfo{person}{Weizhe Yuan}, \bibinfo{person}{Pengfei Liu},
  {and} \bibinfo{person}{Graham Neubig}.} \bibinfo{year}{2021}\natexlab{}.
\newblock \showarticletitle{Can We Automate Scientific Reviewing?}
\newblock \bibinfo{journal}{\emph{CoRR}}  \bibinfo{volume}{abs/2102.00176}
  (\bibinfo{year}{2021}).
\newblock
\urldef\tempurl%
\url{https://arxiv.org/abs/2102.00176}
\showURL{%
\tempurl}


\end{thebibliography}

\appendix
\section{Database Details} \label{sec:app:data}
\begin{figure*}[t]
	    \centering
	    \includegraphics[width=\linewidth]{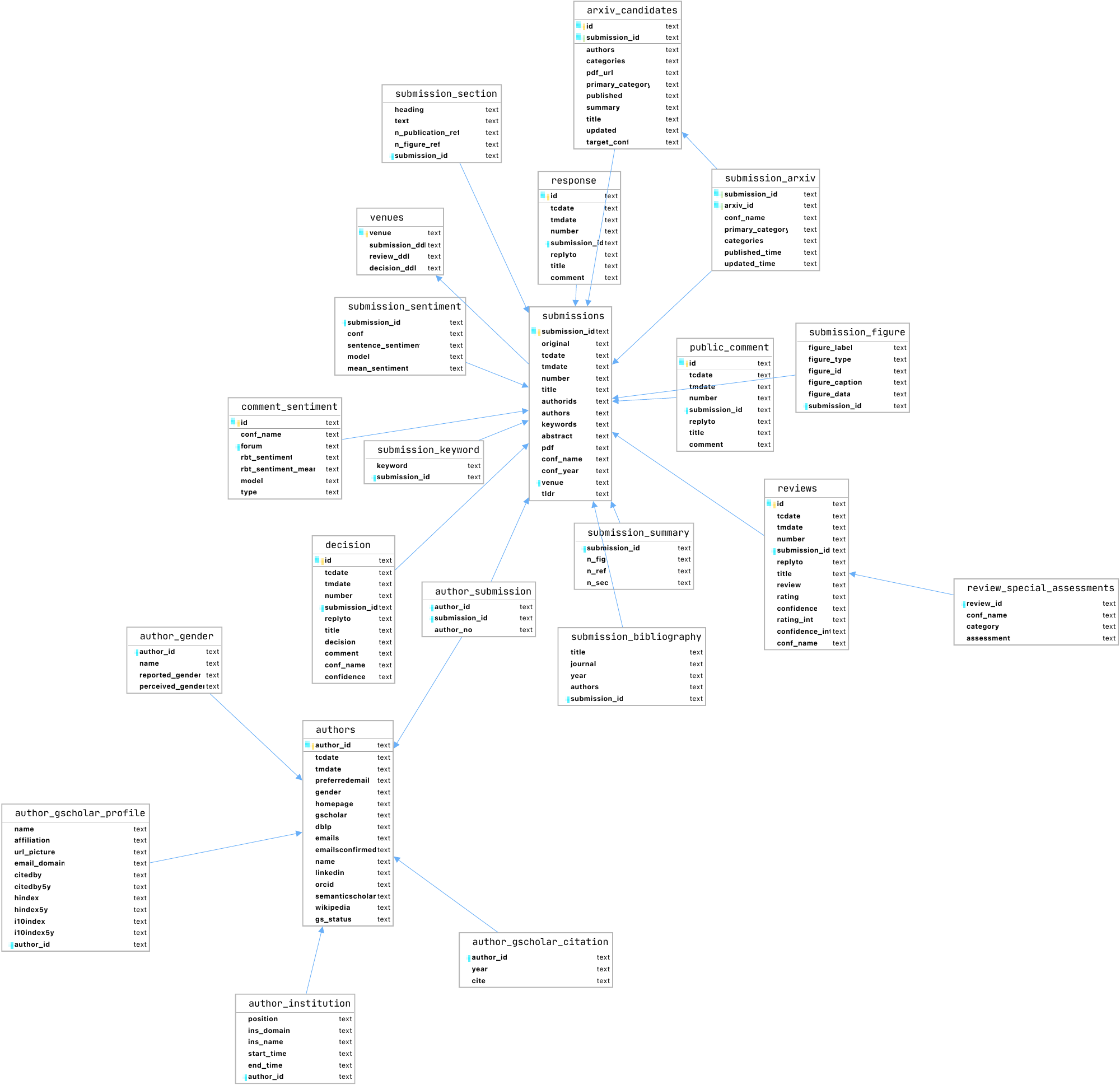}
    \caption{\textbf{ER diagram of the database (CSRanking related entities are omitted).}}
     \label{fig:er:data}
\end{figure*}

The summary of all covariates of the database is
given in \Cref{tab:data:tableone}; we also
include the entity-relation (ER) diagram in \Cref{fig:er:data}.
All collected data are open data, which are granted access by the owner.

\begin{table*}
    \centering
\resizebox{\textwidth}{!}{%
\begin{tabular}{lllllllllll}
\toprule
                           &      & \multicolumn{9}{l}{\bf Conference Year} \\
                           &      &              Missing &            Overall &               2017 &               2018 &               2019 &               2020 &               2021 &               2022 & P-Value \\
\midrule
\midrule
\multicolumn{11}{l}{\bf Submission Covariates} \\
\midrule
n & {} &                      &              10297 &                490 &                910 &               1419 &               2213 &               2595 &               2670 &         \\
decision, n (\%) & Accept (Oral) &                    0 &          169 (1.6) &           15 (3.1) &           23 (2.5) &           24 (1.7) &                    &           53 (2.0) &           54 (2.0) &  <0.001 \\
                           & Accept (Poster) &                      &        3063 (29.7) &         183 (37.3) &         313 (34.4) &         478 (33.7) &         531 (24.0) &         693 (26.7) &         865 (32.4) &         \\
                           & Accept (Spotlight) &                      &          398 (3.9) &                    &                    &                    &          108 (4.9) &          114 (4.4) &          176 (6.6) &         \\
                           & Accept (Talk) &                      &           48 (0.5) &                    &                    &                    &           48 (2.2) &                    &                    &         \\
                           & Invite to Workshop Track &                      &          136 (1.3) &           47 (9.6) &           89 (9.8) &                    &                    &                    &                    &         \\
                           & Reject &                      &        6483 (63.0) &         245 (50.0) &         485 (53.3) &         917 (64.6) &        1526 (69.0) &        1735 (66.9) &        1575 (59.0) &         \\
input\_len, mean (SD) &      &                   62 &   10077.0 (6857.5) &    7304.6 (2810.6) &    8183.9 (3996.5) &    9247.6 (6404.0) &    9483.9 (5411.8) &   10432.7 (7730.6) &   11827.2 (8001.7) &  <0.001 \\
n\_review, mean (SD) &      &                    0 &          3.4 (0.7) &          3.0 (0.5) &          3.0 (0.2) &          3.0 (0.3) &          3.0 (0.4) &          3.9 (0.5) &          3.8 (0.7) &  <0.001 \\
rating\_avg, mean (SD) &      &                    0 &          5.3 (1.4) &          5.7 (1.3) &          5.4 (1.2) &          5.4 (1.2) &          4.4 (1.7) &          5.5 (1.0) &          5.5 (1.3) &  <0.001 \\
rating\_max, mean (SD) &      &                    0 &          6.4 (1.5) &          6.5 (1.5) &          6.4 (1.4) &          6.3 (1.3) &          5.7 (2.0) &          6.6 (1.2) &          6.7 (1.4) &  <0.001 \\
rating\_min, mean (SD) &      &                    0 &          4.2 (1.6) &          4.9 (1.4) &          4.5 (1.4) &          4.6 (1.3) &          3.1 (1.9) &          4.4 (1.2) &          4.3 (1.6) &  <0.001 \\
confidence\_avg, mean (SD) &      &                    0 &          3.6 (0.6) &          3.8 (0.5) &          3.8 (0.5) &          3.8 (0.5) &          3.1 (0.8) &          3.7 (0.5) &          3.7 (0.5) &  <0.001 \\
confidence\_max, mean (SD) &      &                    0 &          4.3 (0.7) &          4.3 (0.6) &          4.4 (0.6) &          4.4 (0.6) &          4.1 (1.0) &          4.4 (0.6) &          4.3 (0.6) &  <0.001 \\
confidence\_min, mean (SD) &      &                    0 &          2.8 (0.9) &          3.3 (0.7) &          3.2 (0.8) &          3.1 (0.8) &          2.1 (1.0) &          3.0 (0.7) &          3.0 (0.7) &  <0.001 \\
sentiment\_avg, mean (SD) &      &                    0 &          0.6 (0.1) &          0.6 (0.1) &          0.6 (0.1) &          0.6 (0.1) &          0.6 (0.1) &          0.6 (0.1) &          0.6 (0.1) &  <0.001 \\
sentiment\_max, mean (SD) &      &                    0 &          0.7 (0.1) &          0.7 (0.1) &          0.7 (0.1) &          0.7 (0.1) &          0.7 (0.1) &          0.7 (0.1) &          0.7 (0.1) &  <0.001 \\
sentiment\_min, mean (SD) &      &                    0 &          0.5 (0.1) &          0.5 (0.1) &          0.5 (0.1) &          0.5 (0.1) &          0.5 (0.1) &          0.5 (0.1) &          0.5 (0.1) &  <0.001 \\
rlen\_avg, mean (SD) &      &                   62 &      585.5 (220.4) &      393.8 (161.2) &      504.4 (188.5) &      552.6 (214.2) &      559.8 (219.6) &      653.4 (222.2) &      620.9 (205.8) &  <0.001 \\
rlen\_max, mean (SD) &      &                   62 &      918.5 (468.8) &      591.7 (287.3) &      768.4 (365.2) &      846.6 (431.7) &      851.9 (442.9) &     1049.6 (517.0) &      995.4 (457.6) &  <0.001 \\
rlen\_min, mean (SD) &      &                   62 &      325.3 (144.0) &      225.4 (121.2) &      281.0 (129.9) &      310.4 (154.2) &      320.6 (143.0) &      357.6 (142.9) &      338.9 (135.1) &  <0.001 \\
sub\_fluency, mean (SD) &      &                   58 &          0.8 (0.0) &          0.9 (0.0) &          0.9 (0.0) &          0.9 (0.0) &          0.9 (0.0) &          0.8 (0.0) &          0.9 (0.0) &  <0.001 \\
n\_fig, mean (SD) &      &                   62 &         12.5 (7.3) &          9.2 (4.9) &          9.5 (4.9) &         11.1 (6.1) &         11.8 (6.4) &         13.3 (7.5) &         14.8 (8.5) &  <0.001 \\
n\_ref, mean (SD) &      &                   62 &        40.3 (17.0) &        30.0 (11.5) &        32.2 (12.7) &        35.6 (14.4) &        38.1 (14.7) &        43.3 (17.5) &        46.3 (18.5) &  <0.001 \\
n\_sec, mean (SD) &      &                   62 &         19.0 (7.2) &         16.2 (5.6) &         16.8 (6.0) &         17.8 (6.3) &         18.3 (6.6) &         19.3 (7.4) &         21.4 (7.9) &  <0.001 \\
arxiv\_first, n (\%) & False &                    0 &        8704 (84.5) &        490 (100.0) &         889 (97.7) &        1359 (95.8) &        1675 (75.7) &        2062 (79.5) &        2229 (83.5) &  <0.001 \\
                           & True &                      &        1593 (15.5) &                    &           21 (2.3) &           60 (4.2) &         538 (24.3) &         533 (20.5) &         441 (16.5) &         \\
n\_author, mean (SD) &      &                    0 &          4.1 (2.0) &          3.7 (1.7) &          3.8 (1.8) &          4.0 (1.9) &          4.1 (1.9) &          4.2 (2.0) &          4.4 (2.1) &  <0.001 \\
any\_reported\_f, n (\%) & False &                    0 &        7547 (73.3) &         405 (82.7) &         726 (79.8) &        1084 (76.4) &        1631 (73.7) &        1859 (71.6) &        1842 (69.0) &  <0.001 \\
                           & True &                      &        2750 (26.7) &          85 (17.3) &         184 (20.2) &         335 (23.6) &         582 (26.3) &         736 (28.4) &         828 (31.0) &         \\
cnt\_reported\_f, mean (SD) &      &                    0 &          0.3 (0.6) &          0.2 (0.5) &          0.2 (0.5) &          0.3 (0.5) &          0.3 (0.6) &          0.3 (0.6) &          0.4 (0.7) &  <0.001 \\
any\_perceived\_f, n (\%) & False &                    0 &        8251 (80.1) &         409 (83.5) &         751 (82.5) &        1160 (81.7) &        1767 (79.8) &        2041 (78.7) &        2123 (79.5) &   0.020 \\
                           & True &                      &        2046 (19.9) &          81 (16.5) &         159 (17.5) &         259 (18.3) &         446 (20.2) &         554 (21.3) &         547 (20.5) &         \\
cnt\_perceived\_f, mean (SD) &      &                    0 &          0.2 (0.5) &          0.2 (0.4) &          0.2 (0.4) &          0.2 (0.5) &          0.2 (0.5) &          0.2 (0.5) &          0.3 (0.6) &  <0.001 \\
demo\_no\_us, n (\%) & False &                    0 &        6191 (60.1) &         312 (63.7) &         586 (64.4) &         910 (64.1) &        1336 (60.4) &        1525 (58.8) &        1522 (57.0) &  <0.001 \\
                           & True &                      &        4106 (39.9) &         178 (36.3) &         324 (35.6) &         509 (35.9) &         877 (39.6) &        1070 (41.2) &        1148 (43.0) &         \\
ins\_rank\_max, mean (SD) &      &                  855 &      118.8 (133.6) &        37.1 (34.6) &        61.0 (60.9) &        83.5 (84.1) &      114.1 (118.0) &      131.0 (134.7) &      159.9 (169.4) &  <0.001 \\
ins\_rank\_avg, mean (SD) &      &                  855 &        73.3 (86.8) &        27.0 (27.5) &        42.1 (46.6) &        54.3 (59.8) &        71.7 (78.5) &        79.2 (85.6) &       95.5 (110.6) &  <0.001 \\
author\_cite\_max, mean (SD) &      &                  700 &  23020.8 (44860.0) &  16049.3 (26747.9) &  20957.1 (35040.7) &  21761.0 (41772.0) &  21904.6 (44486.5) &  27168.5 (49593.9) &  22208.8 (46597.3) &  <0.001 \\
author\_cite\_avg, mean (SD) &      &                  700 &   8692.5 (15223.8) &   7179.0 (10922.5) &   8551.9 (13674.7) &   8575.7 (14236.6) &   9111.5 (18693.2) &   8832.7 (13247.4) &   8581.9 (15584.7) &   0.267 \\
\midrule
\midrule
    \multicolumn{11}{l}{\bf Review Covariates} \\
\midrule
n & {} &                      &          35717 &           1489 &           2748 &           4332 &           6721 &          10026 &          10401 &         \\
rating\_int, mean (SD) &          &                    0 &      5.3 (1.8) &      5.7 (1.6) &      5.4 (1.5) &      5.4 (1.5) &      4.4 (2.2) &      5.5 (1.4) &      5.5 (1.7) &  <0.001 \\
confidence\_int, mean (SD) &          &                    0 &      3.6 (0.9) &      3.8 (0.8) &      3.8 (0.8) &      3.8 (0.8) &      3.1 (1.3) &      3.7 (0.8) &      3.7 (0.8) &  <0.001 \\
sentiment, mean (SD) &          &                    1 &      0.6 (0.1) &      0.6 (0.2) &      0.6 (0.2) &      0.6 (0.1) &      0.6 (0.1) &      0.6 (0.1) &      0.6 (0.1) &  <0.001 \\
review\_len, mean (SD) &          &                  222 &  592.5 (380.0) &  393.8 (253.5) &  504.7 (320.3) &  553.2 (364.4) &  559.7 (367.8) &  653.6 (409.0) &  622.7 (377.0) &  <0.001 \\
review\_type, n (\%) & Borderline &                    0 &   17213 (48.2) &     593 (39.8) &    1181 (43.0) &    1985 (45.8) &    2393 (35.6) &    4932 (49.2) &    6129 (58.9) &  <0.001 \\
                   & Negative &                      &   10649 (29.8) &     392 (26.3) &     825 (30.0) &    1232 (28.4) &    3486 (51.9) &    2533 (25.3) &    2181 (21.0) &         \\
                   & Positive &                      &    7855 (22.0) &     504 (33.8) &     742 (27.0) &    1115 (25.7) &     842 (12.5) &    2561 (25.5) &    2091 (20.1) &         \\
\bottomrule
\toprule
\multicolumn{11}{l}{\bf Author Covariates} \\
\midrule
n & {} &                 &             31780 &              1416 &              2703 &              4286 &              6807 &              7914 &              8654 &         \\
author\_no, mean (SD) &          &               0 &         2.9 (2.0) &         2.6 (1.7) &         2.7 (1.8) &         2.7 (1.8) &         2.8 (1.9) &         2.9 (2.0) &         3.1 (2.4) &  <0.001 \\
reported\_gender, n (\%) & Female &               0 &        2470 (7.8) &          81 (5.7) &         162 (6.0) &         298 (7.0) &         503 (7.4) &         656 (8.3) &         770 (8.9) &  <0.001 \\
                            & Male &                 &      19099 (60.1) &        769 (54.3) &       1565 (57.9) &       2527 (59.0) &       3951 (58.0) &       4763 (60.2) &       5524 (63.8) &         \\
                            & Non-Binary &                 &          18 (0.1) &           1 (0.1) &           2 (0.1) &           2 (0.0) &           2 (0.0) &           5 (0.1) &           6 (0.1) &         \\
                            & Unspecified &                 &      10193 (32.1) &        565 (39.9) &        974 (36.0) &       1459 (34.0) &       2351 (34.5) &       2490 (31.5) &       2354 (27.2) &         \\
perceived\_gender, mean (SD) &          &           10077 &         0.9 (0.3) &         0.9 (0.3) &         0.9 (0.3) &         0.9 (0.3) &         0.9 (0.3) &         0.9 (0.3) &         0.9 (0.3) &  <0.001 \\
country, n (\%) & Australia &            9998 &         318 (1.5) &                   &          18 (1.1) &          42 (1.5) &          72 (1.6) &          93 (1.7) &          93 (1.4) &  <0.001 \\
                            & Austria &                 &          72 (0.3) &           6 (0.8) &          10 (0.6) &           4 (0.1) &           8 (0.2) &          23 (0.4) &          21 (0.3) &         \\
                            & Bangladesh &                 &           1 (0.0) &                   &                   &                   &                   &           1 (0.0) &                   &         \\
                            & Belgium &                 &           9 (0.0) &                   &           1 (0.1) &           3 (0.1) &                   &                   &           5 (0.1) &         \\
                            & Brazil &                 &           2 (0.0) &                   &           1 (0.1) &                   &                   &                   &           1 (0.0) &         \\
                            & Canada &                 &        1111 (5.1) &          55 (6.9) &         111 (6.6) &         167 (5.9) &         215 (4.8) &         258 (4.7) &         305 (4.7) &         \\
                            & Chile &                 &           7 (0.0) &                   &                   &           1 (0.0) &           5 (0.1) &                   &           1 (0.0) &         \\
                            & China &                 &       2443 (11.2) &          10 (1.3) &          75 (4.5) &         196 (6.9) &        479 (10.8) &        655 (11.8) &       1028 (15.8) &         \\
                            & \vdots &                 &       \vdots &          \vdots &          \vdots &         \vdots &        \vdots &        \vdots &       \vdots &         \\
                            & United Kingdom &                 &        1290 (5.9) &          60 (7.6) &         100 (6.0) &         178 (6.3) &         262 (5.9) &         329 (5.9) &         361 (5.5) &         \\
                            & United States &                 &      12465 (57.2) &        590 (74.3) &       1116 (66.8) &       1750 (61.9) &       2562 (57.8) &       3102 (56.0) &       3345 (51.3) &         \\
                            & Viet Nam &                 &           6 (0.0) &                   &                   &                   &                   &           1 (0.0) &           5 (0.1) &         \\
ins\_cum\_cnt, mean (SD) &          &            7717 &      77.3 (124.7) &       13.9 (16.4) &       31.1 (43.6) &       51.1 (72.3) &      69.3 (106.6) &      87.4 (131.5) &     104.1 (155.7) &  <0.001 \\
ins\_per\_year\_rank, mean (SD) &          &            7717 &      77.4 (110.6) &       27.6 (31.3) &       42.4 (53.8) &       55.8 (73.1) &      76.1 (100.6) &      79.6 (106.7) &      99.8 (139.8) &  <0.001 \\
ins\_per\_year\_rank\_pct, mean (SD) &          &            7717 &         0.2 (0.2) &         0.2 (0.2) &         0.2 (0.2) &         0.2 (0.2) &         0.2 (0.2) &         0.2 (0.2) &         0.2 (0.2) &  <0.001 \\
author\_year\_citation, mean (SD) &          &           11503 &  6280.0 (18896.5) &  5212.4 (14413.0) &  5952.3 (16494.4) &  5475.2 (15299.4) &  5986.7 (17614.4) &  7146.3 (21546.8) &  6016.7 (18943.9) &  <0.001 \\
author\_year\_rank, mean (SD) &          &           11503 &         0.5 (0.3) &         0.5 (0.3) &         0.5 (0.3) &         0.5 (0.3) &         0.5 (0.3) &         0.5 (0.3) &         0.5 (0.3) &   0.866 \\
\bottomrule
\end{tabular}
}
    \caption{Summary of the covariates (CSRanking data are omitted).}
    \label{tab:data:tableone}
\end{table*}


\end{document}